\documentclass[12pt,preprint]{aastex}
\usepackage{lscape}
\usepackage{amsmath}
\usepackage{color}

\newcommand{\Fermi}{\textit{Fermi}}
\newcommand{\Swift}{\textit{Swift}}
\newcommand{\RXTE}{\textit{RXTE}}
\newcommand{\ASCA}{\textit{ASCA}}
\newcommand{\Konus}{\textit{Konus-Wind}}
\newcommand{\Suzaku}{\textit{Suzaku}}
\newcommand{\Chandra}{\textit{Chandra}}

\newcommand{\unit}[1]{\ensuremath{\, \mathrm{#1}}}
\newcommand{\scinum}[2]{$#1\times10^{#2}$}

\newcommand{\valerr}[2]{$#1 \pm #2$}

\newcommand{\Tnin}{$T_{\!\unit{90}}$ }
\newcommand{\Tninnos}{$T_{\!\unit{90}}$}
\newcommand{\Tfif}{$T_{\!\unit{50}}$ }
\newcommand{\Tfifnos}{$T_{\!\unit{50}}$}

\newcommand{\RMFITnos}{\textit{RMFIT}}
\newcommand{\Ep}{$E_{\unit{peak}}$ }
\newcommand{\Epnos}{$E_{\unit{peak}}$}

\newcommand{\Jfifteen}{SGR\,J1550$-$5418 }
\newcommand{\Jfifteennos}{SGR\,J1550$-$5418}

\newcommand{\Jzerofive}{SGR\,J0501$+$4516 }
\newcommand{\Jzerofivenos}{SGR\,J0501$+$4516}

\newcommand{\Jzerofour}{SGR\,J0418$+$5729 }
\newcommand{\Jzerofournos}{SGR\,J0418$+$5729}

\newcommand{\SGReighteen}{SGR\,1806$-$20 }
\newcommand{\SGReighteennos}{SGR\,1806$-$20}

\newcommand{\Jeighteen}{SGR\,J1822.3$-$1606 }
\newcommand{\Jeighteennos}{SGR\,J1822.3$-$1606}

\newcommand{\oneEeight}{1E\,1841$-$045 }
\newcommand{\oneEeightnos}{1E\,1841$-$045 }

\newcommand{\AXPfouru}{AXP\,4U 0142$+$61 }
\newcommand{\AXPfourunos}{AXP\,4U 0142$+$61}

\newcommand{\SGRoneE}{AXP\,1E 2259$+$586 }
\newcommand{\SGRoneEnos}{AXP\,1E 2259$+$586}

\newcommand{\SGRMCM}{SGR\,1900$+$14 }
\newcommand{\SGRMDC}{SGR\,1627$-$41 }

\begin{document}

\title{The Five Year \Fermi/GBM Magnetar Burst Catalog}
\author{
A.~C.~Collazzi\altaffilmark{1}, 
C.~Kouveliotou\altaffilmark{2,3}, 
A.~J.~van~der~Horst\altaffilmark{2}, 
G.~A.~Younes\altaffilmark{4,2}, 
Y.~Kaneko\altaffilmark{5}, 
E.~G\"o\u{g}\"u\c{s}\altaffilmark{5}, 
L.~Lin\altaffilmark{6}, 
J.~Granot\altaffilmark{7}, 
M.~H.~Finger\altaffilmark{4}, 
V.~L.~Chaplin\altaffilmark{8}, 
D.~Huppenkothen\altaffilmark{9,10}, 
A.~L.~Watts\altaffilmark{11}, 
A.~von~Kienlin\altaffilmark{12}, 
M.~G.~Baring\altaffilmark{13}, 
D.~Gruber\altaffilmark{14}, 
P.~N.~Bhat\altaffilmark{15}, 
M.~H.~Gibby\altaffilmark{16}, 
N.~Gehrels\altaffilmark{17}, 
J.~McEnery\altaffilmark{17}, 
M.~van~der~Klis\altaffilmark{11}, 
R.~A.~M.~J.~Wijers\altaffilmark{11}
}
\altaffiltext{1}{SciTec, Inc., 100 Wall Street, Princeton, NJ 08540, USA, acollazzi@scitec.com}
\altaffiltext{2}{Department of Physics, The George Washington University, 725 21st Street NW, Washington, DC 20052, USA}
\altaffiltext{3}{Space Science Office, ZP12, NASA/Marshall Space Flight Center, Huntsville, AL 35812, USA}
\altaffiltext{4}{Universities Space Research Association, NSSTC, 320 Sparkman Drive, Huntsville, AL 35805, USA}
\altaffiltext{5}{Sabanc\i~University, Orhanl\i-Tuzla, \.Istanbul 34956, Turkey}
\altaffiltext{6}{Fran\c{c}ois Arago Centre, APC, 10 rue Alice Domon et L\'{e}onie Duquet, F-75205 Paris, France}
\altaffiltext{7}{Department of Natural Sciences, The Open University of Israel, 1 University Road, P.O. Box 808, RaÕanana 43537, Israel}
\altaffiltext{8}{School of Medicine, Vanderbilt University, 1161 21st Ave S, Nashville, TN 37232, USA}
\altaffiltext{9}{Center for Data Science, New York University, 726 Broadway, 7th Floor, New York, NY 10003}
\altaffiltext{10}{Center for Cosmology and Particle Physics, Department of Physics, New York University, 4 Washington Place, New York, NY 10003}
\altaffiltext{11}{Anton Pannekoek Institute, University of Amsterdam, Postbus 94249, 1090 GE Amsterdam, The Netherlands}
\altaffiltext{12}{Max-Planck-Institut f\"{u}r extraterrestrische Physik, Giessenbachstrasse 1, 85748 Garching, Germany}
\altaffiltext{13}{Department of Physics and Astronomy, Rice University, MS-108, P.O. Box 1892, Houston, TX 77251, USA}
\altaffiltext{14}{Planetarium S\"{u}dtirol, Gummer 5, 39053 Karneid, Italy}
\altaffiltext{15}{CSPAR, University of Alabama in Huntsville, 320 Sparkman Dr., Huntsville, AL 35899, USA}
\altaffiltext{16}{Jacobs Technology, Inc., Huntsville, AL, USA}
\altaffiltext{17}{NASA Goddard Space Flight Center, Greenbelt, MD 20771, USA}

\begin{abstract}
Since launch in 2008, the \Fermi\ Gamma-ray Burst Monitor (GBM) has detected many hundreds of bursts from magnetar sources. While the vast majority of these bursts have been attributed to several known magnetars, there is also a small sample of magnetar-like bursts of unknown origin. Here we present the \Fermi/GBM magnetar catalog, giving the results of the temporal and spectral analyses of 440 magnetar bursts with high temporal and spectral resolution. This catalog covers the first five years of GBM magnetar observations, from July 2008 to June 2013. We provide durations, spectral parameters for various models, fluences and peak fluxes for all the bursts, as well as a detailed temporal analysis for \Jfifteen bursts. Finally, we suggest that some of the bursts of unknown origin are associated with the newly discovered magnetar 3XMM\,J$185246.6+0033.7$.
\end{abstract}

\section{Introduction}

In 1979 \citet{Mazets79a,Mazets79b} reported the detection with the Venera satellites of a series of short, energetic events apparently originating from the same directions on the sky. Although their durations would place them in the short gamma-ray burst class, \citet{Mazets79a} argued that the recurrent nature of the emission indicated that we were looking at a new phenomenon. In the following decade, these sources were indeed recognized as a different class of objects, called soft gamma repeaters (SGRs), based on their repeated high-energy burst activity \citep{Golenetskii84,Laros87,Atteia87,Kou87}. In the late 1990s it was discovered that these objects were isolated neutron stars characterized by extremely strong dipole magnetic fields of $\sim10^{14}-10^{15}$~G \citep[][]{Kou98, Kou99}, right along the lines of their theoretical prediction by \citet{D&T92}. Currently, the bulk of the magnetar population comprises mainly two classes of objects: SGRs and Anomalous X-Ray Pulsars (AXPs).  Their distinguishing traits are slow spin periods ($\sim2-12$~s), large period derivatives ($\sim 10^{-13}-10^{-10}$~s~s$^{-1}$), and periods of activity during which they emit multiple bursts in the hard X-ray/soft $\gamma$-ray energy range. With respect to the latter burst characteristic, magnetars seem to fall into two groups: sources that emit a few hundreds to a thousand bursts per active episode and sources emitting only a handful of bursts when active \citep[e.g.,][]{Gogus14}. For detailed reviews on magnetars, see \citet{W&T06} and \citet{Mer08}.

The magnetar discovery rate increased significantly when wide field-of-view monitors with location capabilities became operational, or multiple instruments in space allowed for triangulation of their signals to be performed. The first wave of discoveries in the late 80s was enabled mostly by the {\it Vela} and {\it Prognoz} satellites. This was followed by the synergistic era of the Burst And Transient Source Experiment (BATSE) onboard the {\it Compton Gamma Ray Observatory}, the {\it Rossi X-ray Timing Explorer} ({\it RXTE}), and {\it BeppoSaX} in the 90s; and finally, in the mid to late 2000s, the start of the era of NASA's {\it Swift} and {\it Fermi} observatories. The latter have onboard detectors dedicated to monitoring the high-energy transient sky (the {\it Swift}/Burst Alert Telescope (BAT) and the {\it Fermi}/Gamma-ray Burst Monitor (GBM), respectively), and have been extremely efficient in complementing each other in detecting and identifying new magnetar sources. 

Since its activation in July 2008, GBM has recorded several hundreds of bursts from new and known magnetars, and a few magnetar-like bursts from unknown sources within rather large error box regions. A large number of papers has already been published by the GBM magnetar team on some of these sources with multiple bursts (\Jfifteennos: \citealt{Kaneko10,Lin12,Horst12,Kie12,Huppen14,Younes14}; \Jzerofivenos: \citealt{Watts10,Lin11a,Huppen13}; \Jzerofournos: \citealt{Horst10}). Here we present a comprehensive overview for those magnetars, including bursts that have not been analyzed before, and also for magnetars with only a few events (\SGReighteennos, \Jeighteennos, \AXPfourunos, and \SGRoneEnos). We have compiled the entire GBM magnetar burst sample in a catalog for the first five years of the mission. In Section \ref{data} we describe the GBM data types, selection criteria, and analysis methods. We discuss the properties of bursts from confirmed magnetar sources in Section \ref{sgr_known}. In Section \ref{sgr_unknown} we discuss bursts for which we do not have confirmed source identifications. We conclude in Section \ref{discuss} with a comparison across the source properties and a discussion on the characteristics of all magnetar bursts detected with GBM.

\section{Data Analysis}
\label{data}

GBM is an all-sky monitor with an 8~sr field of view, consisting of 12 NaI detectors, covering an energy range from 8~keV to 1~MeV, and 2 BGO detectors, covering $0.2-40$~MeV \citep{Meegan09}. GBM has three data types, of which two (CTIME and CSPEC) are continuously recorded and one (time-tagged event data, or TTE) recorded only after GBM had triggered on a burst or other selected time intervals during most of the catalog period. This configuration changed after November 2012, and currently GBM is recording TTE data continuously. While CSPEC data have good spectral resolution (128 energy channels over the full energy range), its temporal resolution of 1.024~s is too long for the $\sim0.1$~s long magnetar bursts. The temporal resolution of CTIME data (64~ms) is better suited in this respect, but the spectral resolution (8 energy channels) is not sufficient to perform detailed spectral analysis. TTE data have the highest temporal (2~$\mu$s) and energy (128 energy channels) resolution, and are therefore perfectly suited for magnetar burst analyses. The downside is that this data type is typically only available from $\sim30$~s before a burst trigger to $\sim300$~s afterwards, followed by another $\sim300$ during which no new trigger can take place. For further details on GBM and its data products, see \citet{Meegan09}.

\subsection{Data Selection}

Given the spectral softness of magnetar bursts, with typically no emission above 200~keV, we only use NaI data for our analyses. The 12 NaI detectors are oriented in such a way that GBM can monitor the entire sky, which implies that only a subset of the detectors can see one specific location on the sky at any given time. For every burst we have used only detectors with an angle between the source direction and the detector normal of less than $60^{\circ}$, to ensure a large effective area for these events. We note that in previous studies of GBM bursts the viewing angle varies from 40 to $60^{\circ}$, but this does not have a significant effect on the temporal and spectral results. We have excluded detectors when a source was obstructed by parts of the spacecraft or the Large Area Telescope (LAT) onboard \Fermi. 

We have identified a total of 427 triggered bursts associated with magnetars between July 2008 and June 2013, of which the associations have been established based on localization, detections by other instruments, or source bursting activity. We have also searched for, and included, all untriggered bursts that happened during the 30~s before and 300~s after each trigger, since TTE data were available at these times. Finally, we analyzed 19 magnetar-like bursts during this period, i.e. bursts with durations and spectra similar to those observed in magnetar bursts. Due to the poor positional accuracy of GBM (in particular for spectrally soft events), the crowded areas on the sky where they were found, and the lack of triggers by other gamma-ray instruments, it was not possible to confirm their nature. Several other untriggered bursts were found during the catalog time period, but are not included here since they lack TTE data. In Table~\ref{summary} we show the magnetars included in this catalog, their burst active periods, and the total number of bursts per magnetar used in our analyses.

\subsection{Temporal \& Spectral Analysis}

The majority of bursts presented in this catalog have been published \citep{Horst10,Lin11a,Lin11b,Lin12,Horst12, Kie12,Younes14}. For the remaining bursts we performed temporal and spectral analyses following the techniques and criteria as described in detail in these publications. The \Tnin and \Tfif durations, defined by the time that the cumulative counts rise from 5\% to 95\% and 25\% to 75\%, respectively \citep{Kou93}, were calculated in photon space, i.e. by using the intrinsic (deconvolved) burst spectra instead of the detector
recorded counts to define the durations \citep[for a detailed description, see][]{Lin11a}. Since many of the bursts consist of multiple peaks, we required for two events to qualify as two separate bursts if the time between their peaks is longer than a quarter of the magnetar spin period, and the count rate drops to the background level between the peaks. We excluded the brightest bursts that saturated the High Speed Science Data Bus of GBM from the duration calculations, and for these bursts we only used the non-saturated parts in our spectral analysis. The spectral analyses were performed with the software package \RMFITnos\footnote[1]{R.S. Mallozzi, R.D. Preece, \& M.S. Briggs, "RMFIT, A Lightcurve and Spectral Analysis Tool $\copyright$2008 Robert D. Preece, University of Alabama in Huntsville, 2008} ({\it v4.3}), using detector response matrices generated with {\it GBMRSP} ({\it v2.0}). To generate the response matrices we used the best available positions from other satellites or telescopes for the magnetar bursts in Section~\ref{sgr_known}, and the GBM positions for the bursts of unconfirmed origin in Section~\ref{sgr_unknown}. We fitted the time-integrated spectra over an $8-200$~keV energy range to different spectral models: a single power law (PL), a blackbody (BB), a Comptonized model (COMP), an optically thin thermal bremsstrahlung model (OTTB), and two blackbody functions (BB$+$BB). In each case we used the Castor C-statistic to determine the model significance. For \Jfifteen we also performed time-resolved spectroscopy of the bursts at 4~ms time resolution to determine their peak fluxes. For this purpose we only used the COMP model. Finally, temporal analyses were performed on the whole sample of \Jfifteen bursts, and the methodology and results are discussed in Sections \ref{J1550_BurstPhase} and \ref{J1550_folded}.

\section{Magnetar bursts from confirmed sources}
\label{sgr_known}

\subsection{\Jfifteen}
\label{J1550}

\Jfifteen (RA = $15^{\rm{h}}50^{\rm{m}}54^{\rm{s}}.11$, Dec = $-54^{\circ}18^{\prime}23^{\prime\prime}.7$; \citealt{Camilo07}) became active in October 2008, when it emitted several hundreds of magnetar-like bursts \citep{Palmer09, Kou09}; this activity ceased April 2009. The source was discovered by the \textit{Einstein} satellite \citep{L&M81}, named 1E~1547.0$-$5408, and identified as a magnetar candidate following observations with \textit{XMM-Newton} and the \textit{Chandra} X-Ray Observatory \citep{G&G07}. It was later confirmed as a magnetar with a spin period of 2.07~s, a period derivative of \scinum{2.32}{-11}~s/s, and an inferred magnetic field of \scinum{2.2}{14}~G (\citealt{Camilo07}; for a detailed description of the history of \Jfifteennos, see \citealt{Horst12} and \citealt{Kie12}). Although \Jfifteen was initially classified as an AXP, after this extremely prolific activity very similar to ones from SGRs 1806$-$20, 1900$+$14, and 1627$-$41, it was reclassified as an SGR and renamed \Jfifteen \citep{Palmer09, Kou09}. The source distance has been estimated to be $\sim5$~kpc \citep{G&G07,Tiengo10}.

Throughout this section, we will refer to the activity of \Jfifteen in terms of three active periods: October 2008 \citep{Kie12}, January 2009 \citep{Horst12}, and  March - April 2009 \citep{Kie12}.  \citet{Kie12} studied the GBM burst data for the first and third periods and found that the spectra of the former were best fit with a single BB, while the latter were best fit with an OTTB. \citet{Horst12} studied the bursts of the first two days of the second period, and found that their spectra were fit equally well with a COMP, an OTTB, and a BB$+$BB model. \citet{Lin12} later refined the results of \citet{Horst12}, using bursts from the second period that were also seen with the \textit{Swift}/X-Ray Telescope (XRT), and established that a BB$+$BB model best fit these joint spectra. Taken together these three studies suggest that the first outburst was significantly different than the other two periods, which exhibited similar spectral and temporal properties. \citet{Younes14} presented time-resolved spectroscopy for the brightest bursts over all three periods, and obtained an estimate of the lower limit on the internal magnetic field. Besides the analysis of these bursts, \citet{Kaneko10} presented spectral and temporal analysis of enhanced persistent emission during the onset of the second bursting period, resulting in the discovery of the smallest hot spot ever measured for a magnetar. Furthermore, recent detailed variability analysis has revealed candidate quasi-periodic oscillations (QPOs) in bursts during the second emission period \citep{Huppen14}.

We searched for additional bursts in the GBM data during the gaps between the previously studied bursting periods, and found 66 previously unstudied bursts that temporally best belong to the second period (spanning 22~January $-$ 24~February 2009). These 66 bursts include 5 untriggered bursts for which we have TTE data, which were found with the search algorithm described in \citet{Kaneko10}. Between 24 February and 22 March there were no bursts observed from \Jfifteennos, although the source was still visible in the X-rays (see e.g., \citealt{S&K11}).

\subsubsection{Temporal and Spectral Evolution}
\label{J1550_TempSpec}

\begin{figure}
\begin{center}
\includegraphics[scale=0.35]{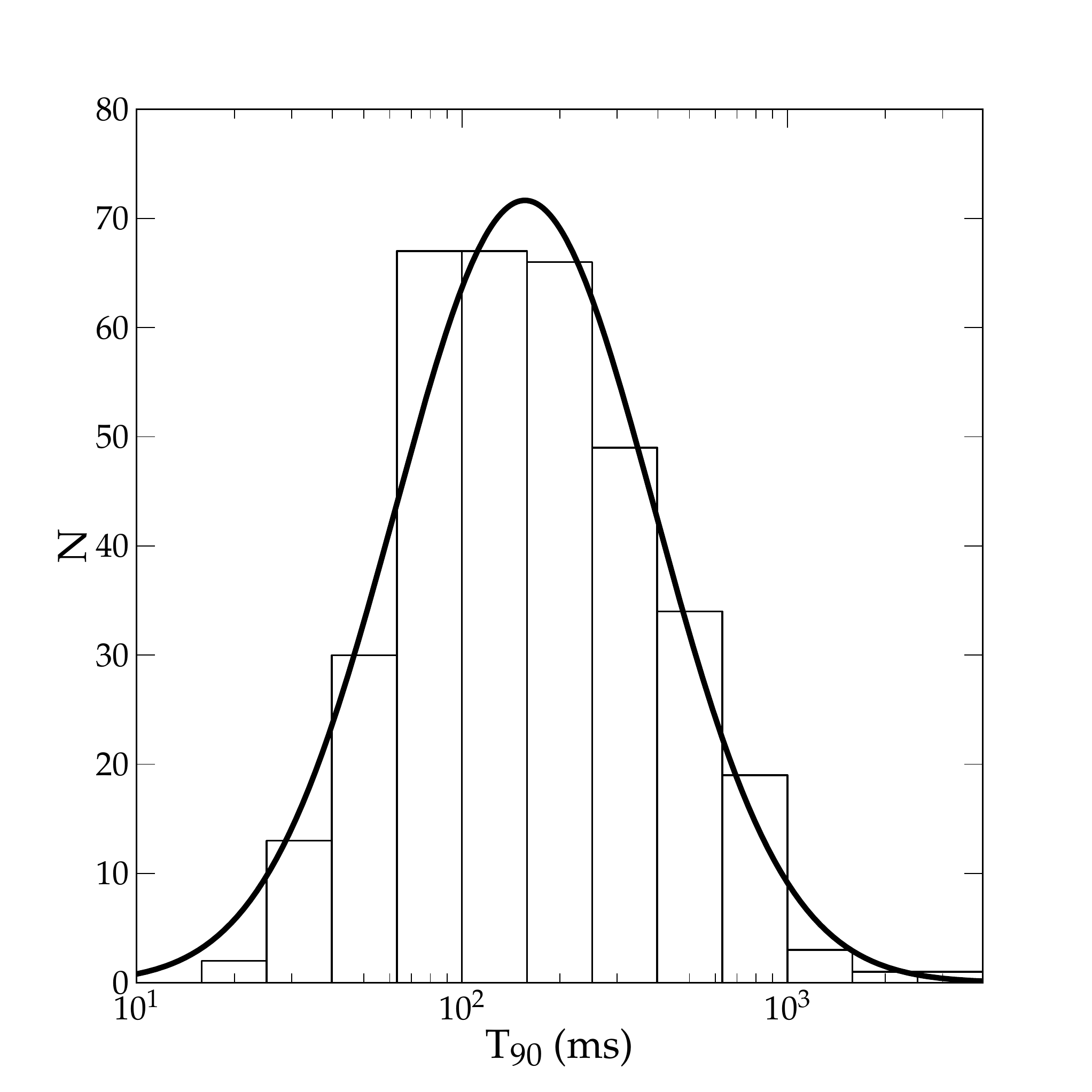}
\includegraphics[scale=0.35]{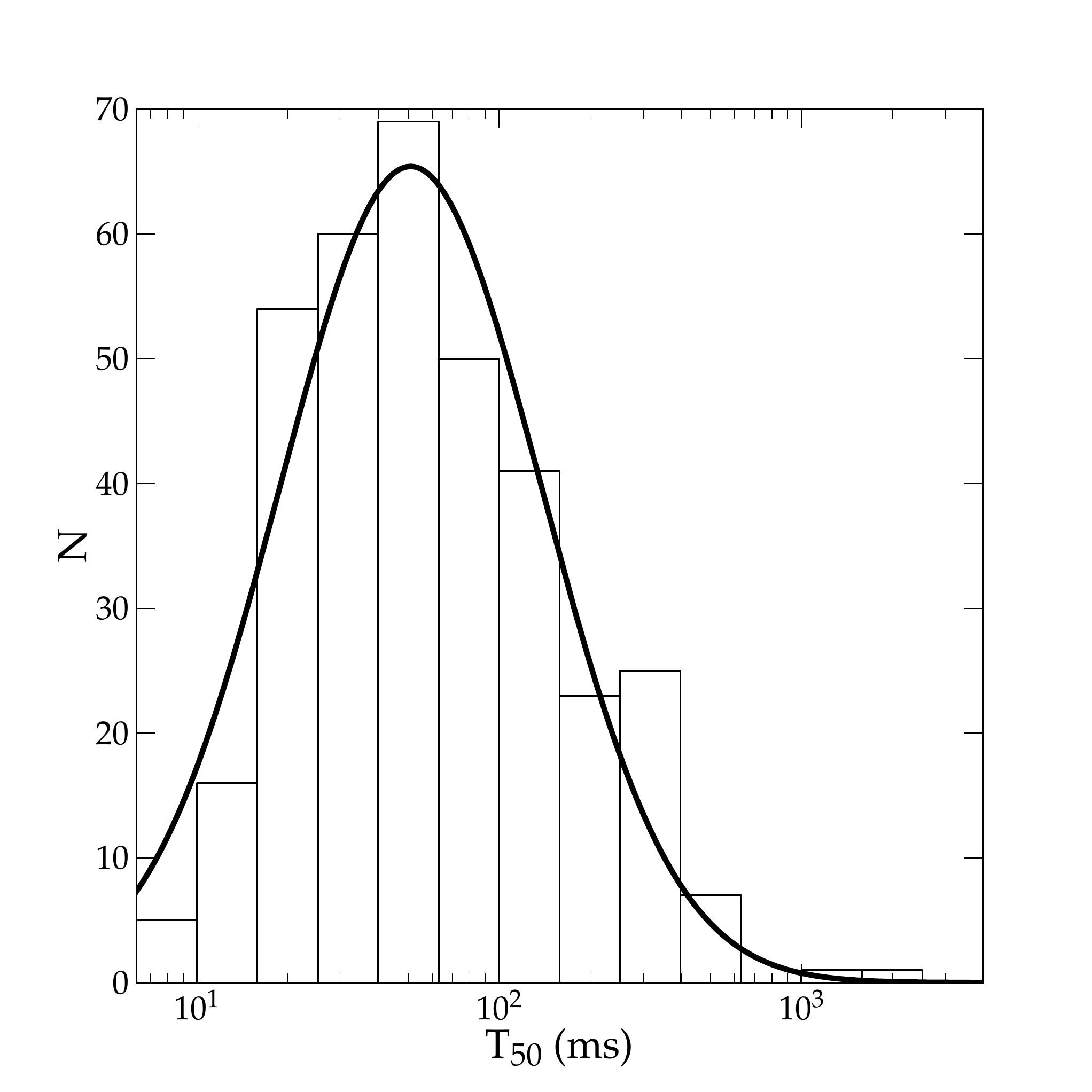}
\end{center}
\caption{Distribution of \Tnin (left panel) and \Tfif (right panel) for all \Jfifteen bursts. A log-normal fit (black line) to the histograms provides $\mu =$\valerr{2.19}{0.02} log(ms) and $\sigma = $\valerr{0.40}{0.02} log(ms) for \Tninnos, and $\mu =$\valerr{1.71}{0.04} log(ms) and $\sigma = $\valerr{0.44}{0.03} log(ms) for \Tfifnos.}
\label{1550_dur_full}
\end{figure}

\begin{figure}
\begin{center}
\includegraphics[scale=0.4]{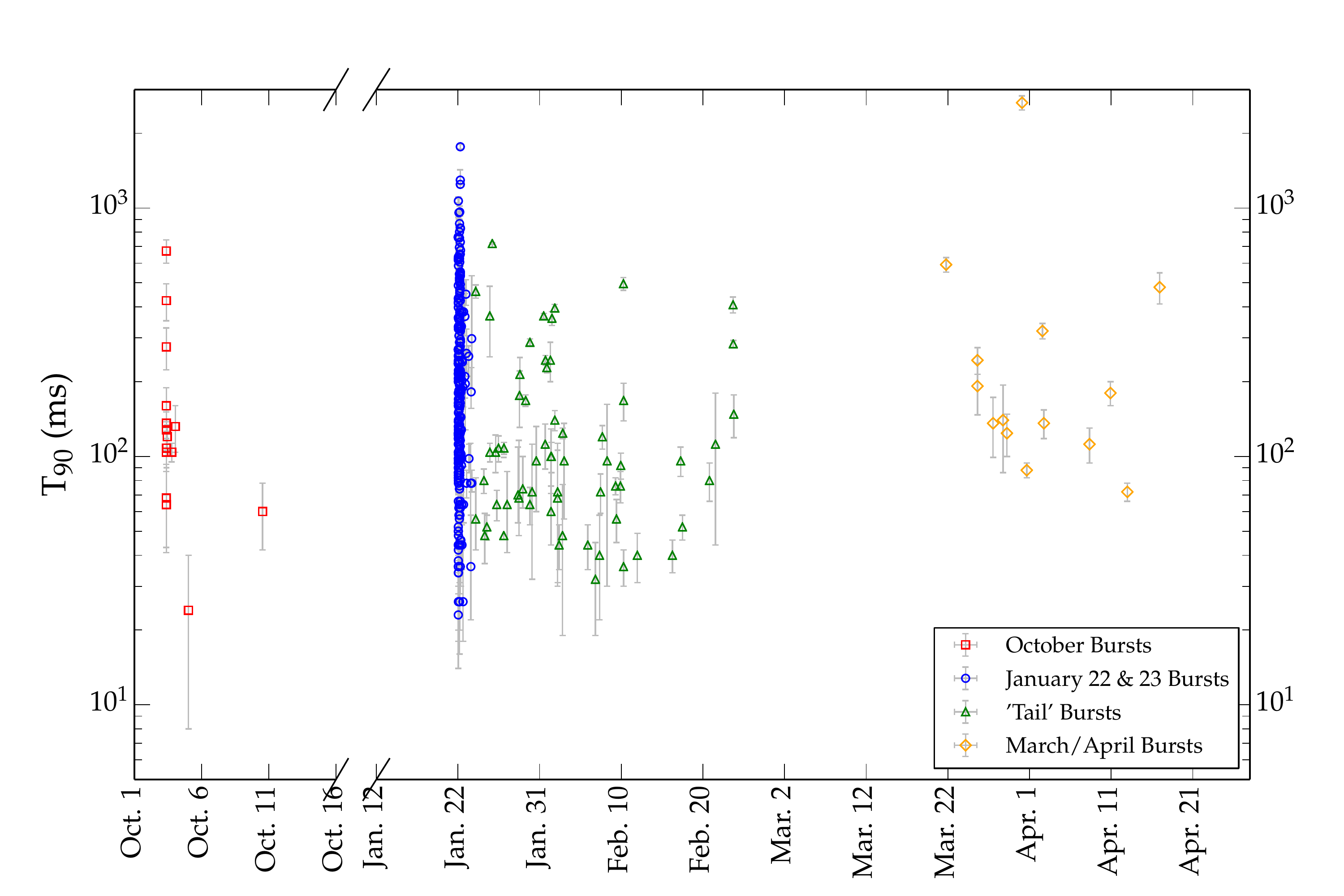}
\includegraphics[scale=0.4]{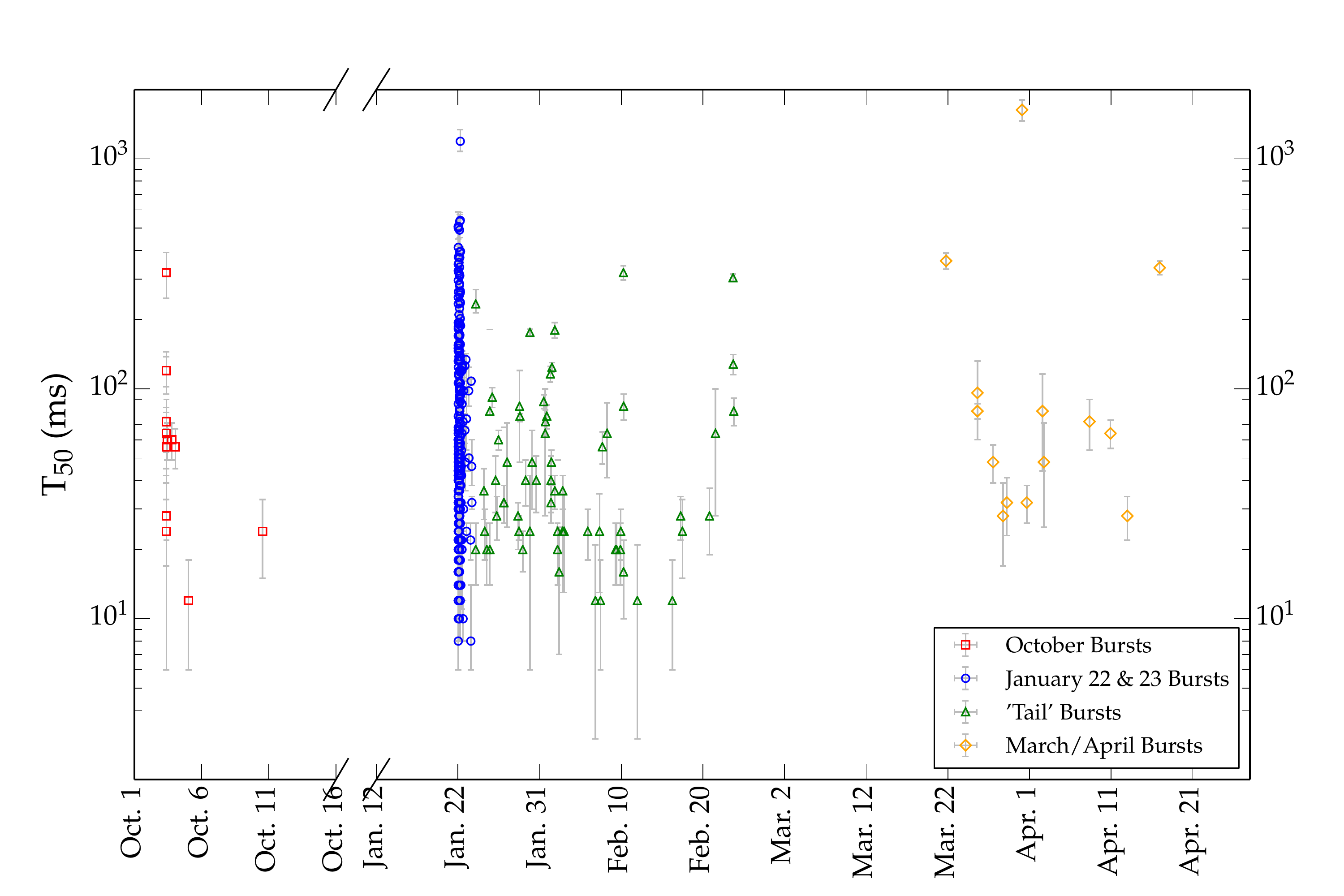}
\end{center}
\caption{\Tnin and \Tfif for \Jfifteen over the course of all outburst periods. The 'tail' bursts in this figure, and the following ones, are bursts detected between January 24 and February 24. We find no trends in these duration over time.}
\label{1550_duration_time}
\end{figure}

We study here the statistical properties of all bursts from \Jfifteen detected with GBM and the evolution of the burst activity across the entire $2008-2009$ source activation period. Figure~\ref{1550_dur_full} displays the \Tnin and \Tfif duration distributions of 354 unsaturated events (see also Table~\ref{1550_time}). We fit these distributions with a log-normal function (solid line) and find that the \Tnin (\Tfifnos) duration centers at $155\pm10$~ms ($51\pm5$~ms), with a logarithmic width of $\sim 0.4$. The duration distribution of the bursts from \Jfifteen is similar to what we have seen in other magnetars \citep[e.g.,][]{Woods99,Gogus01,Gav04,Esp08,Lin11a}. Figure~\ref{1550_duration_time} shows the evolution of  \Tnin and \Tfif over time. We find no significant trend in the evolution of the burst durations over the entire outburst period of roughly 7 months. We note that the longest event on Figure~\ref{1550_duration_time} is very faint with a peak flux of $9.2\times10^{-7}$ erg/cm$^2$/s (see also \citealt{Lin13} for a sample of very dim but long \Jfifteen bursts detected by \Swift/XRT).

\begin{figure}
\begin{center}
\includegraphics[scale=0.4]{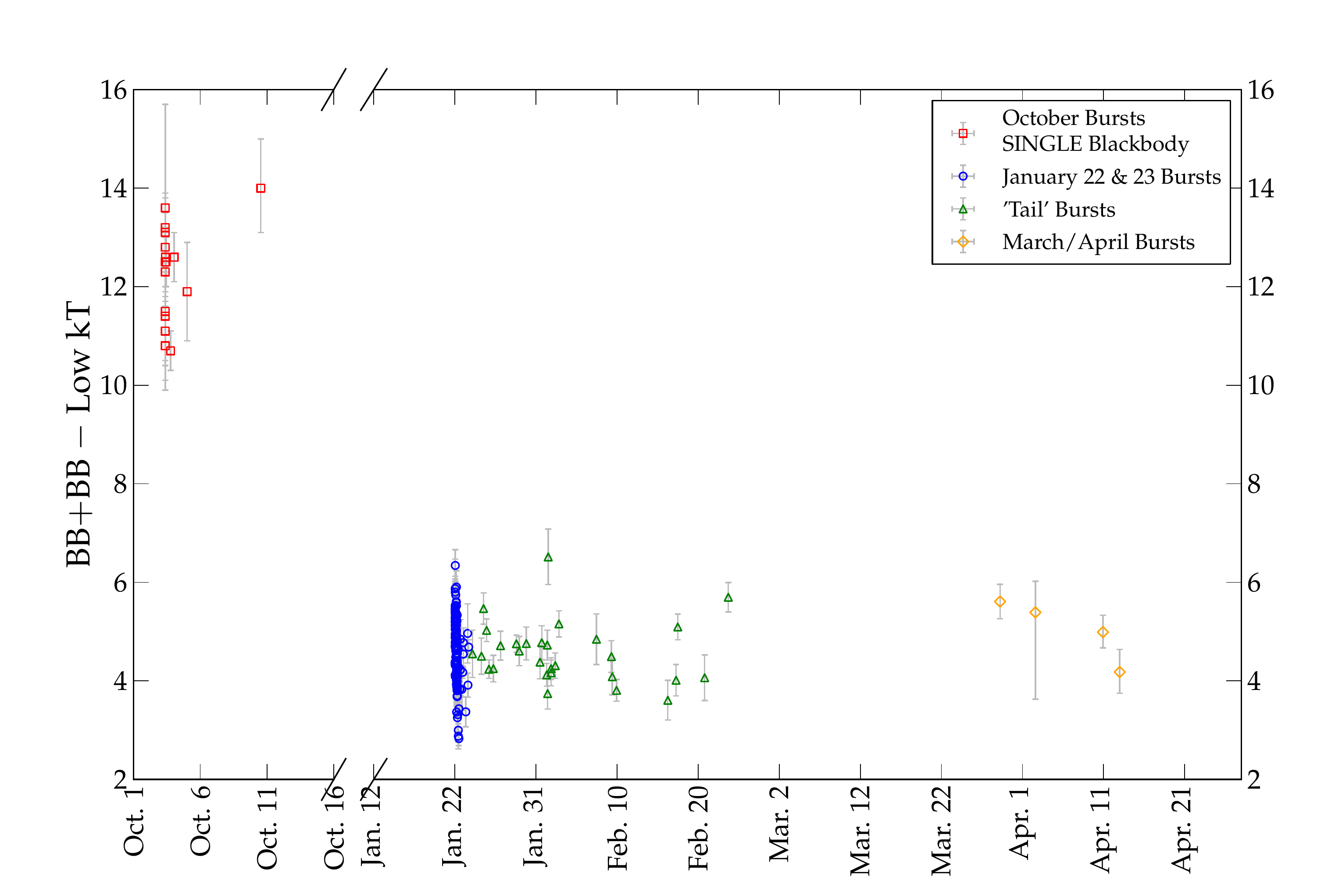}
\includegraphics[scale=0.4]{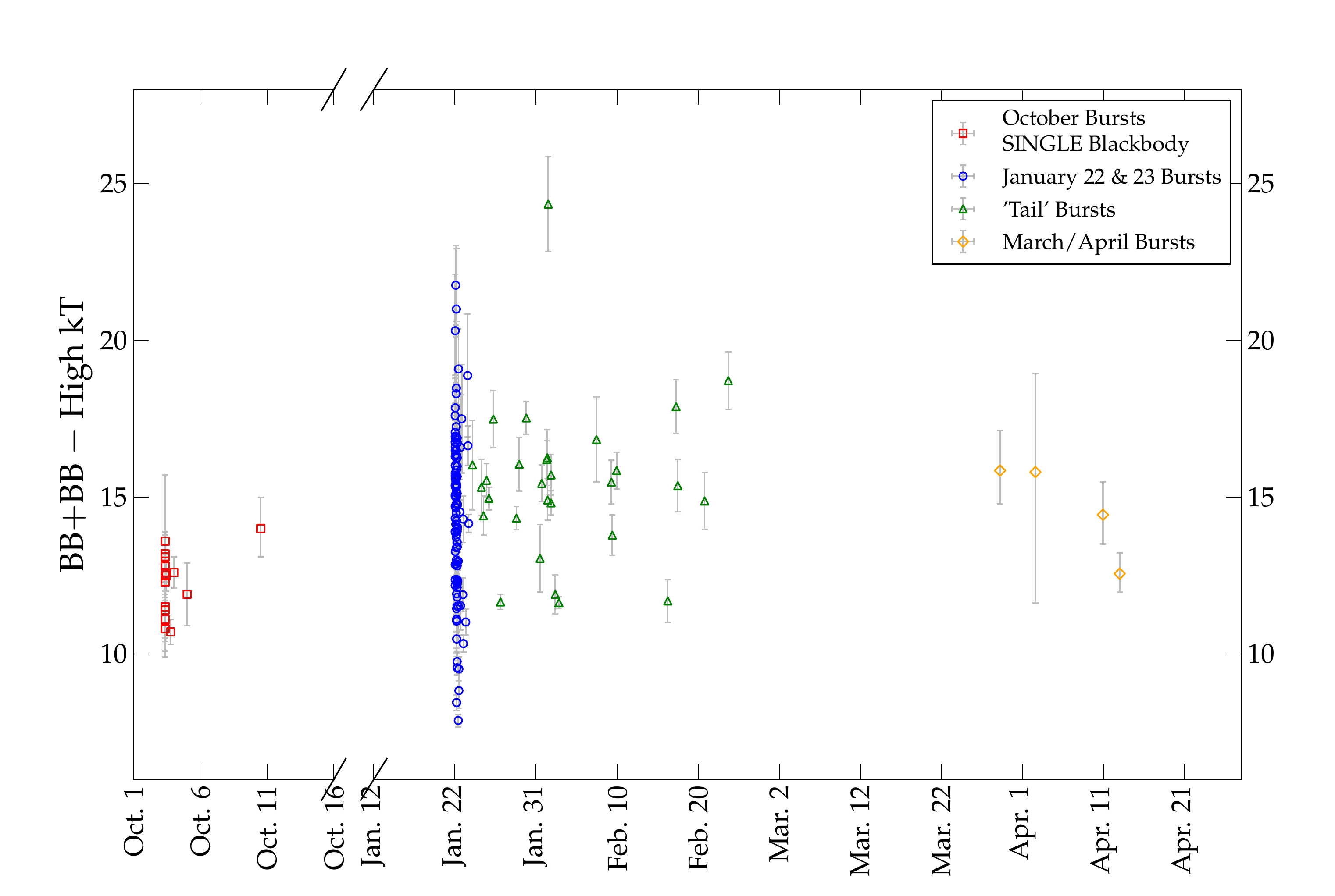}
\end{center}
\caption{Evolution of the low (top panel) and high (bottom panel) BB temperatures from the BB$+$BB spectral model over time for \Jfifteen bursts. For the October bursts we use the values for the single BB fits in both panels (red squares). There is a clear division between these single BB temperatures and the low temperatures of the BB$+$BB model. However, the former temperatures are similar to the hot BB temperatures of the BB$+$BB model.}
\label{1550_param_time}
\end{figure}

\begin{figure}
\begin{center}
\includegraphics[scale=0.4]{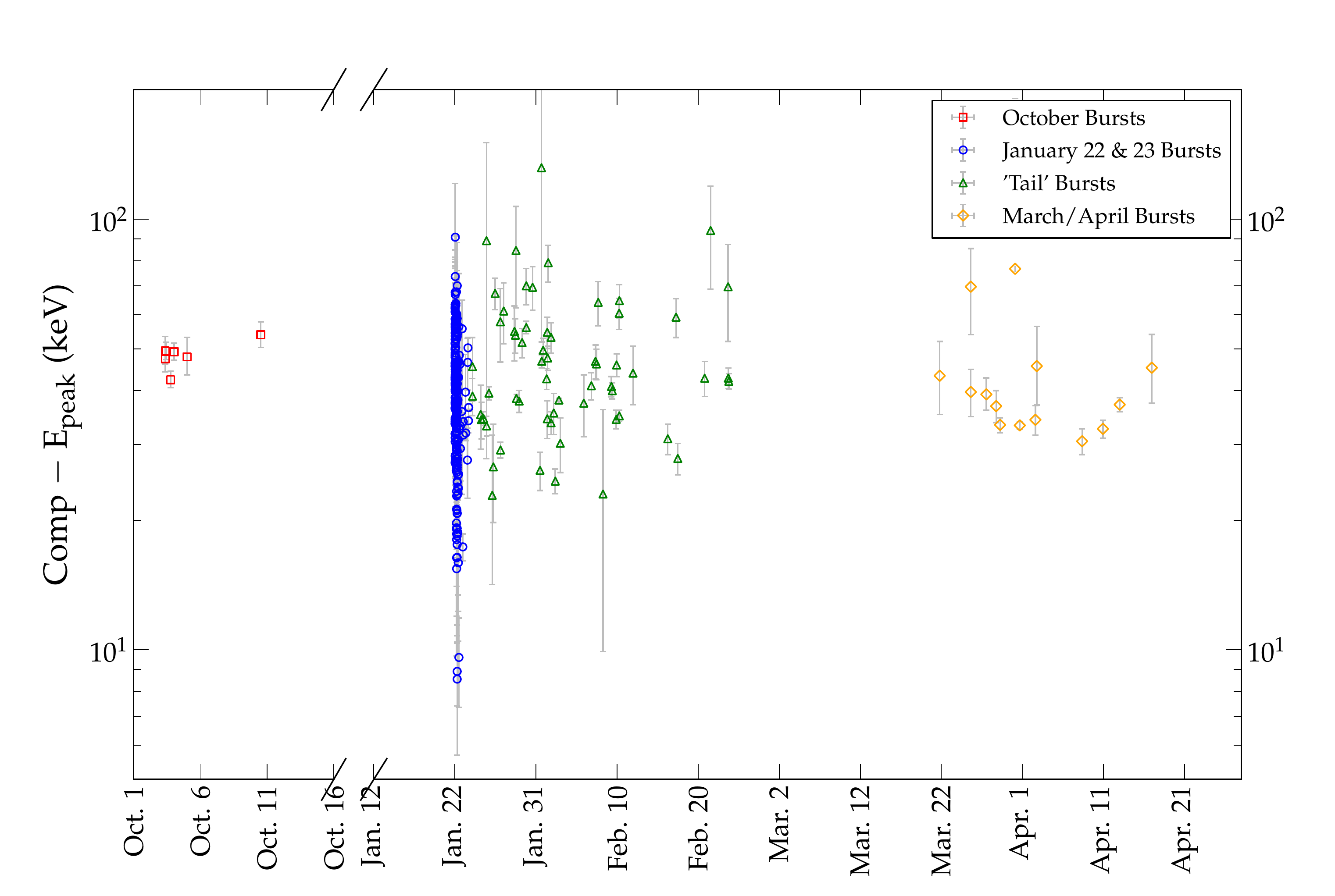}
\includegraphics[scale=0.4]{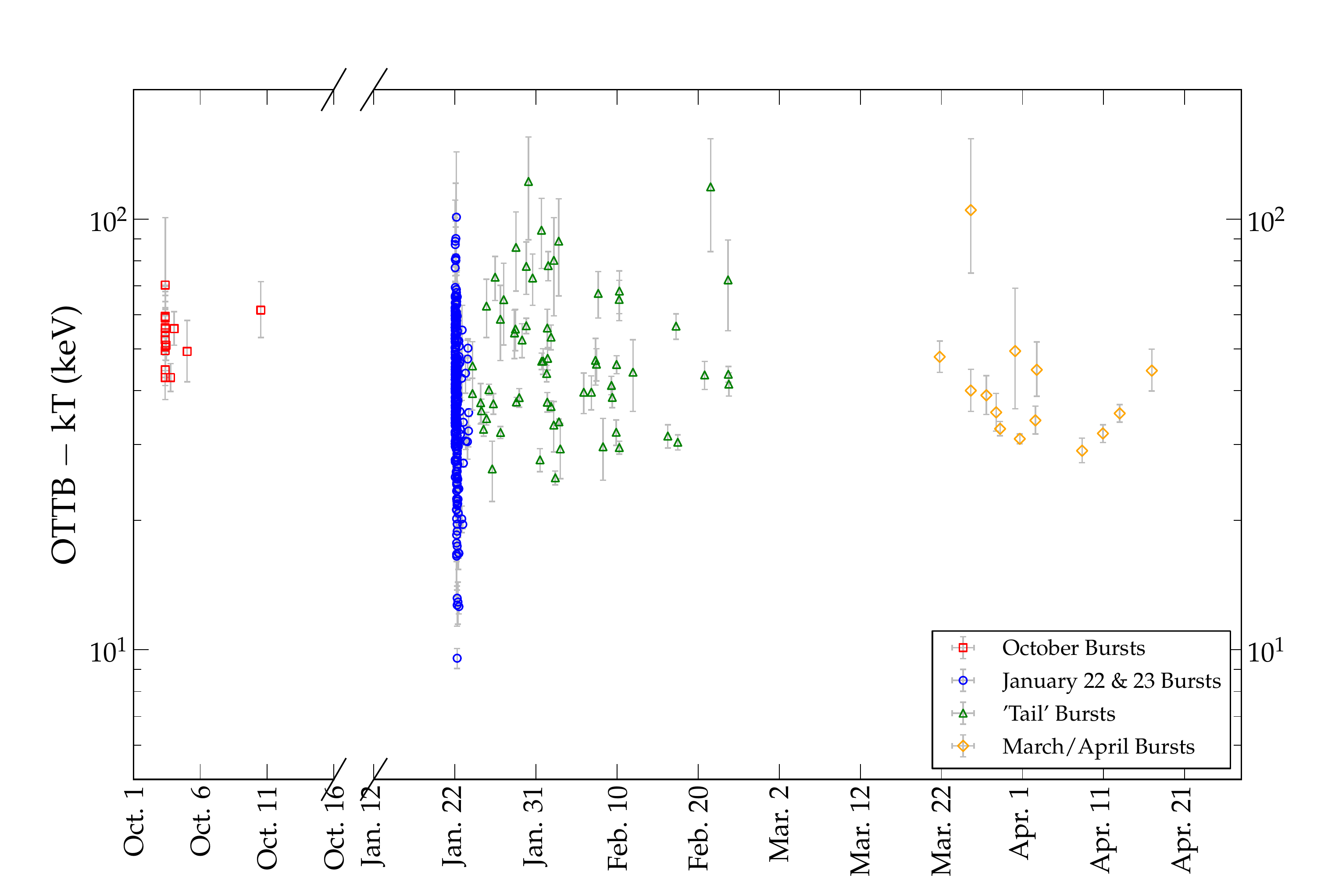}
\end{center}
\caption{Spectral peak as measured by the Comptonized model (top panel, \Epnos) and OTTB model (bottom panel, $kT$) for \Jfifteennos. Overall, the values are in agreement with each other, and they are both consistent over time.}
\label{1550_param_time2}
\end{figure}

\begin{figure}
\begin{center}
\includegraphics[scale=0.4]{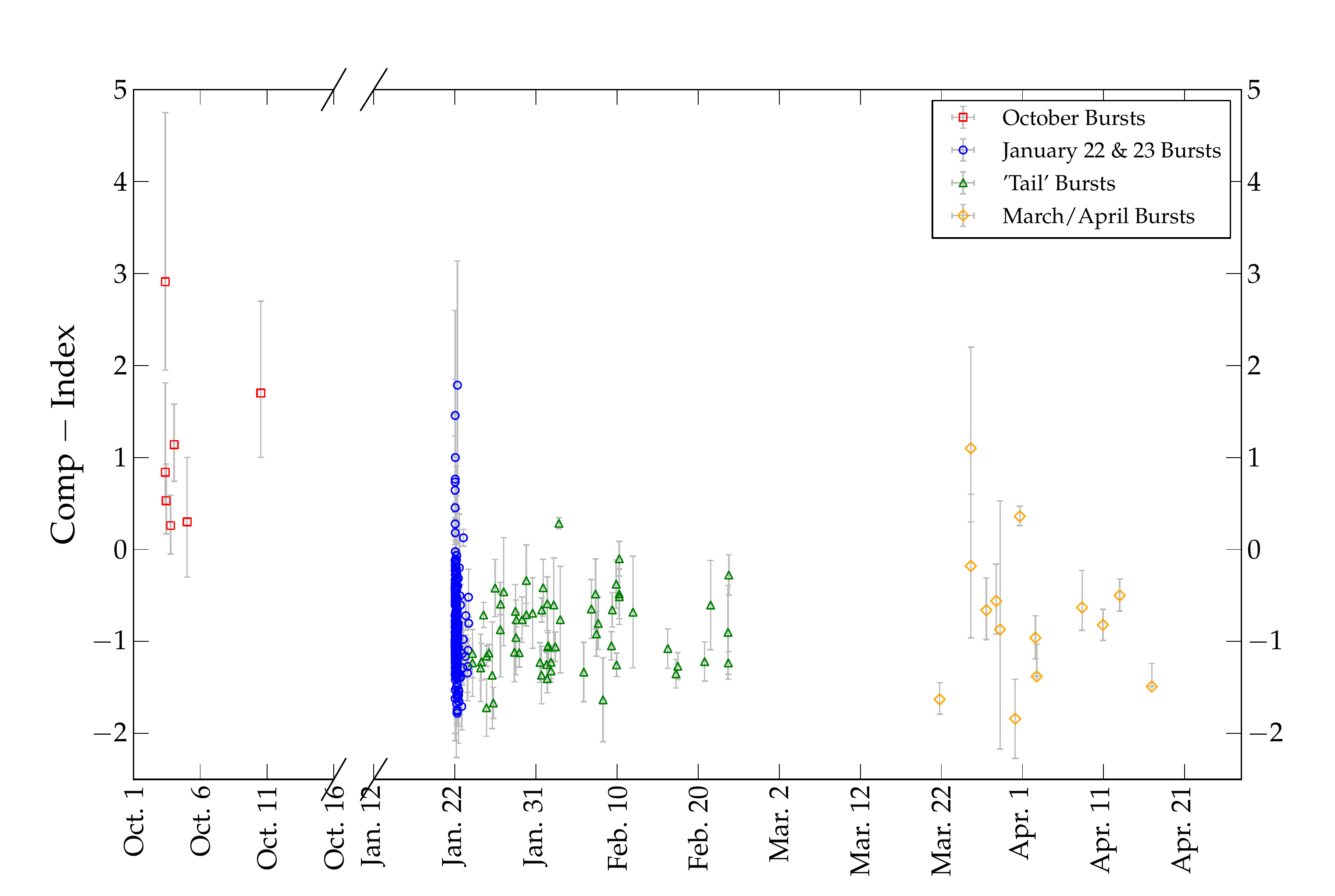}
\end{center}
\caption{Power-law index of the COMP model for \Jfifteen bursts. The bursts from the October period are clearly different compared to the rest of the sample, confirming the spectral evolution between burst active periods seen with the BB$+$BB model fits.}
\label{1550_param_time3}
\end{figure}

In Figures \ref{1550_param_time}$-$\ref{1550_param_time3} we present the evolution of time-integrated spectral parameters from the OTTB, COMP and BB$+$BB models for the entire sample of \Jfifteen bursts \citep[Table~\ref{1550_spec};][]{Horst12,Kie12}. In Figure~\ref{1550_param_time} we plot the temperatures of our BB$+$BB fits for the second and third periods along with the temperature of the single BB that best fit the first activity period. The latter is included in both panels (high and low $kT$) for comparison purposes. As is evident from the lower panel (high $kT$), the temperature of the single BB is similar to the high $kT$ of the BB$+$BB model, which means that the lower $kT$ BB is absent in the first activity period. This confirms the suggestion of \citet{Kie12} that there is a spectral evolution from the first to the other burst active periods. We note that it would be possible for the lower $kT$ BB to be present below the detection limit of GBM, but this would still indicate a significant change in the spectral shape between the different periods. The burst peak energies, derived with the COMP (\Epnos) and OTTB ($kT$) models, are presented in Figure~\ref{1550_param_time2} in the upper and lower panel, respectively.  We observe no significant evolution across the outburst.  However, the evolution of the spectral index of the COMP model (Figure~\ref{1550_param_time3}) confirms that the bursts of the first burst period are significantly harder.

\begin{figure}
\begin{center}
\includegraphics[scale=0.35]{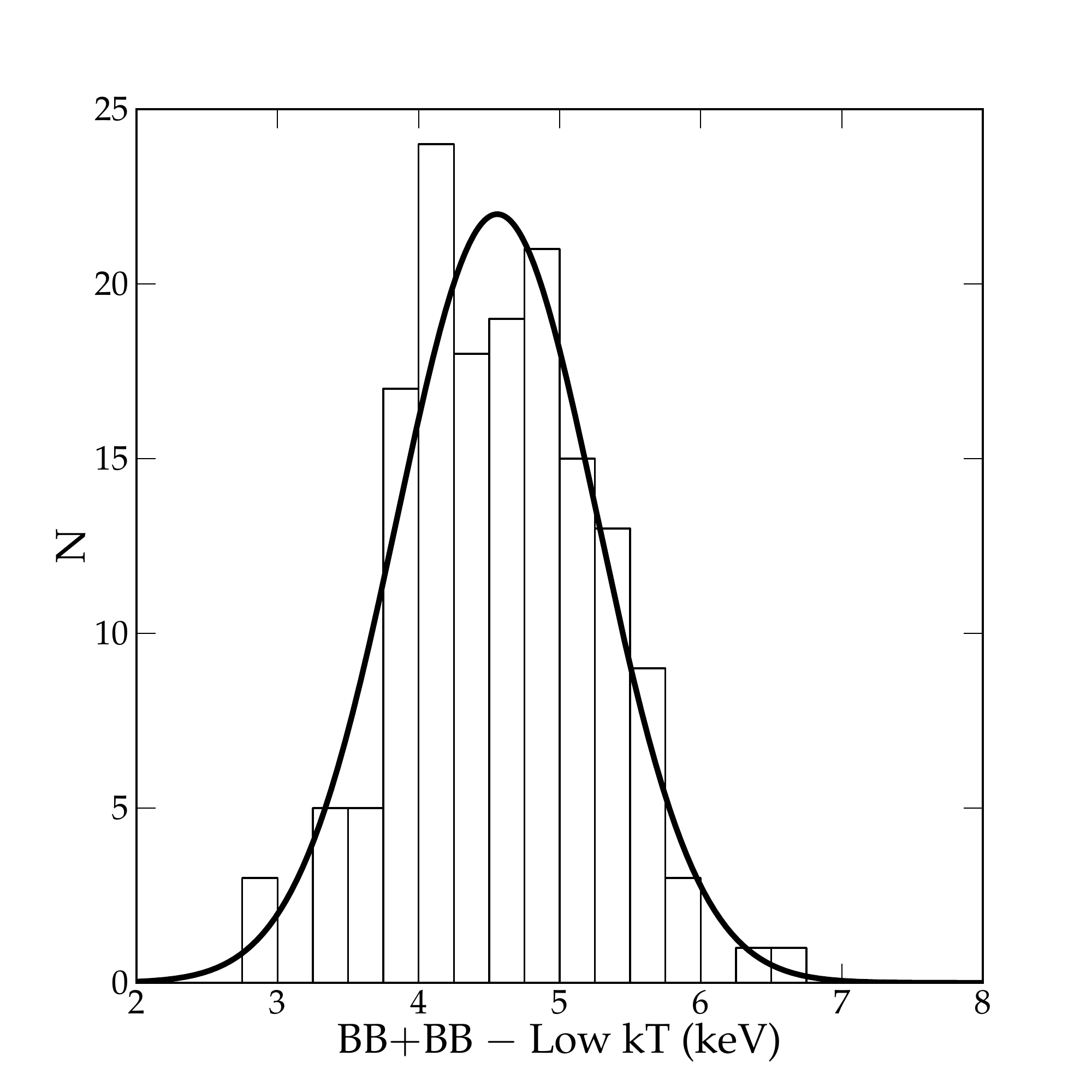}
\includegraphics[scale=0.35]{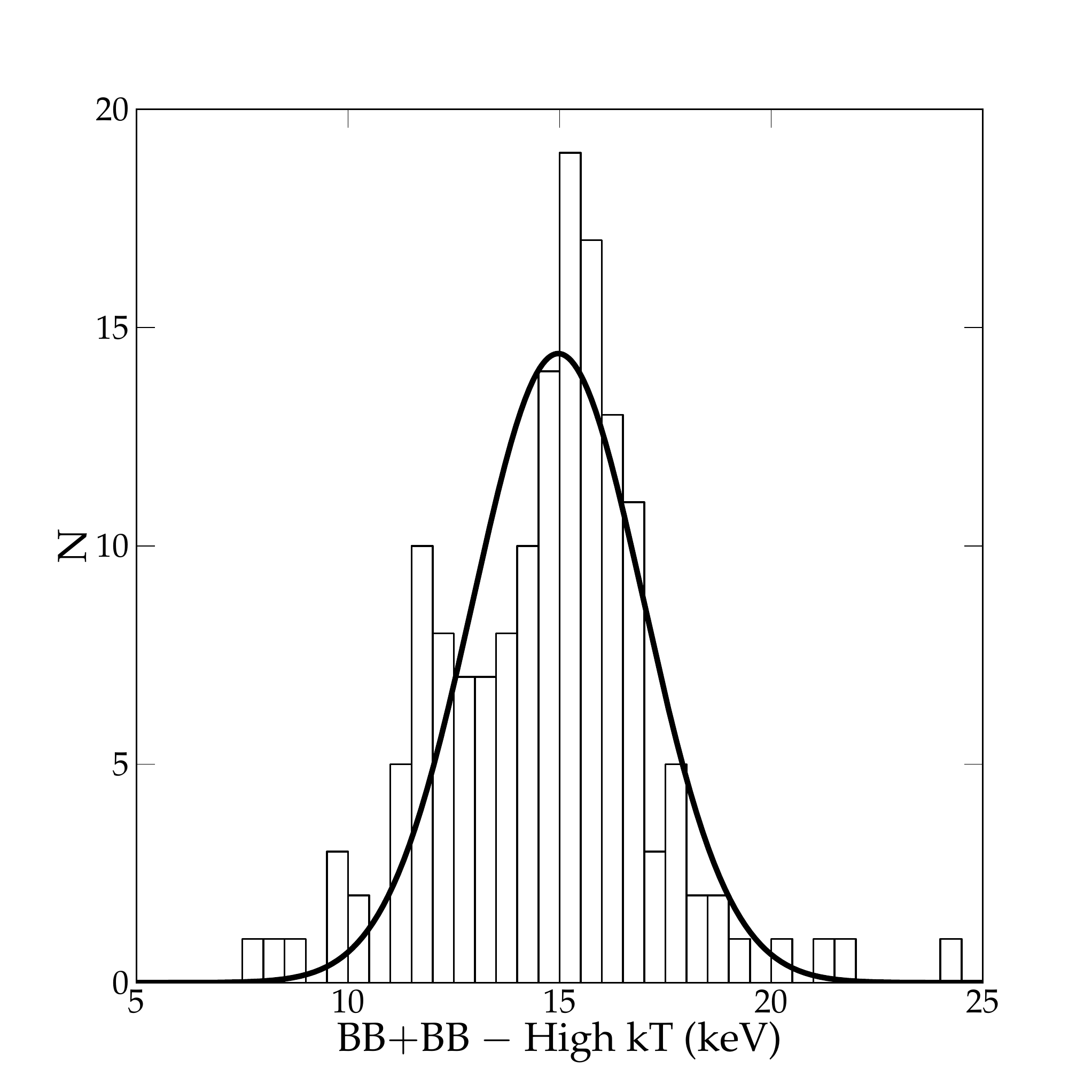}
\end{center}
\caption{Low (left panel) and high (right panel) BB temperatures for all bursts of \Jfifteennos. We fit these histograms to a Gaussian (black line) with $\mu =$ \valerr{4.55}{0.05}~keV and $\sigma =$ \valerr{0.71}{0.05}~keV for the low $kT$ BB, and $\mu =$ \valerr{14.96}{0.16}~keV and $\sigma =$ \valerr{2.03}{0.16}~keV for the high $kT$ BB.}
\label{1550_param_hist}
\end{figure}

\begin{figure}
\begin{center}
\includegraphics[scale=0.35]{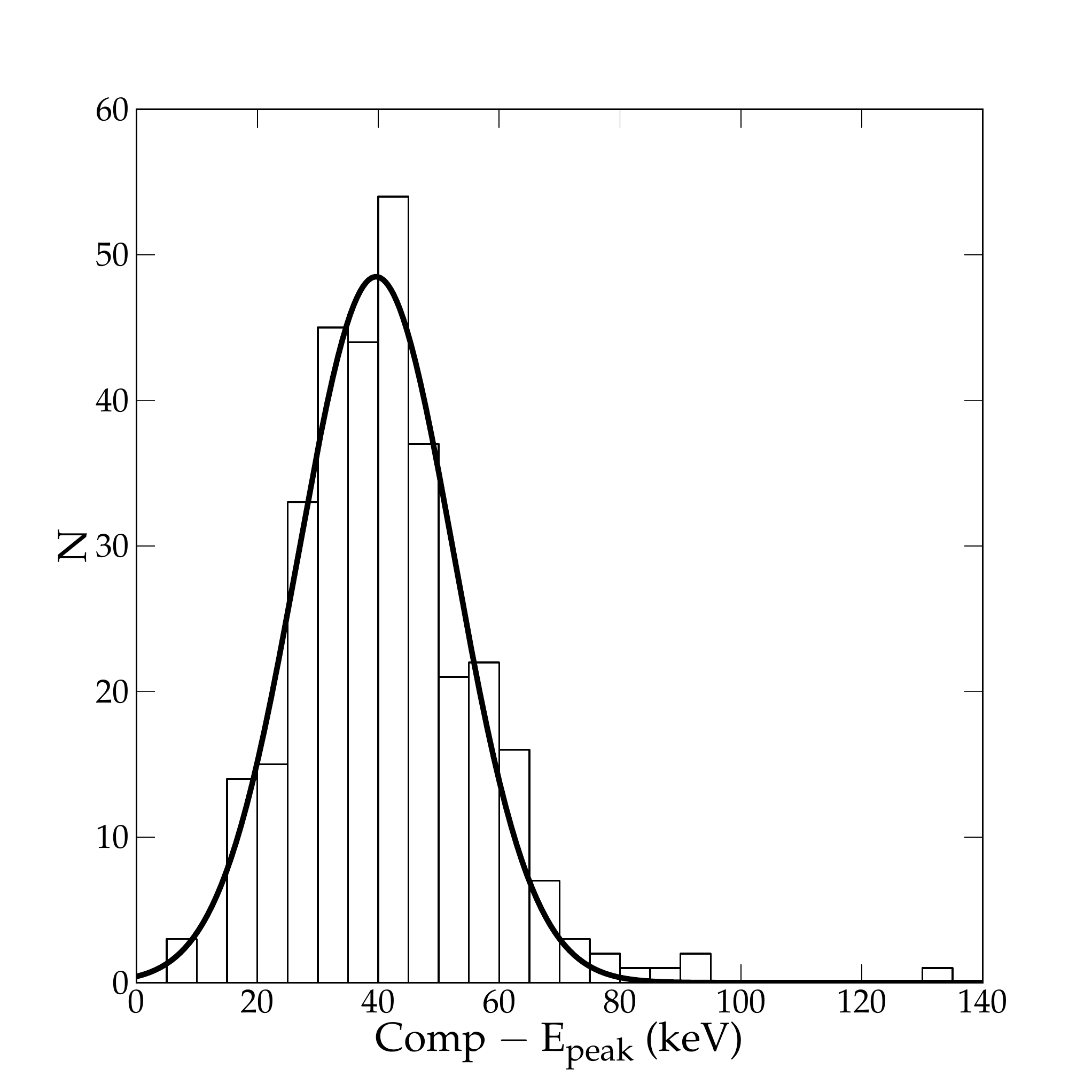}
\includegraphics[scale=0.35]{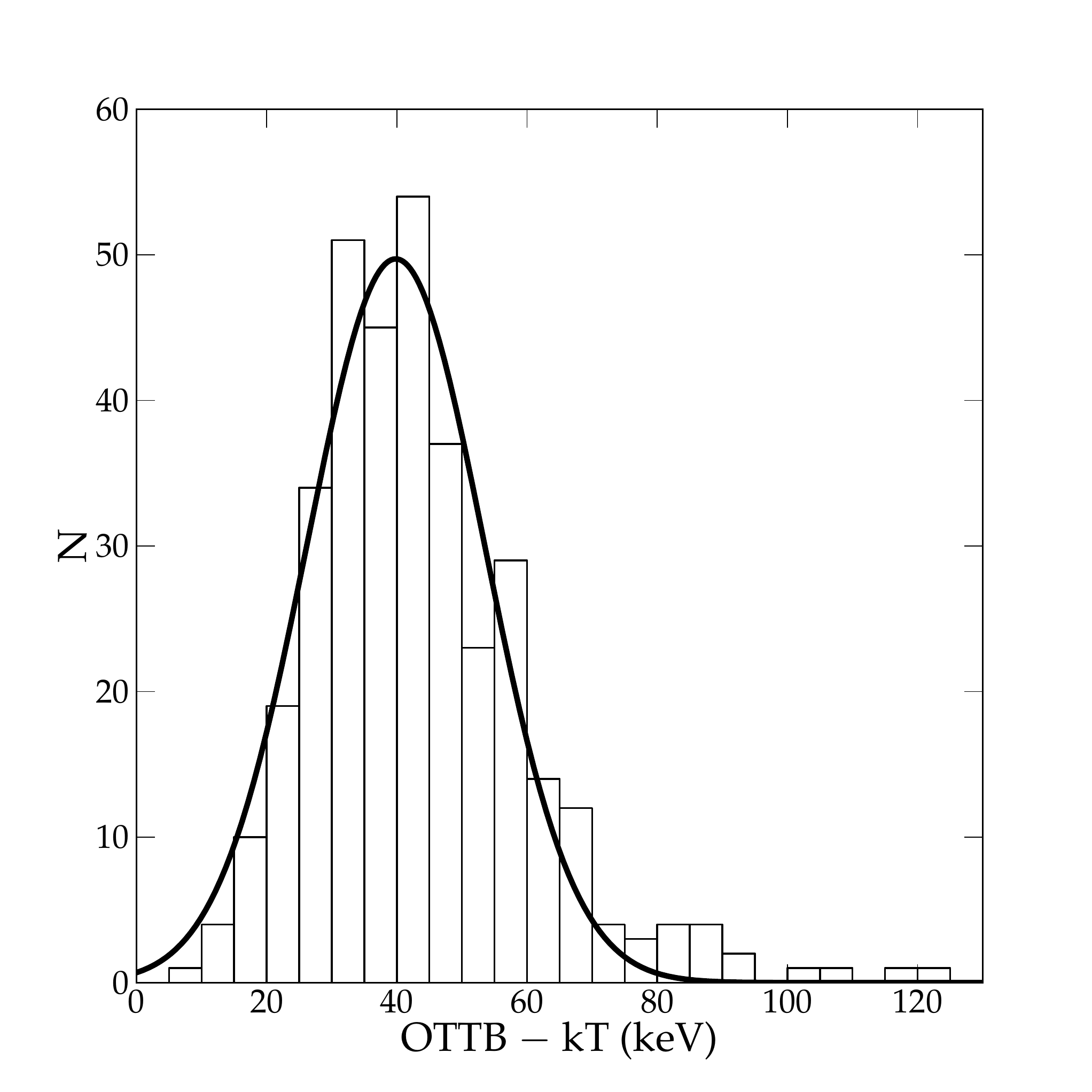}
\end{center}
\caption{\Ep as measured by the COMP and OTTB models for \Jfifteen bursts. We fit these histograms to Gaussians (black line).  For the COMP model, we find \Ep to have $\mu =$ \valerr{39.6}{0.6}~keV and $\sigma =$\valerr{12.9}{0.6}~keV. For the OTTB model, we find $kT$ to have $\mu =$ \valerr{39.8}{0.7}~keV and $\sigma =$\valerr{13.7}{0.7}~keV.}
\label{1550_param_hist2}
\end{figure}

\begin{figure}
\begin{center}
\includegraphics[scale=0.35]{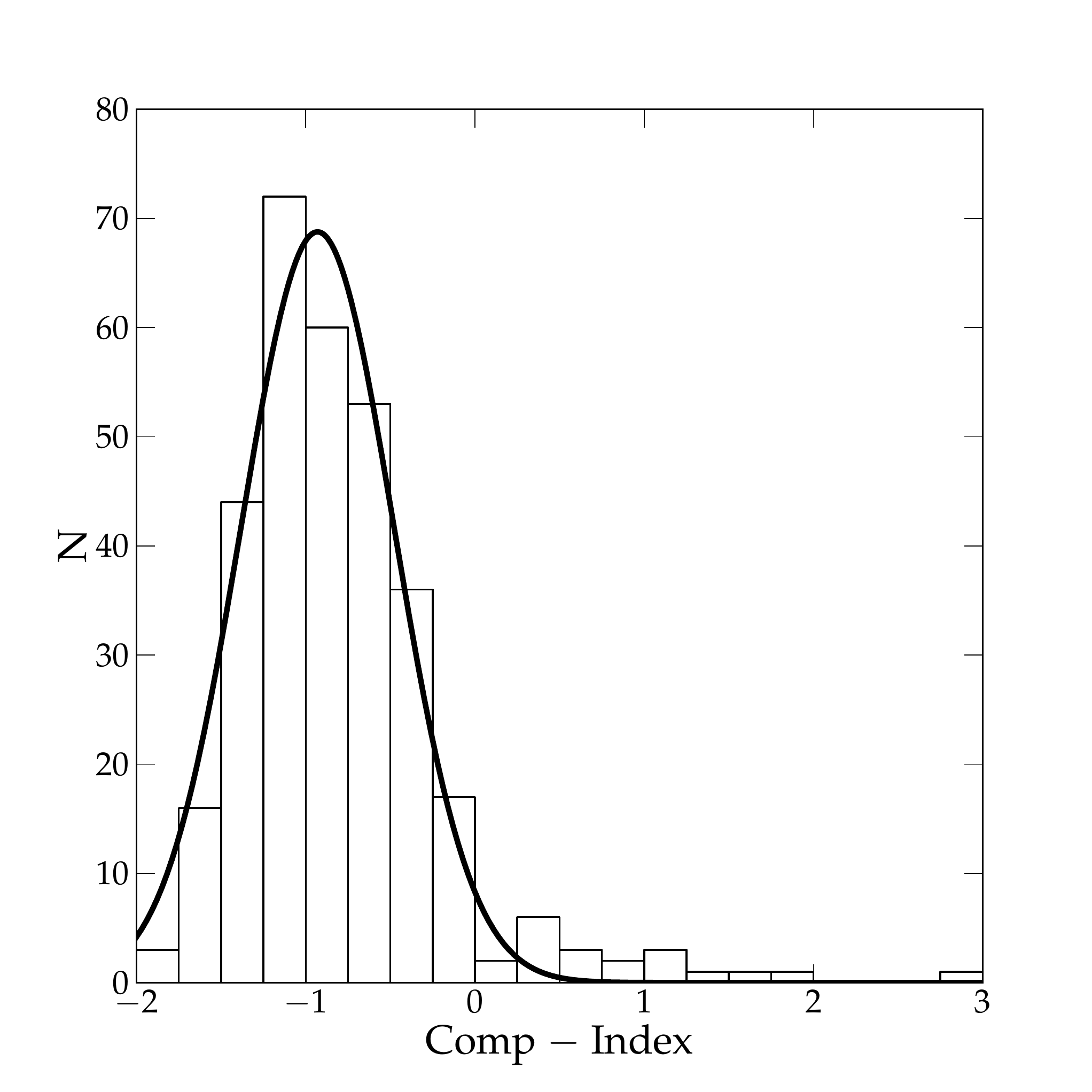}
\end{center}
\caption{Power-law spectral index as measured by the COMP model for all bursts of \Jfifteennos.  The distribution was fit to a Gaussian (black line) with $\mu =$ \valerr{-0.93}{0.02} and $\sigma =$ \valerr{0.45}{0.02}.}
\label{1550_param_hist3}
\end{figure}

The distributions of the spectral parameters discussed above are shown in Figures \ref{1550_param_hist}$-$\ref{1550_param_hist3}. The solid lines represent Gaussian fits, giving similar results to our prior analyses of individual periods (Table~\ref{1550_burst_info}).  We find that the BB$+$BB low and high temperatures center at $\sim4.6$~keV and $\sim15.0$~keV, respectively, with the width of the higher $kT$ distribution about three times broader than the lower $kT$ one. The apparent sub-peak around 12 $kT$ for the high-$kT$ distribution is caused largely by bursts from the October activity period. The peak energies (Figure~\ref{1550_param_hist2}) from the COMP and OTTB models agree with each other, with an average of $\sim40$~keV.  Finally, we find the COMP power-law index distribution to be narrow with an average of \valerr{-0.93}{0.02}  (Figure~\ref{1550_param_hist3}). The significant tail excesses in the OTTB $kT$ and the COMP index distributions, above $\sim80$~keV and 0, respectively, are due to events from all three periods. 

\subsubsection{Energetics}
\label{J1550_Energetics}

The fluence range of the  \Jfifteen bursts covers over three orders of magnitude, from $\sim$\scinum{7}{-9}~erg/cm$^2$ to at least $\sim$\scinum{1}{-5}~erg/cm$^2$ ($8-200$ keV). The corresponding total energy range is $\sim$ \scinum{2}{37} $d_{5}^{2}$ erg to $\sim$ \scinum{3}{40} $d_{5}^{2}$ erg (with $d_{5}$ the distance to the source divided by its estimated distance of 5~kpc). 
Compared to the other prolific magnetars, these bursts are brighter than the ones reported for \SGReighteen (\scinum{1.2}{-10} $-$ \scinum{1.9}{-7} erg/cm$^2$; \scinum{3.0}{36} $-$ \scinum{4.9}{39} erg; for $E>25$~keV; \citealt{Gogus01}) and \SGRMCM (\scinum{1.2}{-10} $-$ \scinum{3.3}{-7} erg/cm$^2$; \scinum{7}{35} $-$ \scinum{2}{39} erg; for $E>25$~keV; \citealt{Gogus01}).  \citet{Woods99} give the peak luminosity range of \SGRMDC bursts to be $10^{39} - 10^{42}$ erg/s ($E > 25$ keV), which corresponds (assuming a burst average duration of 0.1~s) to an energy range of $10^{38} - 10^{41}$~erg, indicating that these bursts were at the higher end of the energy distribution. However, since these data were obtained with a higher threshold instrument with limited trigger algorithm options for very short events ({\it CGRO}/BATSE), the non-detection of a fainter subset may well have been an instrumental effect. \Jfifteen bursts are an order of magnitude fainter than those of \Jzerofive (\scinum{4}{-8} $-$ \scinum{2}{-5} erg/cm$^2$; $2.0\times10^{37} - 1.0\times10^{40}$ erg; for $8-200$ keV; \citealt{Lin11a}).  Finally, all SGR bursts are significantly more energetic when compared to bursts from AXP\,1E$2259+586$, which range between \scinum{5}{34} and \scinum{7}{36} erg ($2-60$~keV; \citealt{Gav04}).

Next we derive a lower limit on the total energy fluence emitted by bursts from the source.  The limit is taking into account that we could not include bursts that occurred when \textit{Fermi} was in the South Atlantic Anomaly, the Earth occulted the source, or no TTE data existed.  For this limit we include the unsaturated parts of the saturated bursts across all outburst periods.  We get a value of \scinum{2.2}{-4} erg/cm$^{2}$, which corresponds to a (lower-limit) energy release of \scinum{6.6}{41} $d_{5}^{2}$ erg ($8-200$ keV). This is 4 orders of magnitude above the upper limit derived for bursts from AXP\,1E$2259+586$ ($2-60$ keV; \citealt{Gav04}). 

\begin{figure}
\begin{center}
\includegraphics[scale=0.5]{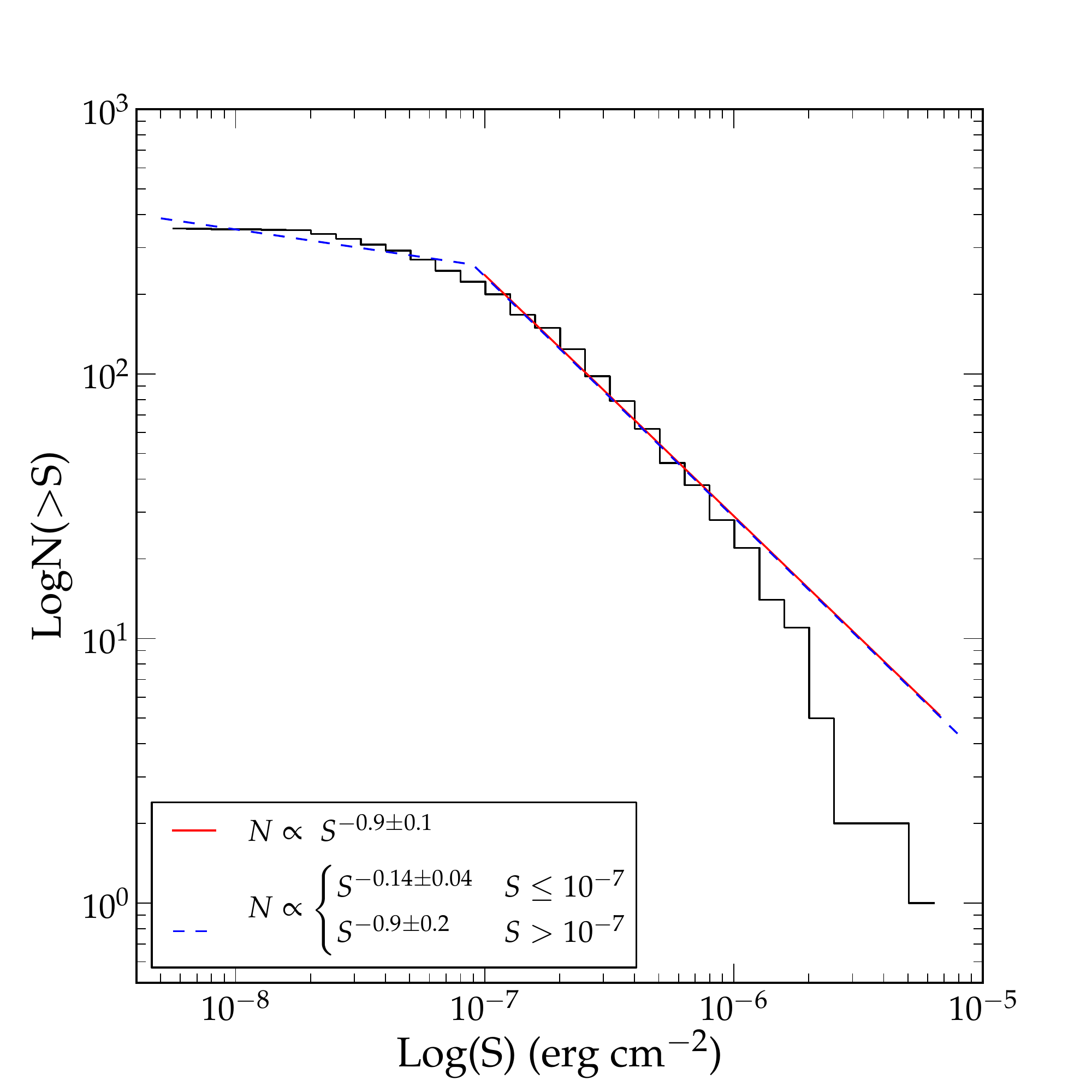}
\end{center}
\caption{Log N($>$S) $-$ Log (S) diagram of \Jfifteen bursts. The data are fit to a single PL (red solid line) and a broken  PL (blue dashed line). For the single PL, we only fit the data with S $>$ \scinum{1}{-7}~erg/cm$^2$.}
\label{1550_logNlogS}
\end{figure}

In Figure~\ref{1550_logNlogS} we present a Log N($>$S) $-$ Log (S) diagram for all 354 unsaturated events, where $S$ is the energy fluence ($8-200$\,keV). We fit this histogram with both a PL and a broken PL. For the single PL we only fit the bins  $\ge$\scinum{1}{-7} erg/cm$^2$, obtaining an index of $-0.9\pm0.1$.  We find that the broken PL fits the data best with a break at \scinum{1}{-7} erg/cm$^2$, and indices of $-0.14\pm0.04$ and $-0.9\pm0.1$ below and above the break, respectively. The PL slope above the break is steeper than what has been found for other magnetars with a large sample of bursts, ranging from $-0.5$ to $-0.7$ \citep{Gogus99,Woods99,Aptekar01,Gogus01,Gav04}, although a similar slope has been found for a large sample of \SGReighteen bursts \citep{Gotz06}. The PL slope below the break is clearly much shallower, which is due to the lower sensitivity of the instrument to faint bursts. Regarding the PL slope at high fluences, the energy and fluence ranges that are probed in our \Jfifteen sample differ from those in other magnetar samples, making a clear direct comparison unfeasible.

\begin{figure}
\begin{center}
\includegraphics[scale=0.4]{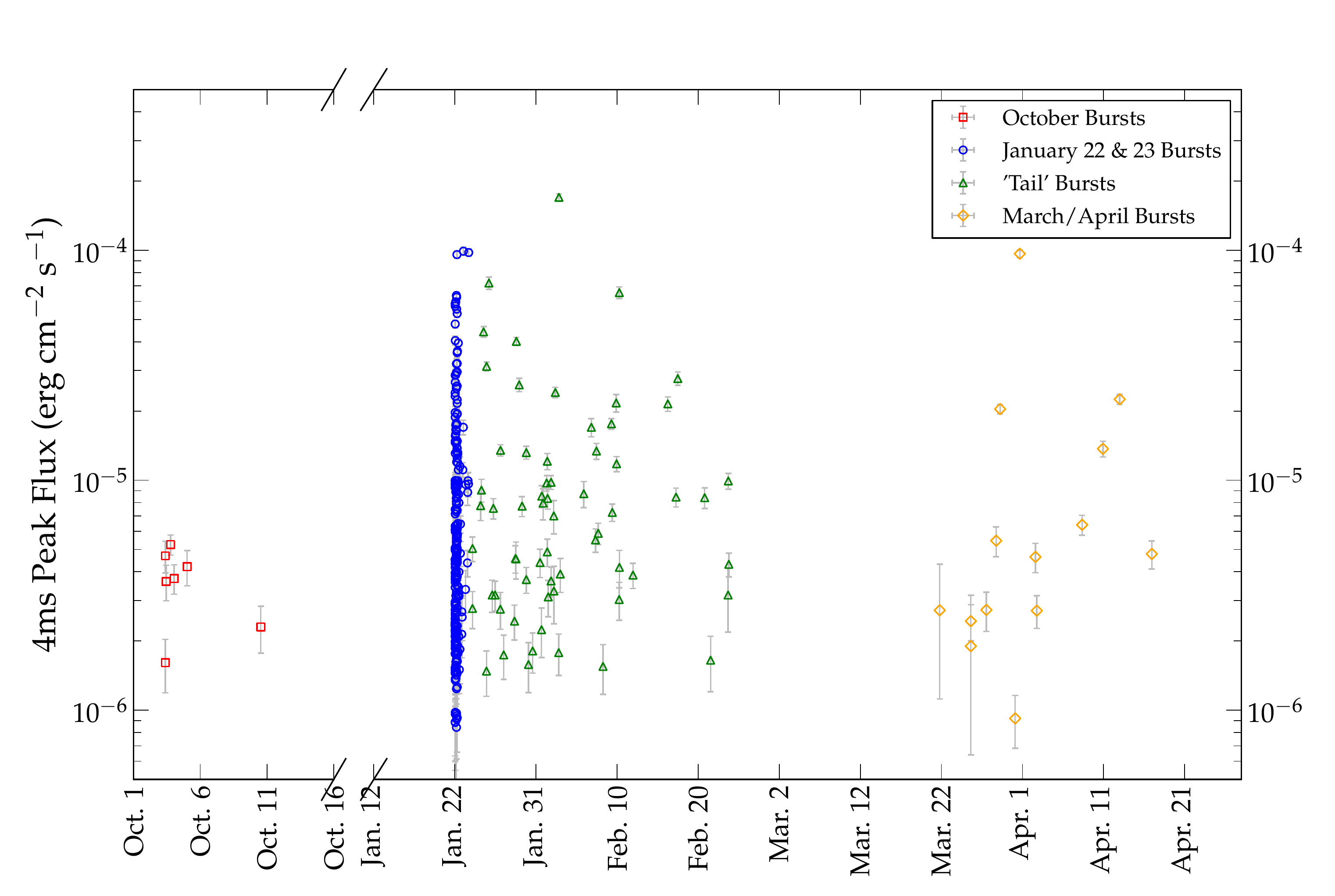}
\end{center}
\caption{Evolution of the 4~ms peak flux of all \Jfifteen bursts as measured by the COMP model.  The October bursts were faint compared to the other periods, but there is no other distinct evolution in the peak flux of the bursts over time.}
\label{1550_4ms_pk_all}
\end{figure}

Finally, we find that the 4~ms peak fluxes of \Jfifteen bursts range from $\sim$\scinum{8}{-7} to $\sim$\scinum{2}{-4} erg/s/cm$^{2}$, corresponding to a peak luminosity range of $\sim$\scinum{5}{38} to $\sim$\scinum{1}{41} erg/s. The 4~ms fluxes extend the peak flux range to about a full order of magnitude higher than was covered with the average fluxes from this source \citep{Lin12, Horst12, Kie12}.  Figure~\ref{1550_4ms_pk_all} shows the evolution of peak fluxes across the outburst and demonstrates the uniformly low peak fluxes of the events of the first period.

\subsubsection{Correlations}
\label{J1550_Correlations}

\begin{figure}
\begin{center}
\includegraphics[scale=0.4]{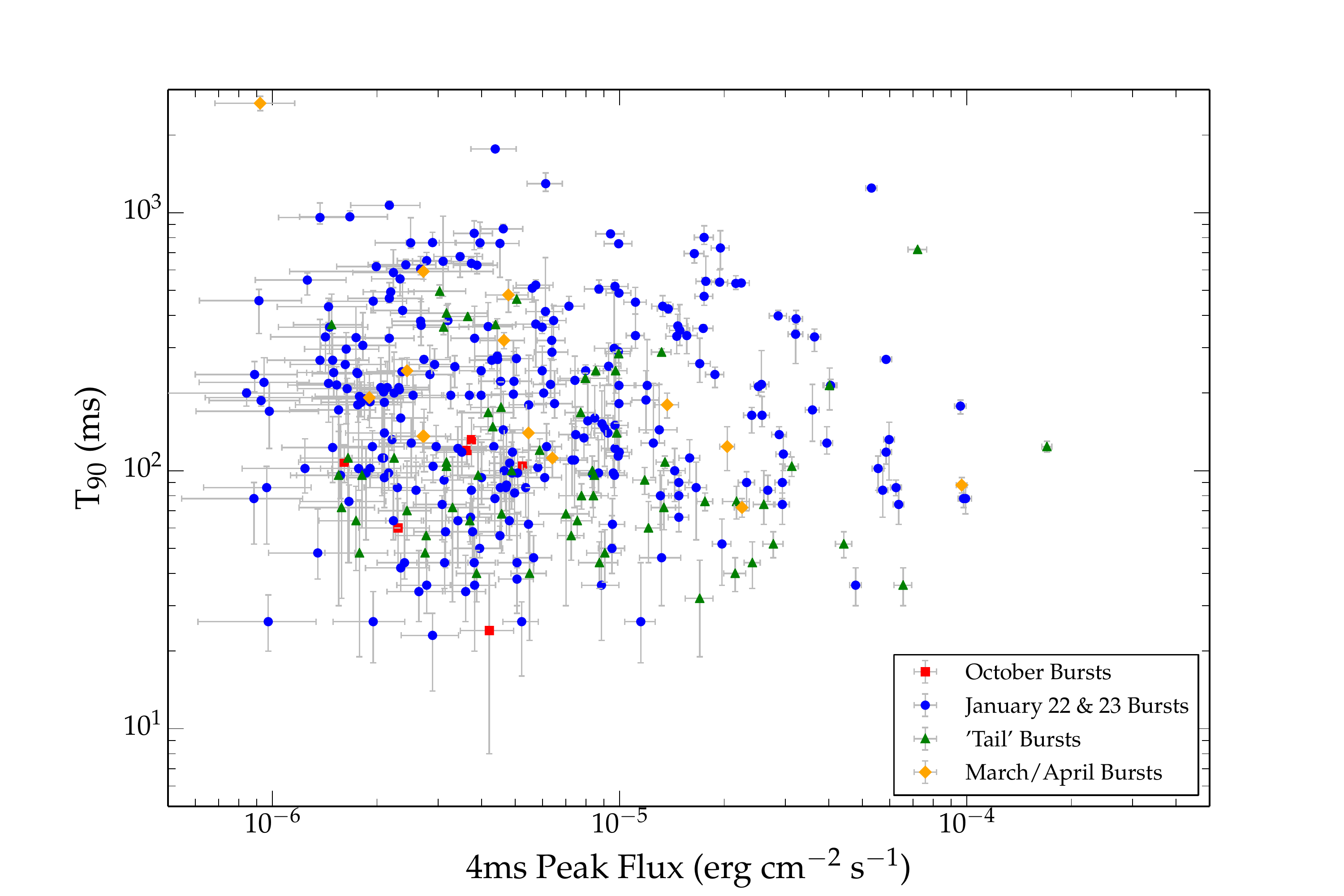}
\includegraphics[scale=0.4]{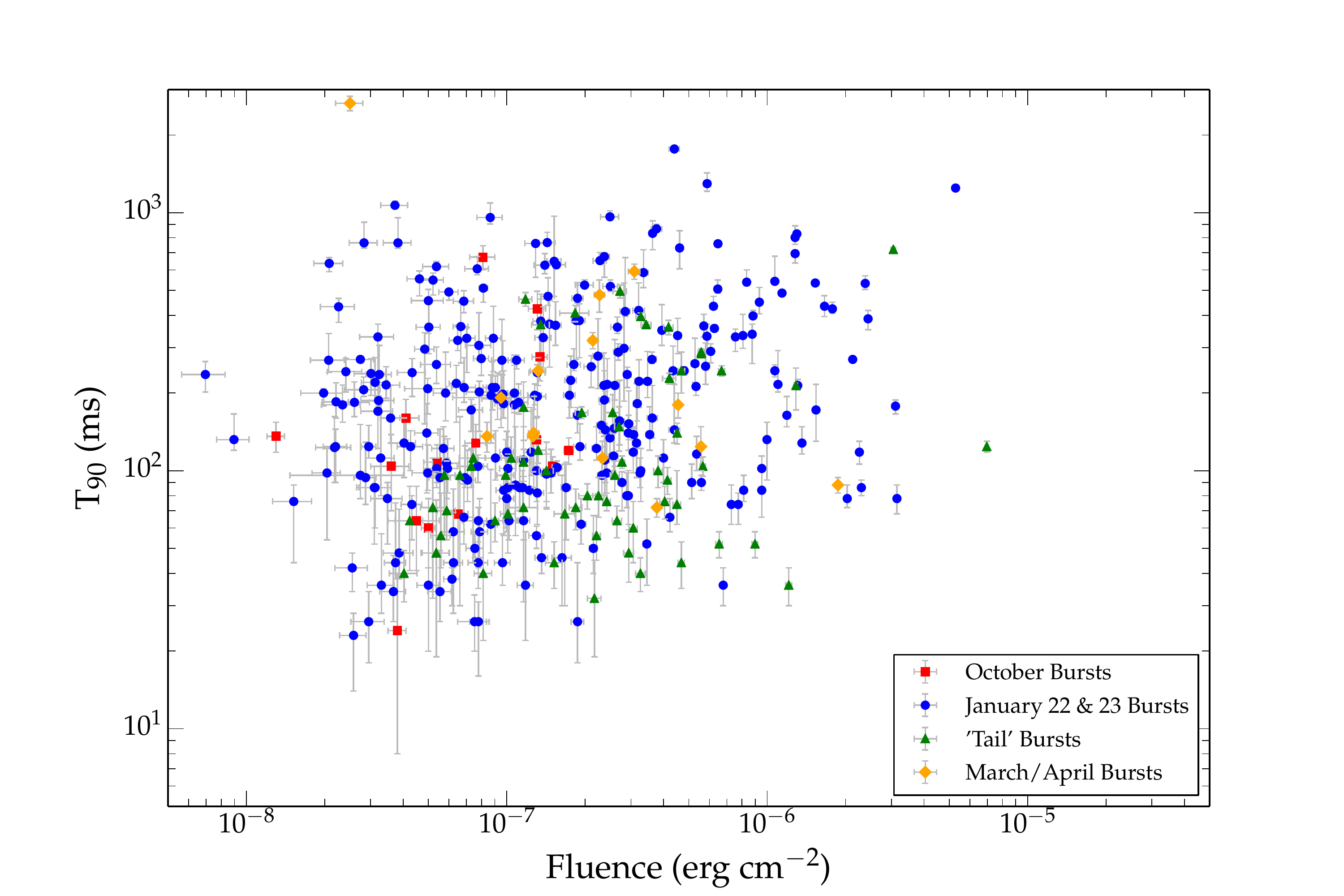}
\end{center}
\caption{\Tnin duration versus peak flux (top panel) and fluence (bottom panel) for all \Jfifteen bursts.  There is a marginally significant correlation between \Tnin and the fluence, but not between \Tnin and the peak flux.}
\label{1550_t90_v_fluflux}
\end{figure}

\begin{figure}
\begin{center}
\includegraphics[scale=0.4]{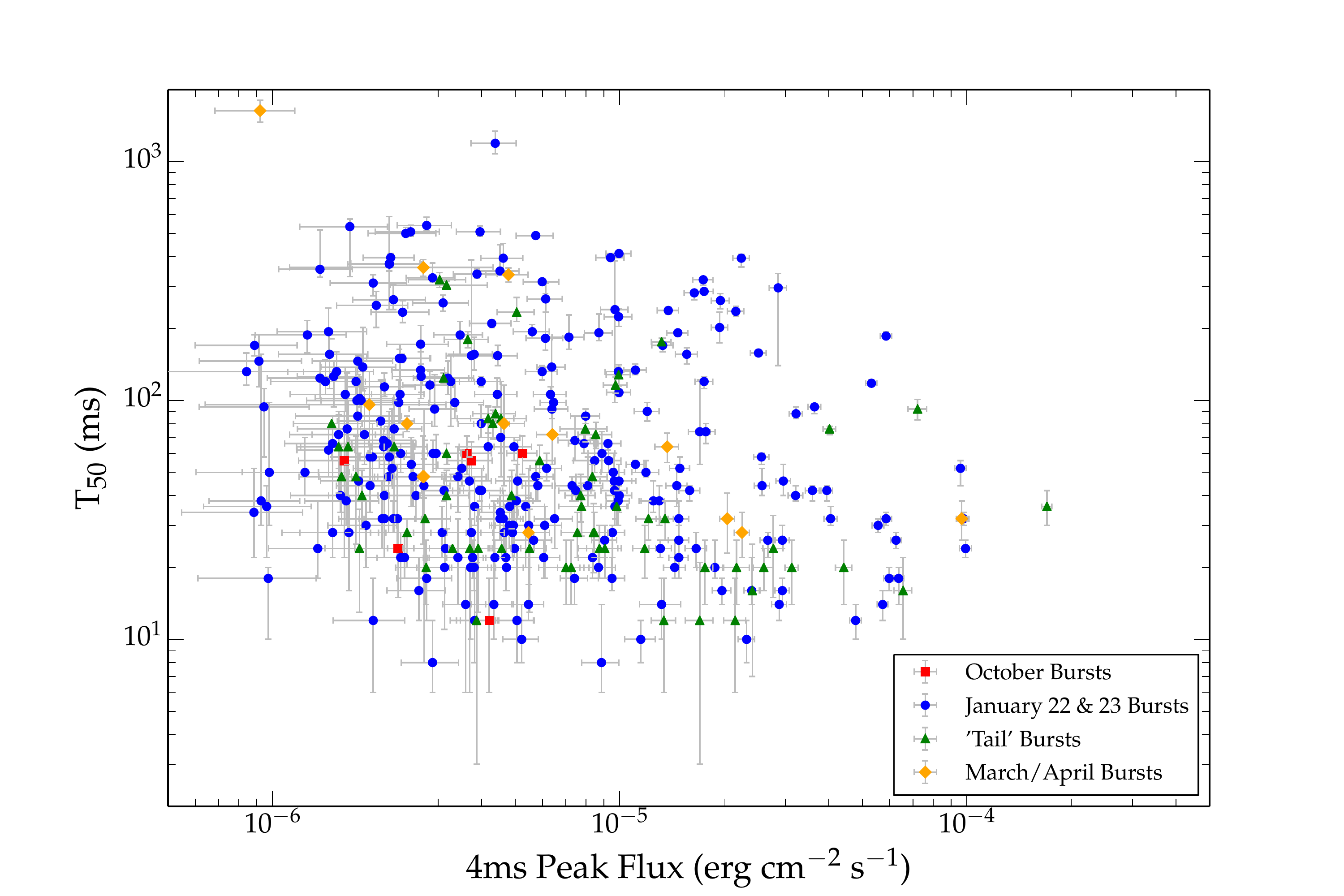}
\includegraphics[scale=0.4]{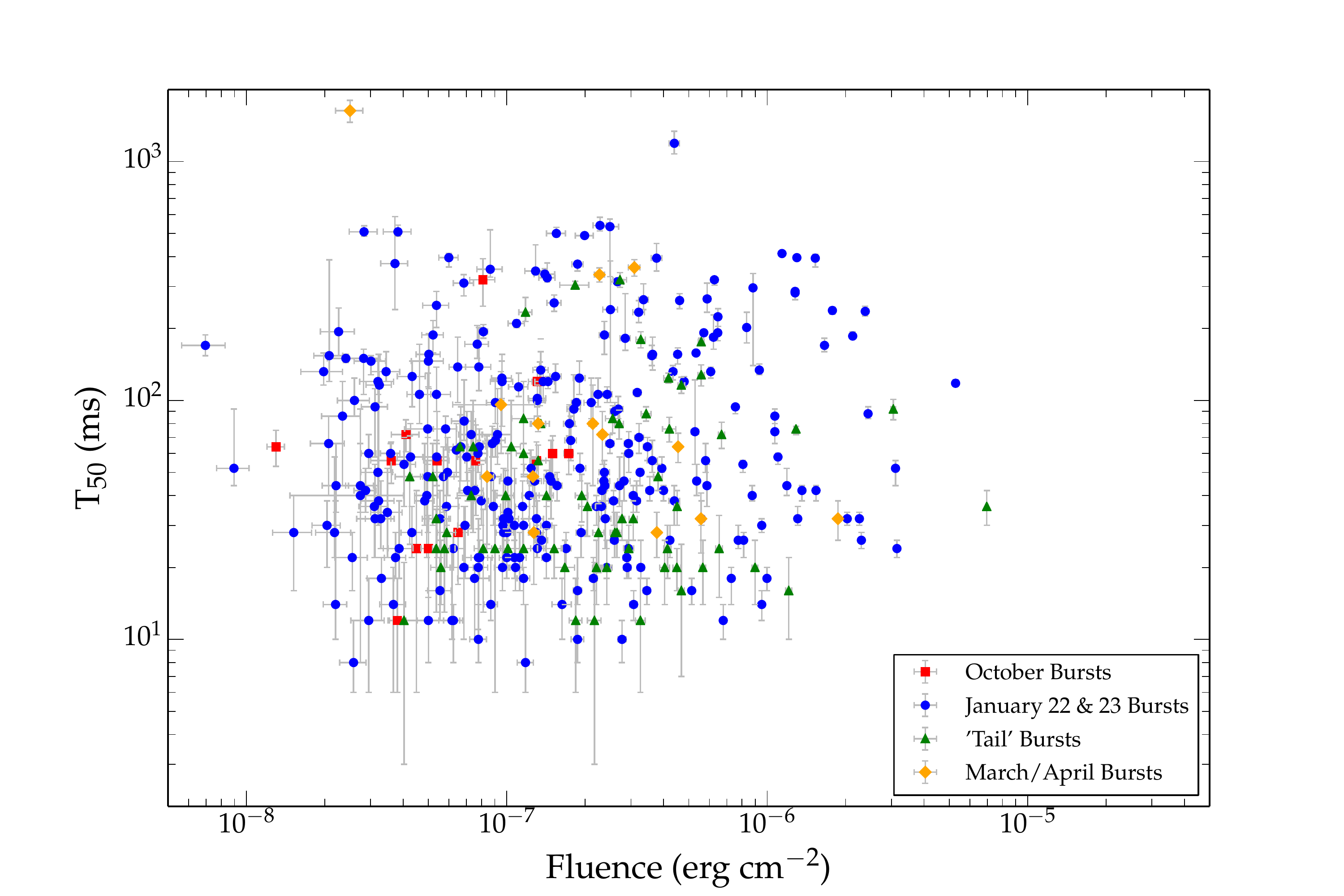}
\end{center}
\caption{\Tfif duration versus peak flux (top panel) and fluence (bottom panel) for all \Jfifteen bursts.  There is a marginally significant correlation between \Tfif and the peak flux, but not between \Tfif and the fluence.}
\label{1550_t50_v_fluflux}
\end{figure}

\begin{figure}
\begin{center}
\includegraphics[scale=0.35]{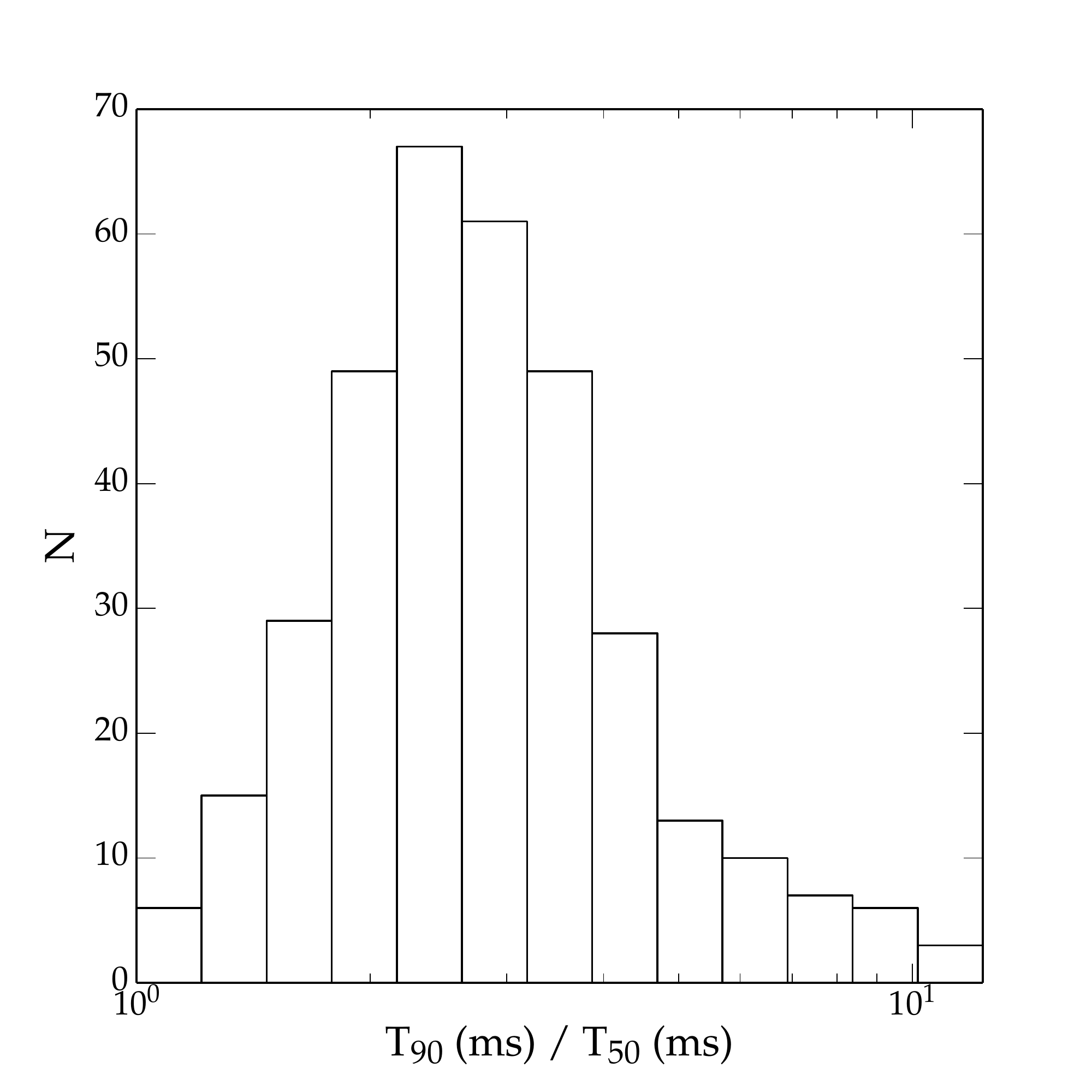}
\end{center}
\caption{Histogram of the \Tnin / \Tfif ratio for \Jfifteen bursts, showing that the distribution is broad, extending a factor of $\sim10$.}
\label{1550_hist_t90_t50_ratio}
\end{figure}

We searched for correlations among burst properties across the entire burst activity from the source, using the Spearman rank order correlation test. We considered multiple combinations of parameters and estimated the Spearman coefficient and probability for each; the results are listed in Table~\ref{1550_cor_table}. In most cases, the significance of the correlation was very low, but we identified the following significant correlations: fluence $-$ peak flux, \Ep $-$ fluence, and \Ep $-$ peak flux (all determined using the COMP model). Figure~\ref{1550_flu_v_pkflux} exhibits the strong correlation between fluence and peak flux; the data are fit with a PL with an index \valerr{1.2}{0.4}.  The \Ep $-$ fluence correlation is moderately weaker than the one previously reported for bursts of only the second period \citep{Horst12}, but there still is a very small chance probability of \scinum{9.7}{-14}.  We previously found a marginal correlation between \Tnin and fluence \citep{Horst12}; here we find a weaker \Tnin $-$ Fluence correlation (chance probability of \scinum{2.5}{-4}), and no correlation between \Tnin and peak flux (Figure~\ref{1550_t90_v_fluflux}). However, we do find a marginal anti-correlation (chance probability of \scinum{7.1}{-6}) between \Tfif and peak flux (see also Figure~\ref{1550_t50_v_fluflux}). This can be explained by the fact that for bursts of which the light curve is dominated by one very bright peak, the \Tfif is likely to encompass only that bright peak and not include any lower level activity that is part of that burst (while the latter is included in the \Tninnos). This is further illustrated in Figure~\ref{1550_hist_t90_t50_ratio}, where we show a histogram of the \Tnin / \Tfif ratio. This histogram peaks between 2 and 3, but has a broad tail extending above 10, which shows that there is indeed a large amount of bursts that consist of a bright peak and relatively long time span of lower level emission.

\begin{figure}
\begin{center}
\includegraphics[scale=0.4]{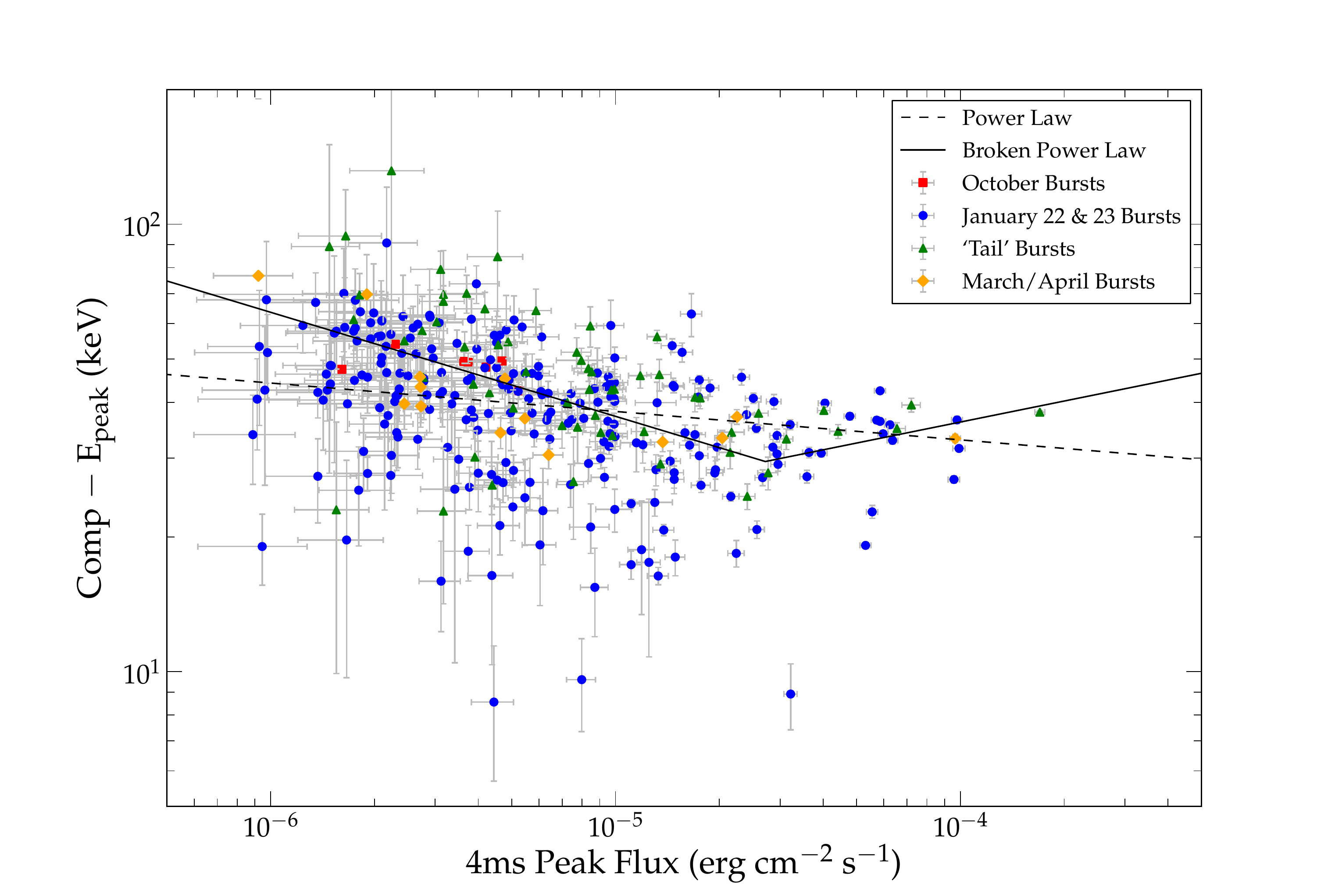}
\end{center}
\caption{\Ep from the COMP model versus the 4ms peak flux of \Jfifteen bursts. We fit this correlation with a PL (dashed line) and a broken PL (solid line).}
\label{1550_ep_v_fluflux}
\end{figure}

We estimated peak fluxes at a 4~ms time scale, and investigated the \Ep $-$ peak flux relation at this time scale for the first time in this study (Figure~\ref{1550_ep_v_fluflux}). We find that this correlation is considerably stronger than the one using the average fluxes, reported previously by \citet{Horst12}. We fit the data to both a PL and a broken PL.  The PL index is \valerr{-0.06}{0.01}, and the broken PL indices are \valerr{-0.23}{0.03} and \valerr{0.16}{0.03} with the break at \scinum{2.7}{-5} erg/s/cm$^{2}$. The break value is almost an order of magnitude higher than the one estimated using the average flux values \citep{Horst12}, but closer to the break flux determined in time-resolved spectroscopy of the brightest \Jfifteen bursts \citep{Younes14}. The latter is also comparable to the break flux estimated for the time-resolved spectroscopy of bright bursts from \Jzerofive \citep{Lin11a}. We note that we find a considerably stronger correlation among the bursts below the break (Spearman rank $= -0.422$, chance probability of \scinum{5.9}{-14}) than the one reported in \citet{Horst12}.  For the bursts above the break, we again see an improved but still insignificant correlation (chance probability of 0.17) as compared to one found when using the average flux (chance probability of 0.64).

\subsubsection{Burst Peak Arrival Times - Pulse Phase Correlation Analysis}
\label{J1550_BurstPhase}

We have studied the burst peak arrival times with respect to the spin phase of the source. To determine a burst peak, we fit for each burst the time-resolved spectra accumulated over 4~ms bins to a COMP model using \RMFITnos. The binning process was repeated for different bin edges to account for windowing effects.  We assigned the time of the center of the brightest 4~ms as the peak in each burst. All times were barycenter corrected using the JPL ephemeris file as provided by HEASOFT-6.11.1; the phase of each burst peak was estimated using the spin ephemeris of \citet{Dib12}.

\begin{figure}
\begin{center}
\includegraphics[scale=0.5]{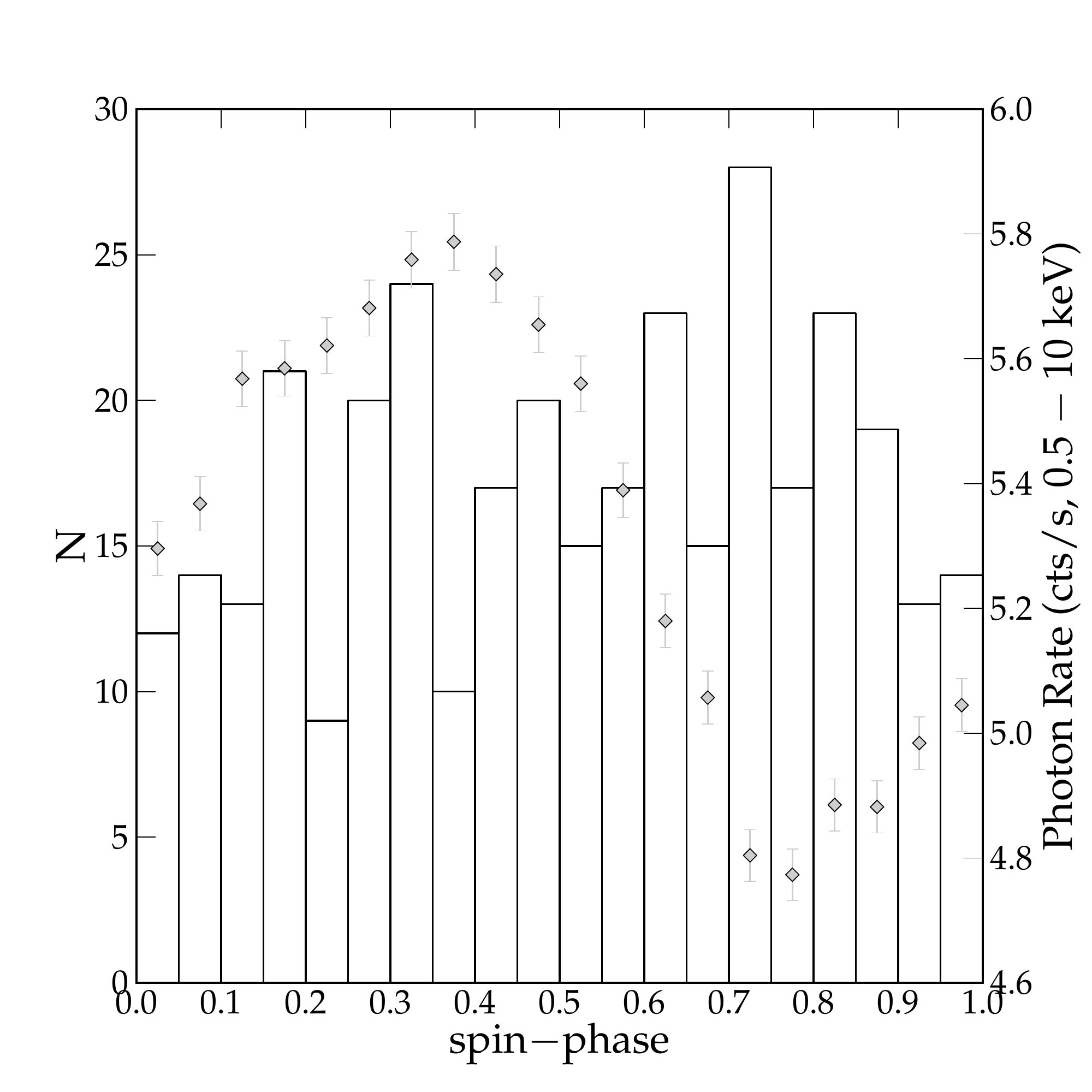}
\end{center}
\caption{Distribution of the phase of the peak of all the \Jfifteen bursts. We do not find a preferred spin phase for burst peaks, consistent with what has been found for other magnetars.}
\label{1550_phasehist_all}
\end{figure}

\begin{figure}
\begin{center}
\includegraphics[scale=0.4]{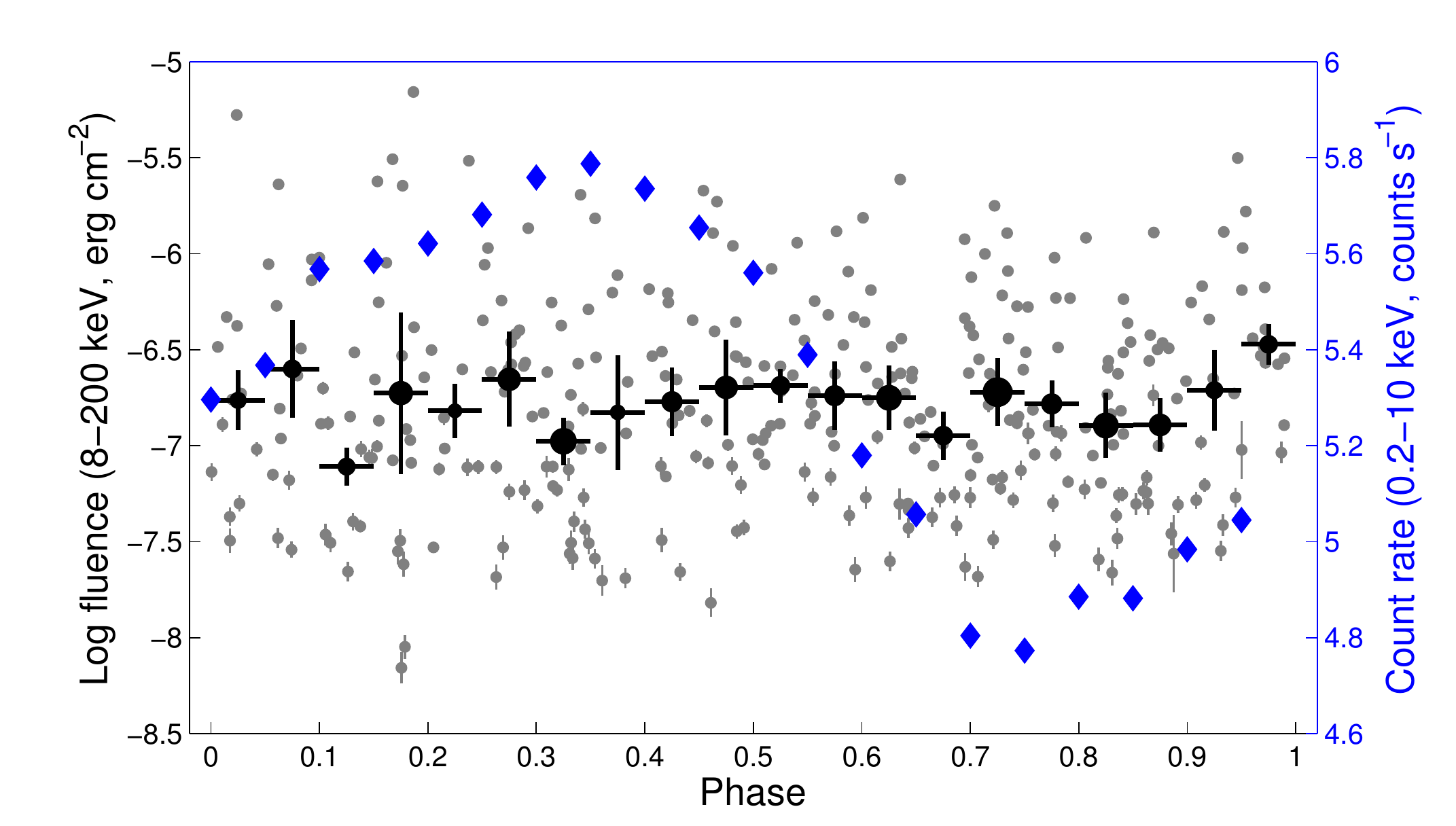}
\includegraphics[scale=0.4]{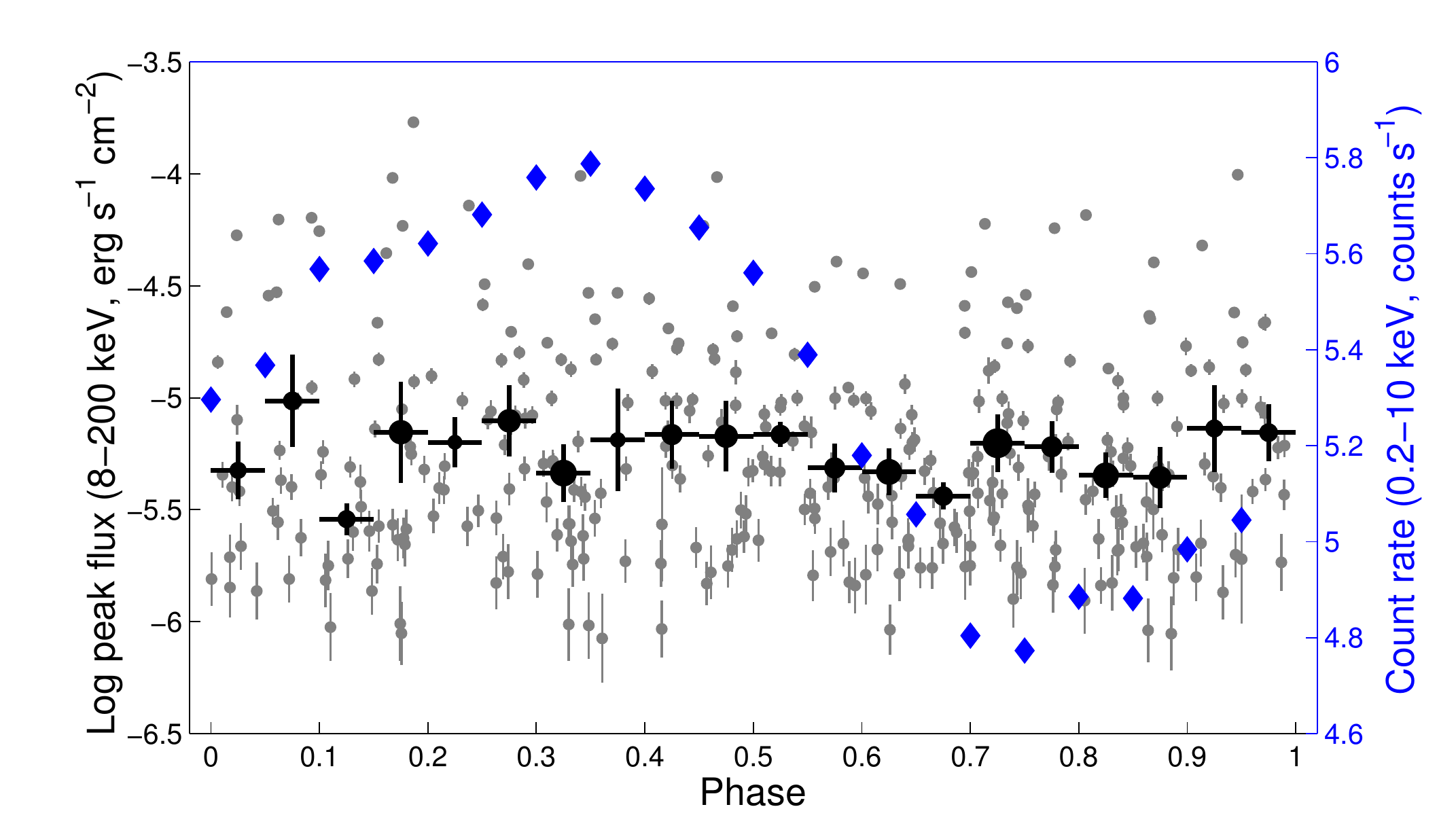}
\includegraphics[scale=0.4]{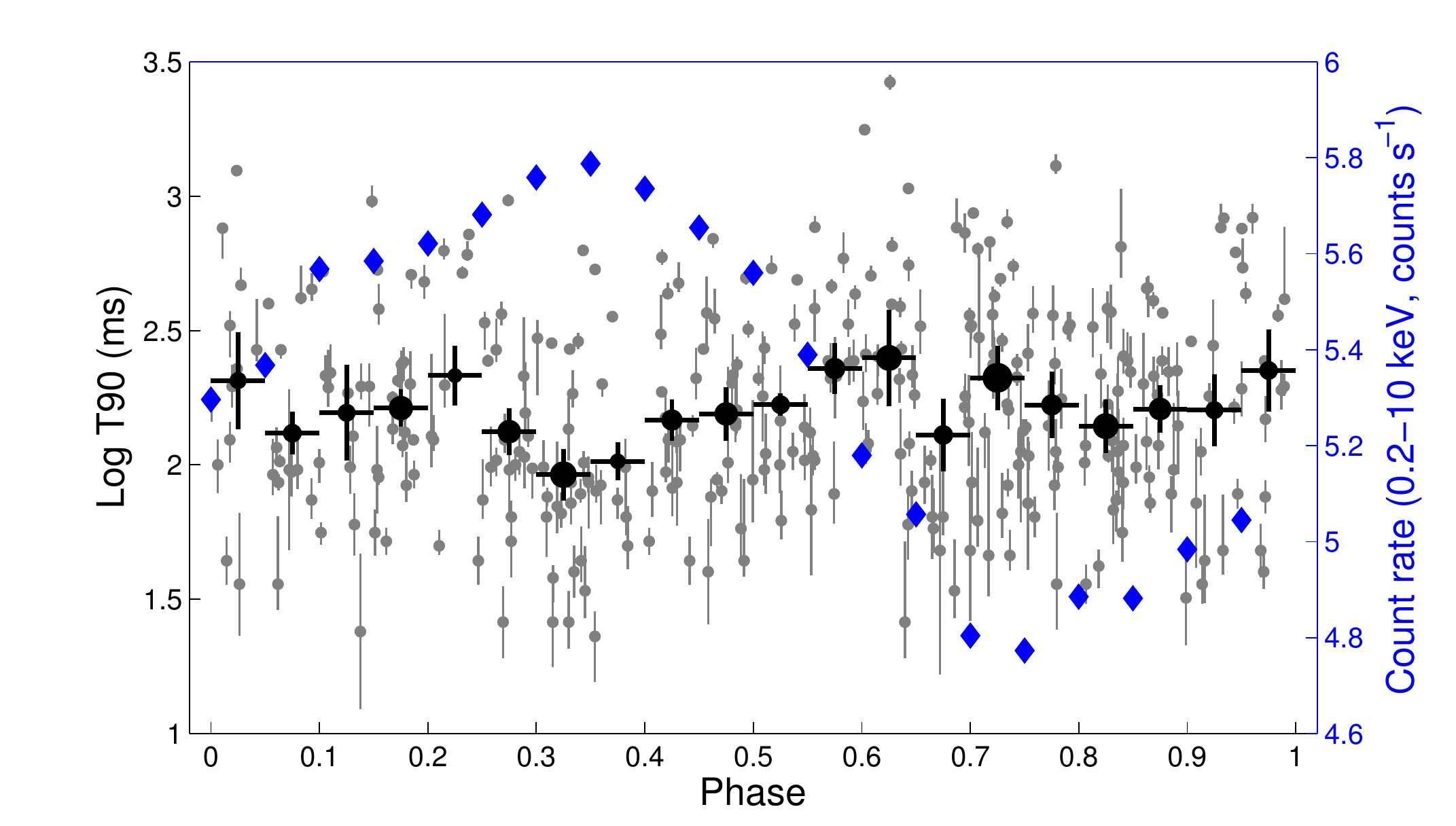}
\end{center}
\caption{Fluence (top panel), peak flux (middle panel), and \Tnin (bottom panel) of \Jfifteen bursts compared to the spin phase. The gray dots represent each individual burst, and the black dots represent the weighted average of these gray dots grouped by phase. The size of the black dots represent how many bursts are in each group. The blue diamonds indicate the 0.2$-$10~keV pulse profile \citep{Lin12}. There are no correlations between the fluence, peak flux or \Tnin and the spin phase.}
\label{enhanced_fldvp}
\end{figure}

Figure~\ref{1550_phasehist_all} shows a histogram of all burst peak phases, with the \Jfifteen pulse profile as measured with {\it XMM-Newton} \citep{Lin12} overplotted. We find that the peak arrival times are distributed uniformly across the spin phase (average bin height: N = 17 events, $\sigma= 5$ events). We tested for a correlation between the peak flux, fluence, and burst duration against spin phase. Figure~\ref{enhanced_fldvp} shows the individual points, their weighted averages by 0.05 phase bins, and the \Jfifteen pulse profile. We find no correlation between these three burst properties with spin phase.
 
\subsubsection{Burst Profile Epoch Folding}
\label{J1550_folded}

\begin{figure}
\begin{center}
\includegraphics[scale=0.4]{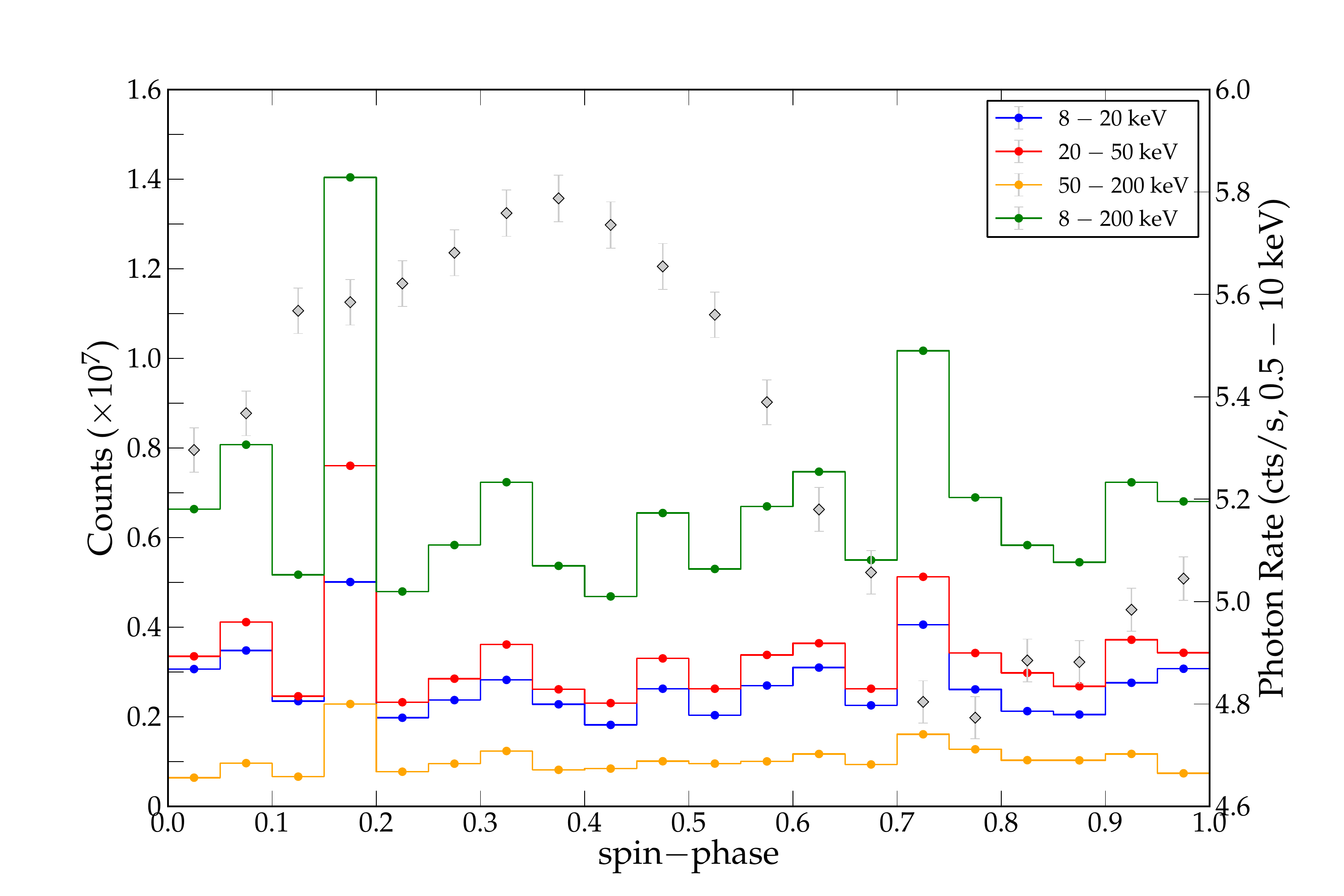}
\end{center}
\caption{Folded lightcurve of all bursts from \Jfifteen in 3 channels: 8$-$20~keV (blue), 20$-$50~keV (red), and 50$-$200~keV (yellow). The sum of all three channels is shown in green. There is a possible peak at the bin that spans phase 0.15$-$0.20. This peak, however, is not significant (see Section~\ref{J1550_folded}).}
\label{folded_lc}
\end{figure}

\citet{S&K11} found, using \textit{Swift}/XRT data ($0.2 - 10$ keV), that the epoch folded light curves had one  (fairly smooth) peak, at $0.45-0.65$ of the pulse phase. However, this was not found in an analysis of \textit{Swift}/XRT bursts by \citet{Lin12}. We have performed a similar analysis with the GBM data to check this at higher energies. We epoch folded the background-subtracted light curves of each burst using the ephemeris of \citet{Dib12}. We then added all folded light curves together to create the pulse profiles shown in Figure~\ref{folded_lc}. We display the folded profiles in four energy bands: $8-20$ keV, $20-50$ keV, $50-200$ keV, and $8 - 200$ keV together with the pulse profile in $0.5 - 10$ keV. We notice a peak at phase $0.15 - 0.20$, which appears in all energy bands and is not associated with the pulse maximum or minimum. 

\begin{figure}
\begin{center}
\includegraphics[scale=0.4]{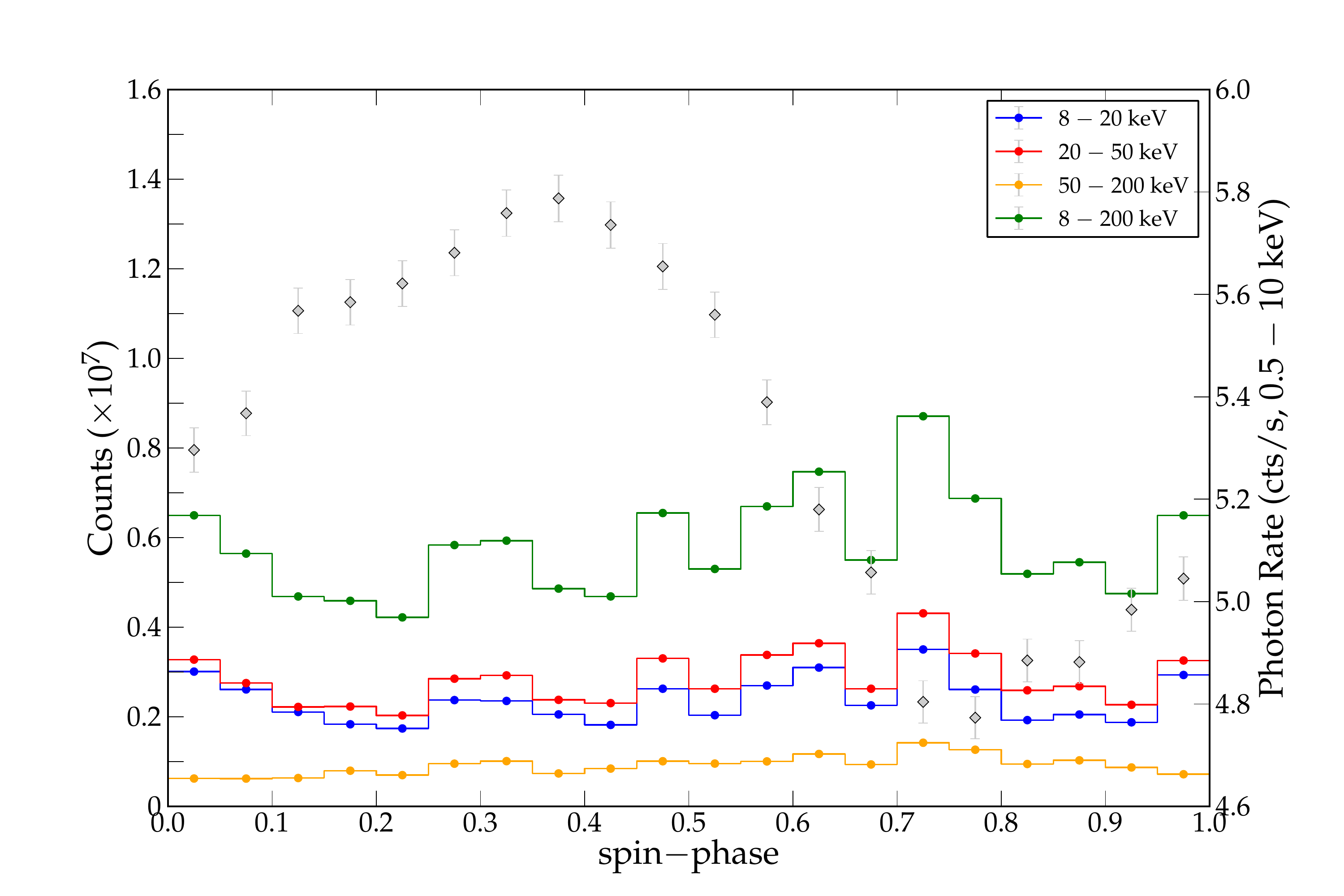}
\includegraphics[scale=0.4]{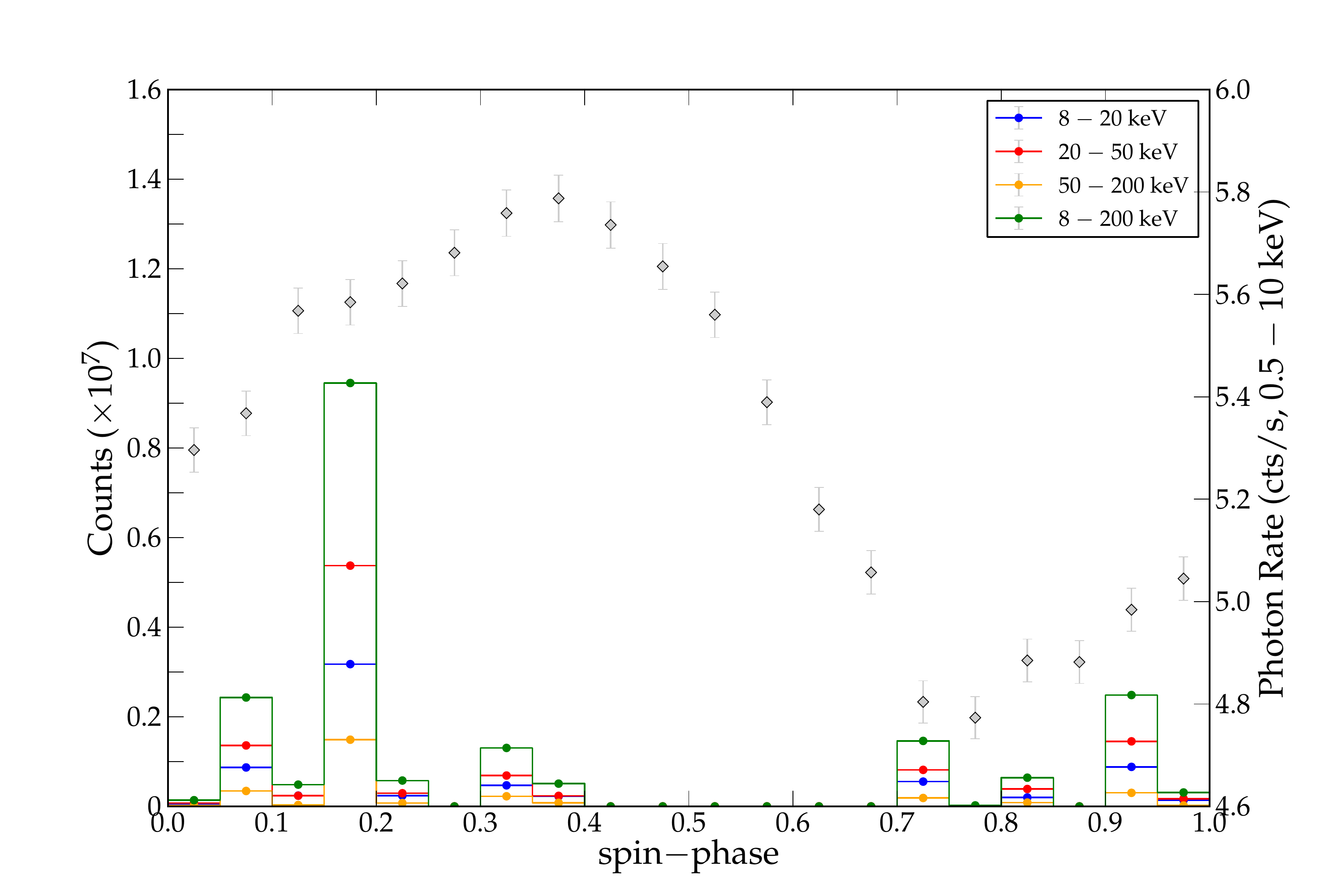}
\end{center}
\caption{Folded lightcurve of all bursts from \Jfifteen without the ten brightest ones (top panel), and folded light curve of only the ten brightest ones; color coding the same as in Figure~\ref{folded_lc}.  The peak at phase bin 0.15$-$0.20 in Figure~\ref{folded_lc} is not apparent in the top panel, but clearly present in the bottom panel.}
\label{folded_lc_10}
\end{figure}

We have studied the significance of this peak by calculating its deviation from the mean as $\sigma_{i} = \left(x_{max} - \bar{x}\right) / \sigma_{19}$, where $x_{max}$ is the value of the peak, and $\bar{x}$ is the average of the histogram without that maximum bin. For the $8-200$ keV energy range, we find $\sigma_{i} \sim 5.9$. To test for the significance of this deviation, we performed a Monte Carlo simulation, randomly shifting all \Jfifteen light curves in phase space by a random number chosen from a uniform distribution. After each shift, we created a new count-phase histogram and recalculated $\sigma_{i}$. This process was repeated 10,000 times, and we find that 1.24\% of the time, $\sigma_{i}$ would be at least as high as it is in the real data. This corresponds to a  $\sim 2.5\sigma$ result, rendering this feature insignificant. Interestingly, we find that if we remove the top ten brightest bursts from the histogram, the peak disappears (Figure~\ref{folded_lc_10}), suggesting that the brightest bursts are preferring a certain spin phase, a fact that may have been buried in the arrival time analysis. We tested this peak in the same way as described above and find that in this case, $\sigma_{i}\sim11.4$. However, a Monte Carlo simulation on the brightest 10 bursts shows that this peak is only a $\sim2.95\sigma$ result. Therefore, our findings confirm the results of \citet{Lin12} that there is no strong evidence for a correlation between the burst counts and spin phase.

\subsection{\Jzerofive}
\label{J0501}

\Jzerofive was discovered with \Swift/BAT \citep{H&S08, Barthelmy08} when it emitted a magnetar-like burst on 2008 August 22, which also triggered GBM \citep{Lin11a}. The entire source outburst lasted for about 2 weeks, with multiple bursts observed with \Swift, GBM, \RXTE, \Konus, and \Suzaku\ \citep{Enoto09, Aptekar09, Kumar10, Lin11a}. \citet{Gogus08} found the spin-period with \RXTE: $\sim5.76$s; the combined \RXTE and \Swift/XRT data revealed a spin-down rate of $\dot{P} =$ \scinum{1.5}{-11}~s/s, which corresponds to a magnetic field of \scinum{2.0}{14}~G \citep{Woods08, Rea09, Gogus10}. The most accurate location of \Jzerofive was found with \Chandra\ to be RA = $05^{\rm{h}}01^{\rm{m}}06^{\rm{s}}.76$, Dec = $+45^{\circ}16^{\prime}13^{\prime\prime}.92$ with an uncertainty of $0^{\prime\prime}.11$ \citep[$1\sigma$;][]{Gogus10}. The source distance has been estimated to be $\sim2$~kpc \citep{Xu06,Lin11a}.

\begin{figure}
\begin{center}
\includegraphics[scale=0.5]{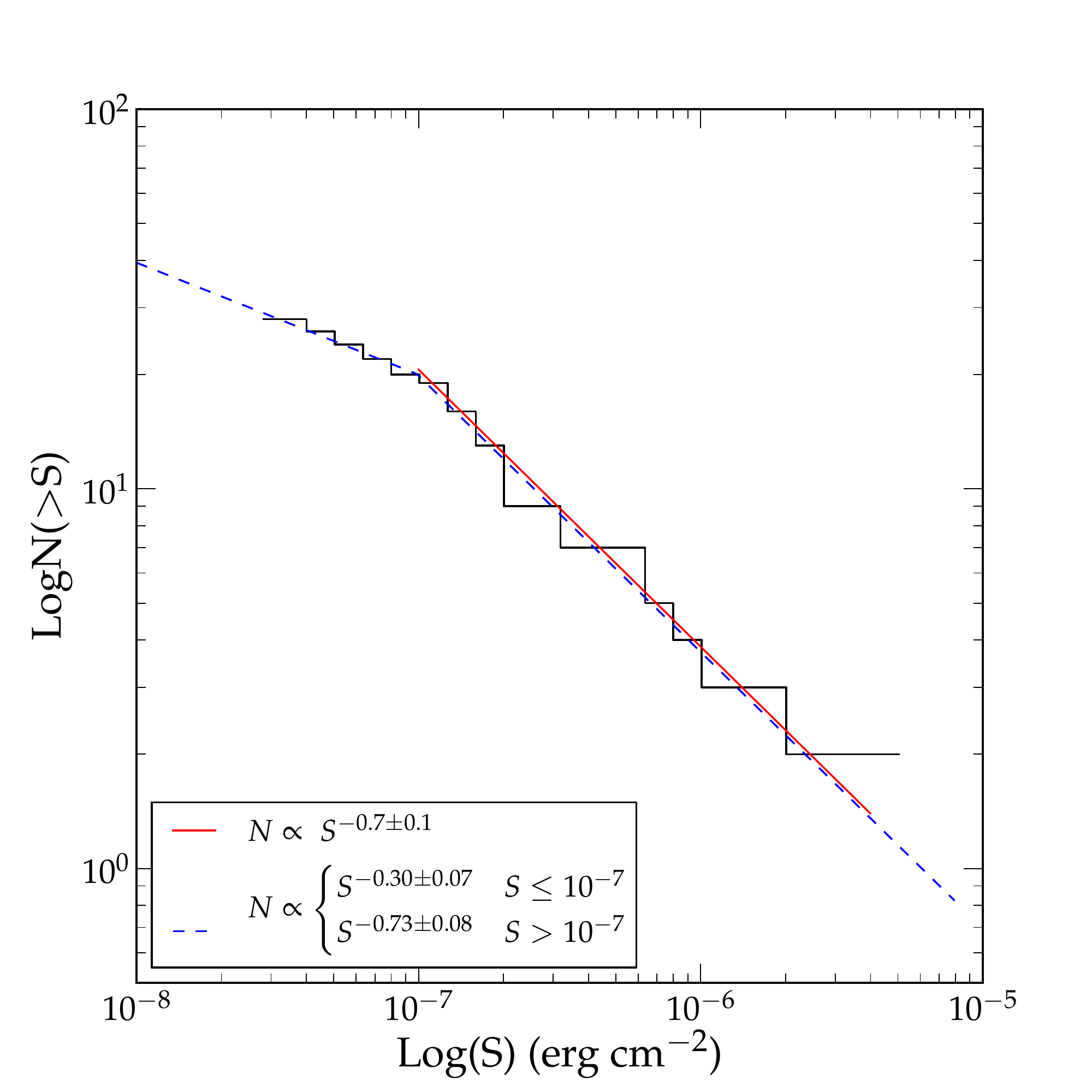}
\end{center}
\caption{Log N($>$S) $-$ Log (S) diagram of \Jzerofivenos. The data are fit to a single PL (red solid line) and a broken  PL (blue dashed line). For the single PL, we only fit the data with S $>$ \scinum{1}{-7}~erg/cm$^2$.}
\label{0501_logNlogS}
\end{figure}

GBM triggered on 26 bursts during the entire outburst interval; an untriggered search as described in \citet{Kaneko10} yielded 3 more bursts with TTE data, resulting in a total of 29 bursts. Two of these bursts saturated the High Speed Science Data Bus of GBM and were excluded from the detailed temporal and spectral analyses which are described in \citet{Lin11a}. In Tables~\ref{0501_durations}$-$\ref{0501_Spec} we provide the \Tnin and \Tfif durations in photon space, and the spectral fitting parameters for these bursts. We display in Figure~\ref{0501_logNlogS} the Log N($>$S) $-$ Log (S) diagram for \Jzerofivenos bursts, computed in this study. We fit the histogram with both a PL and a broken PL. For the former we only fit the bins $\ge$ \scinum{1}{-7}~erg/cm$^2$, yielding an index of $-0.7\pm0.1$. The broken PL indices are $-0.30\pm0.07$ and $-0.73\pm0.08$ below and above a break at $10^{-7}$ erg/cm$^2$, respectively. We note that in contrast to \Jfifteennos, the single PL index and the high-fluence index for the broken PL are consistent with indices found for other magnetars \citep[Section~\ref{J1550_Energetics};][]{Gogus99,Woods99,Gogus01,Gav04}. 

Finally, we note that a detailed variability analysis of the GBM bursts of \Jzerofive did not result in a significant detection of QPOs \citep{Huppen13}.

\subsection{\oneEeight}
\label{1841}

\oneEeight is a magnetar located at RA = $18^{\rm{h}}41^{\rm{m}}19^{\rm{s}}.343$, Dec = $-04^{\circ}56^{\prime}11^{\prime\prime}.16$ \citep[from \Chandra\ observations][]{Wachter04}, and at a distance of $\sim8.5^{+1.3}_{-1.0}$~kpc \citep{T&L08}. \ASCA\ observations have revealed a period of $\sim11.8$~s \citep{V&G97}; the spin down rate of $\dot{P} =$ \scinum{4.16}{-11}~s/s has been determined through \ASCA, \RXTE, and {\it BeppoSaX} observations \citep{Gotthelf99, Gotthelf02}. \Swift/BAT was first triggered by one burst on 6 May 2010 \citep{K&SH10}. This was followed by a period of activity between 8 February and 2 July 2011. During this time, \Swift\ observed 4 bursts \citep[e.g.,][]{Barthelmy11, Rowlinson11, Melandri11}. \Fermi/GBM triggered on magnetar-like events six times during this time interval, with one of these triggers being one of the four \Swift\ detections. The other five detections have no concurrent observations with other instruments. In addition, an untriggered burst search was performed over all of February 2011 and 10 June $-$ 6 July 2011 using the algorithm described in \citet{Kaneko10}. This search yielded only one burst, but one for which there is no TTE data, so it is excluded from analyses. We argue that despite the lack of concurrent observations, given the coincidence in time and position of these bursts, it is highly likely that they originate from \oneEeight \citep[see][]{Lin11b}.

Detailed spectral and and temporal analyses of these GBM and \Swift\ bursts are described in \citet{Lin11b}.  Here we provide in Tables~\ref{1841_durations}$-$\ref{1841_Spec} the \Tnin and \Tfif durations in photon space, and the spectral parameters for these bursts. New in this catalog compared to previous studies, we provide OTTB time-integrated spectral fits and 4~ms peak fluxes for these bursts (see Table~\ref{1841_Spec}).

\subsection{Bursts from other known magnetar sources}
\label{other_known}

In this section, we cover 7 bursts each originating from a known source as confirmed with \Swift\ observations, but with not enough bursts to perform a statistical analysis as with the magnetars in the previous sections. These include two bursts from a new magnetar source,  \Jzerofournos, discovered with GBM and also observed with \Swift/BAT and Konus-RF \citep{Horst09}; the GBM data were published in \citet{Horst10} and are also included in Tables~\ref{known_table1}$-$\ref{known_table2}. The other events originated from \SGReighteen \citep[contemporaneous \Swift/BAT detection;][]{Bhat10}; \Jeighteennos, which is a new source discovered and localized by \Swift\ \citep{Cummings11, Pagani11, K&K11};  \AXPfourunos, for which GBM observed a burst simultaneous with a \Swift/BAT burst \citep{Palmer13, Bhat11}; \SGRoneEnos, for which a single burst was observed with a location consistent with the source during an increased activity of the source's persistent emission \citep{Foley12, Archibald13}; \Jfifteennos, which emitted a single burst in June 2013, which was also detected with \Swift/BAT \citep{Holland13}, and was comparable in brightness to the bursts in the October 2008 period. For all bursts we performed spectral and temporal analyses following the methods described in Section~\ref{data}. The results of these analyses are presented in Tables~\ref{known_table1}$-$\ref{known_table2}.

\section{Magnetar Bursts of unconfirmed origin}
\label{sgr_unknown}

\begin{figure}
\begin{center}
\includegraphics[scale=0.7]{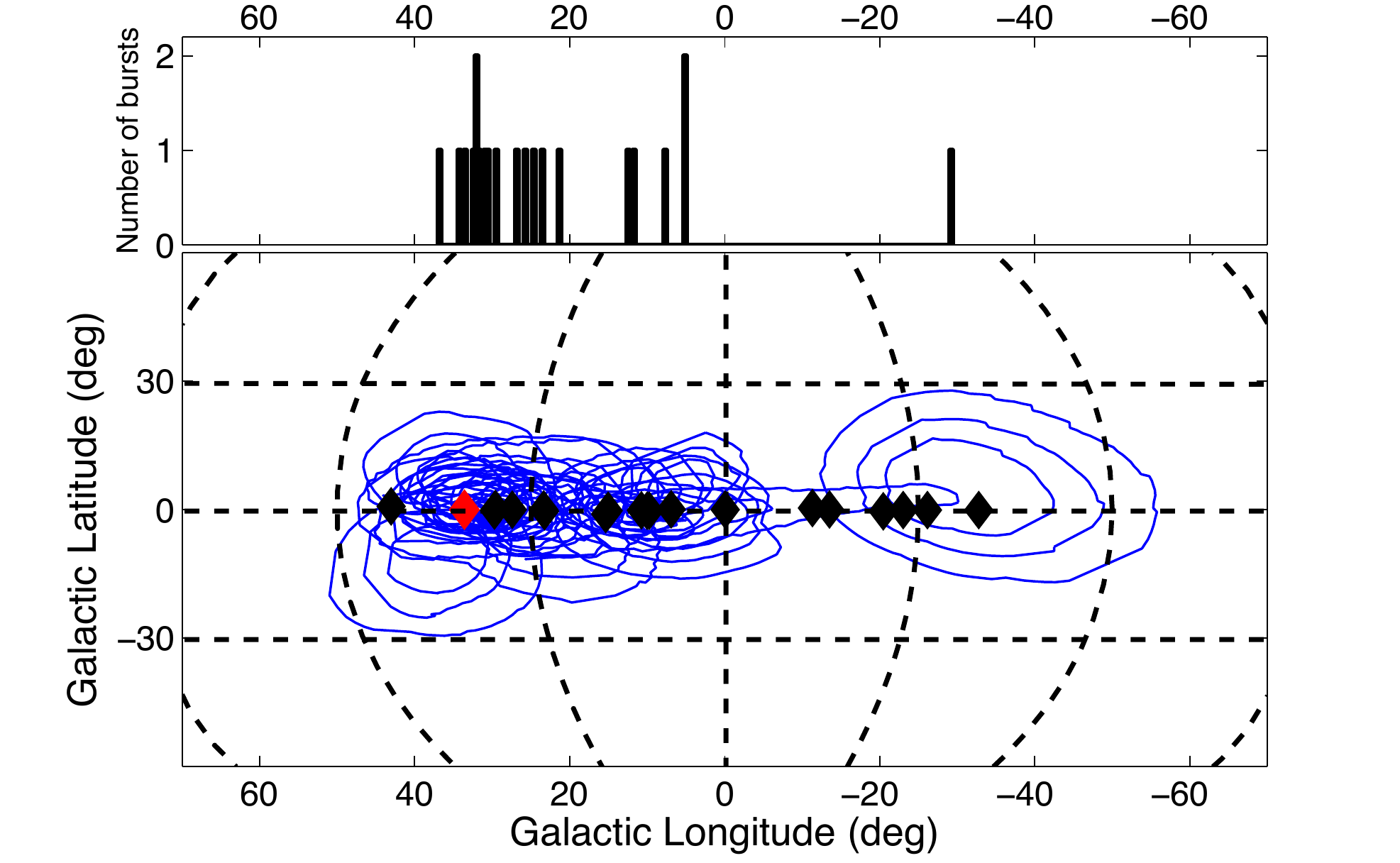} \\
\end{center}
\caption{Locations in galactic coordinates of magnetar-like bursts with unconfirmed origin, displayed as 1, 2, and 3$\sigma$ statistical error contours (blue lines). The known magnetars and magnetar candidates are shown as black diamonds, and the magnetar 3XMM\,J$185246.6+0033.7$ \citep{Zhou13} is highlighted in red. The histogram in the top panel is for the centroids of the GBM location contours.}
\label{unknown_LOCs}
\end{figure}

GBM triggered on 19 bursts with magnetar-like temporal and spectral properties. Several of these bursts, however, were located in the vicinity of the Galactic Center, which is a very crowded region including many confirmed magnetars; see Table~\ref{unknown_table1} for the burst locations. Due to the large error boxes of these bursts, we were not able to identify their origin unambiguously, also because no other instruments with better localization capabilities recorded any of them. Figure~\ref{unknown_LOCs} shows the locations of all 19 bursts with their 1, 2 and 3$\sigma$ statistical error contours together with the locations of the known magnetars and magnetar candidates \citep[][]{O&K13}. All but one of the unknown bursts are in the same general location on the sky. The upper panel of Figure~\ref{unknown_LOCs} displays the distribution of the contour centroids of the burst locations in Galactic longitude. We notice a concentration of 8 events between $30^{\circ}$ and $35^{\circ}$. Recently, \citet{Zhou13} reported the discovery with {\it XMM-Newton} of a new magnetar, 3XMM\,J$185246.6+0033.7$ (red diamond on Figure~\ref{unknown_LOCs}). The source location ($l = 33^{\circ}34^{\prime} 45^{\prime\prime}.1, b = -00^{\circ}02^{\prime} 39^{\prime\prime}.9$) coincides with the highest density of unknown magnetar bursts, strongly suggesting that at least several bursts originate from this source (in particular those detected in 2013). 

For this sample of bursts we have also performed temporal and spectral analyses, of which the results are shown in Tables~\ref{unknown_table1} and \ref{unknown_table2}. Their analysis procedure is the same as we used for the bursts in Section \ref{other_known}, albeit using the locations determined with the GBM data.

\section{Discussion \& Conclusions}
\label{discuss}

We have compiled here the 5-year GBM magnetar burst catalog comprising 446 events of which 19 are from unknown sources and 427 are distributed across seven known sources as shown in Table~\ref{summary}. We display here the results of previously published temporal and spectral analyses  for 357 events and include new results for the remaining 89. All these data are compiled here to provide a single reference catalog to facilitate large scale analyses of magnetar bursts for the community. The capabilities of \Fermi/GBM allow for detailed spectral and temporal characterizations of magnetar bursts, which can be used to further our understanding of these bursts.

\begin{figure}
\begin{center}
\includegraphics[scale=0.4]{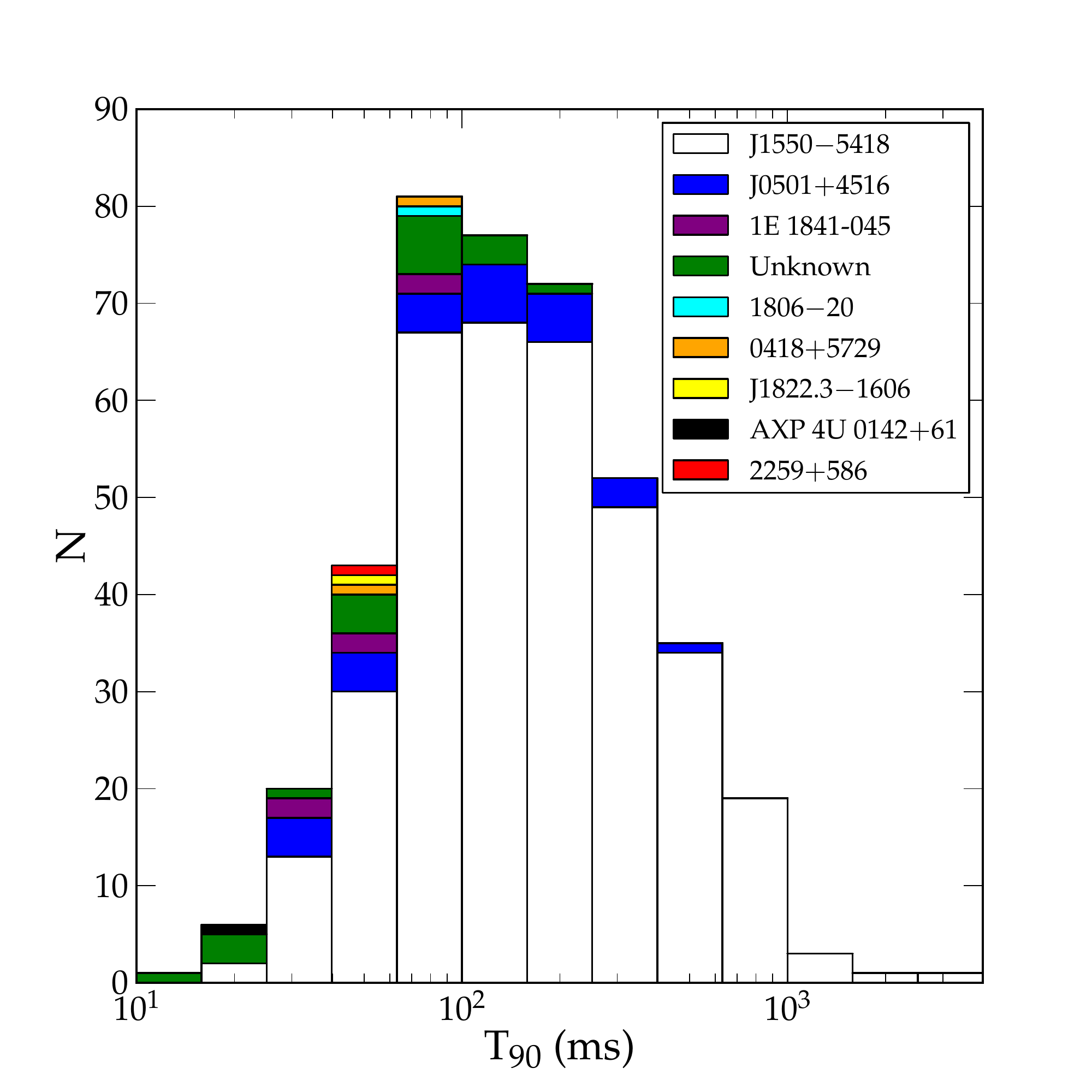} \\
\includegraphics[scale=0.4]{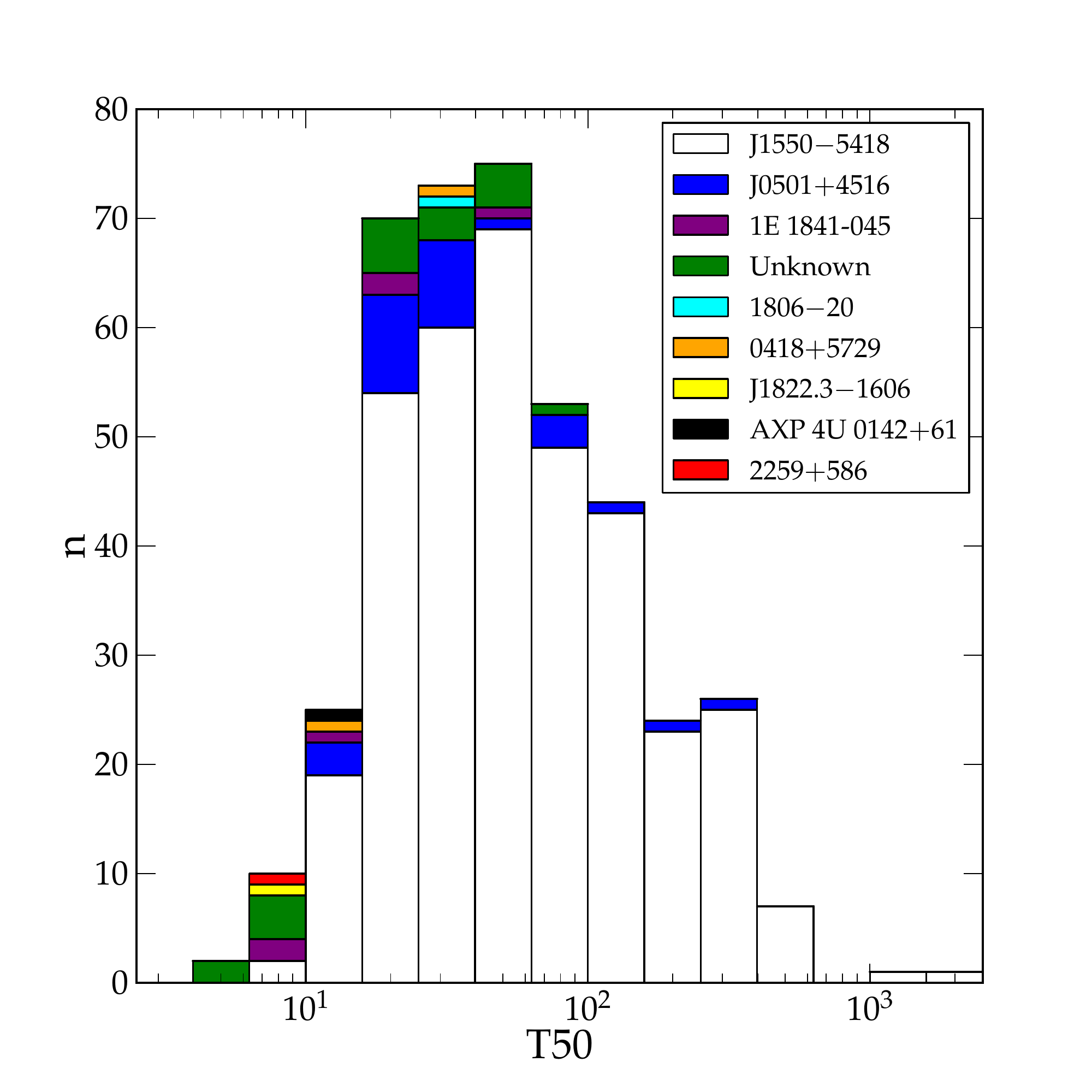}
\end{center}
\caption{Distributions of the \Tnin and \Tfif durations for all magnetar bursts in this catalog.}
\label{allSGRs_duration}
\end{figure}

\begin{figure}
\begin{center}
\includegraphics[scale=0.4]{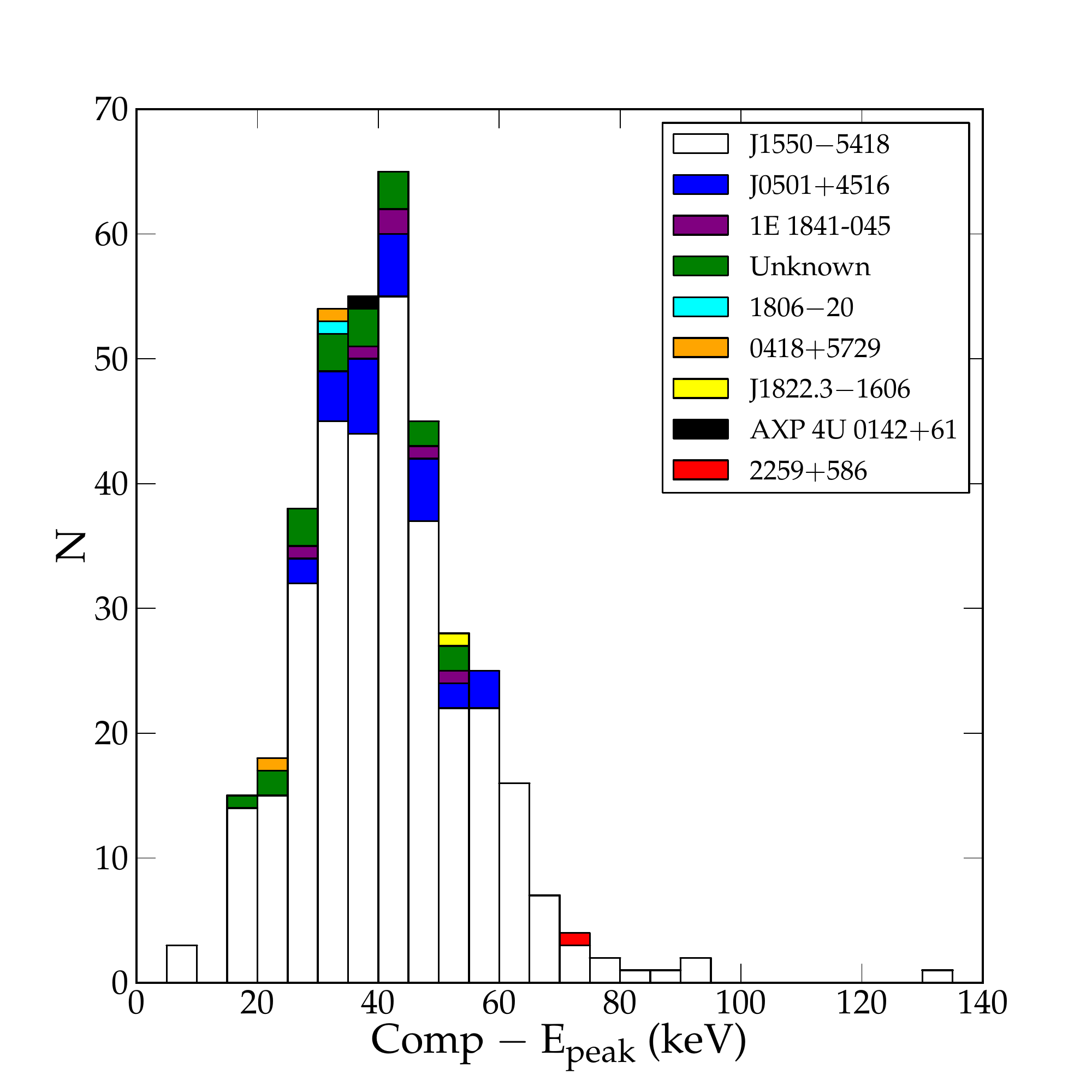}\\ 
\includegraphics[scale=0.4]{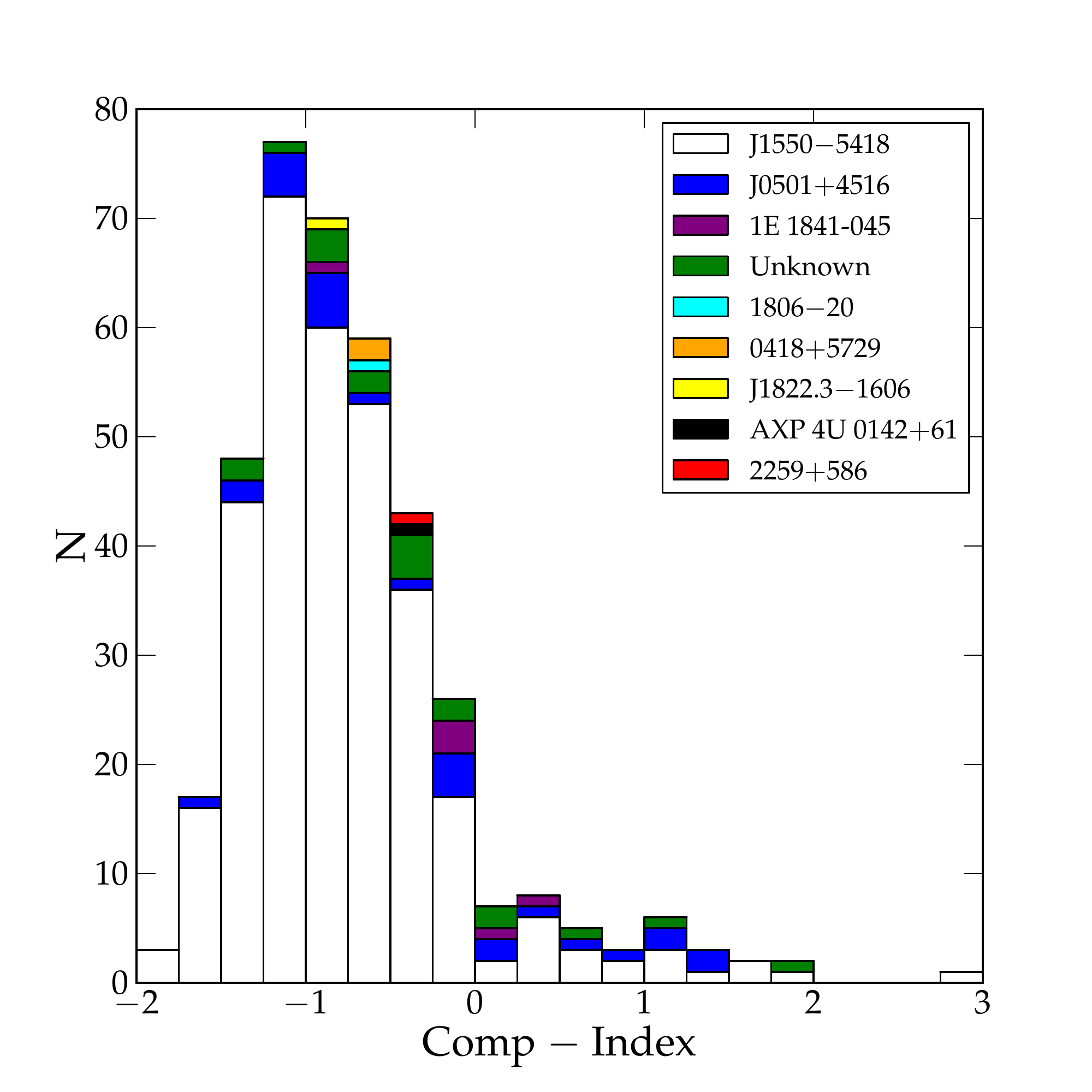}
\end{center}
\caption{Distributions of \Ep and the spectral index of the COMP model for all magnetar bursts in this catalog.}
\label{allSGRs_comp}
\end{figure}

\begin{figure}
\begin{center}
\includegraphics[scale=0.4]{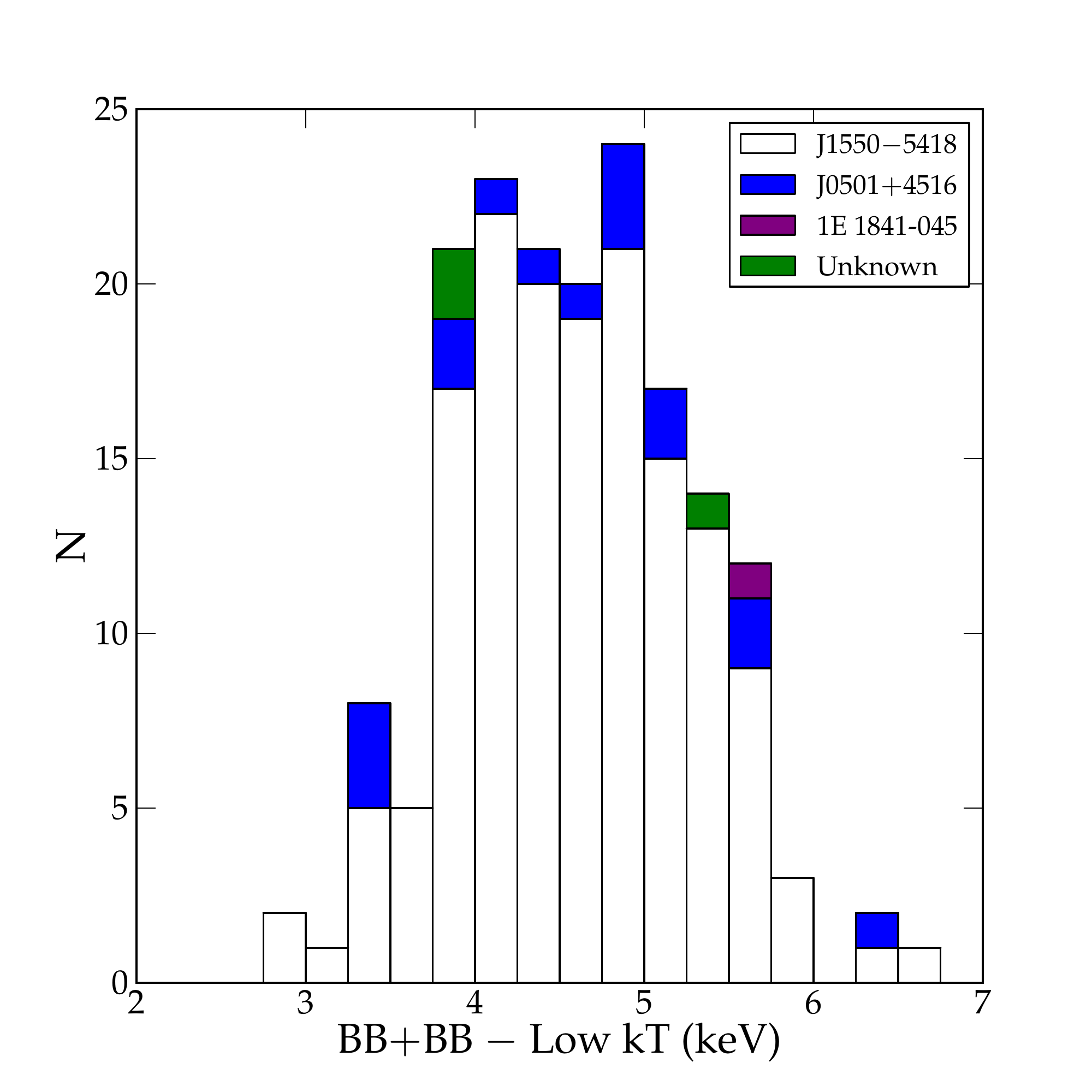} \\
\includegraphics[scale=0.4]{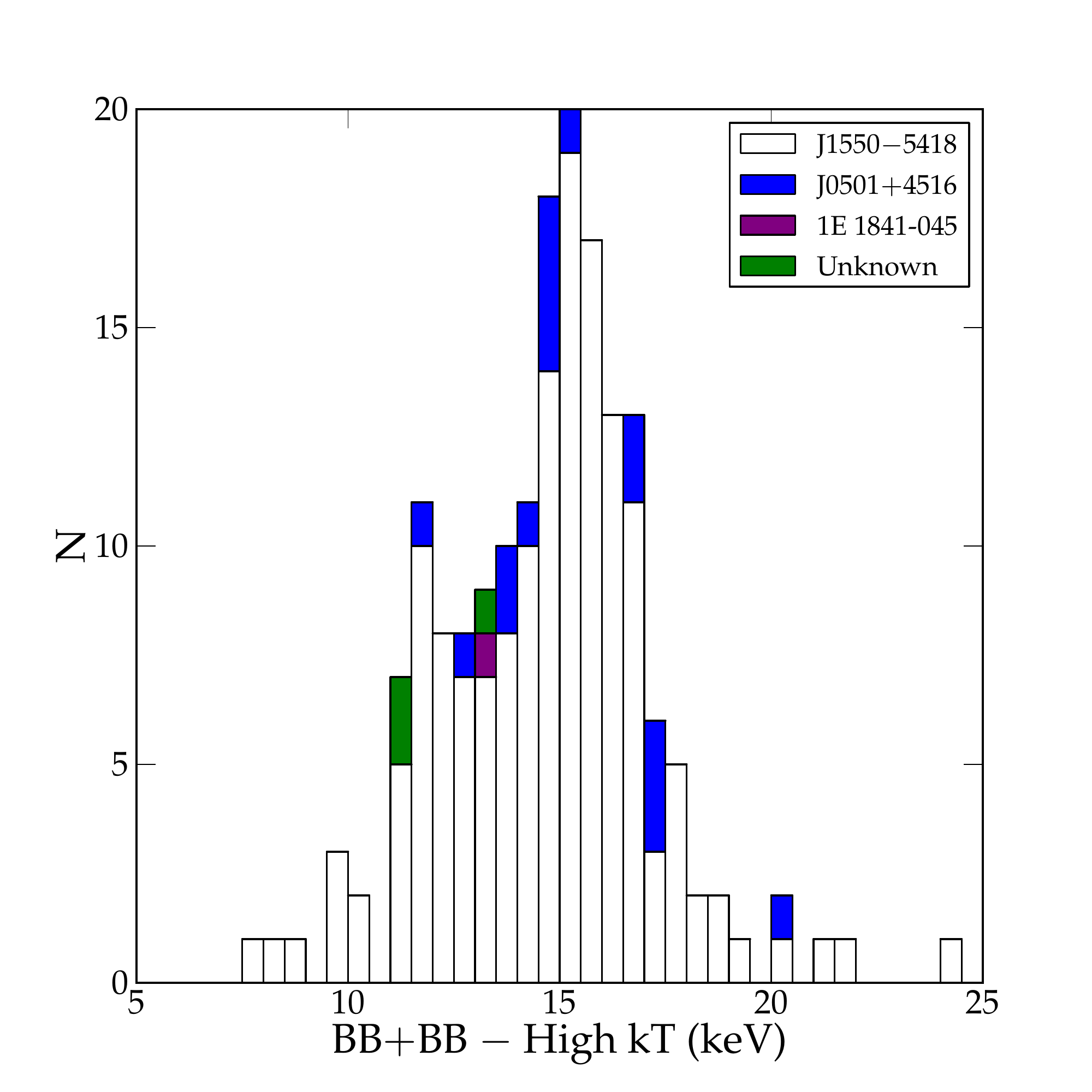}
\end{center}
\caption{Distributions of the low- and high-temperature BB for all magnetar bursts in this catalog.}
\label{allSGRs_bbbb}
\end{figure}

Figures \ref{allSGRs_duration} $-$ \ref{allSGRs_bbbb} combine durations and spectral parameters for all 446 events. Each event is color coded per source, and the 19 unknown magnetar bursts are given the same color. Although \Jfifteen bursts dominate the sample, we note several similarities in the spectral and temporal parameters across the bursts, regardless of their source. With respect to durations, all known source events tend to center around  \Tnin$\sim100$~ms (Figure~\ref{allSGRs_duration}). The unknown event durations tend to be shorter than most magnetar bursts, with an average of $T_{\unit{90}}\sim61$~ms. However, the uncertainties on the individual durations are also considerably high, largely due to the low intensities of some of these bursts, hampering any definitive conclusions on the possible differences in burst durations.

\begin{figure}
\begin{center}
\includegraphics[scale=0.4]{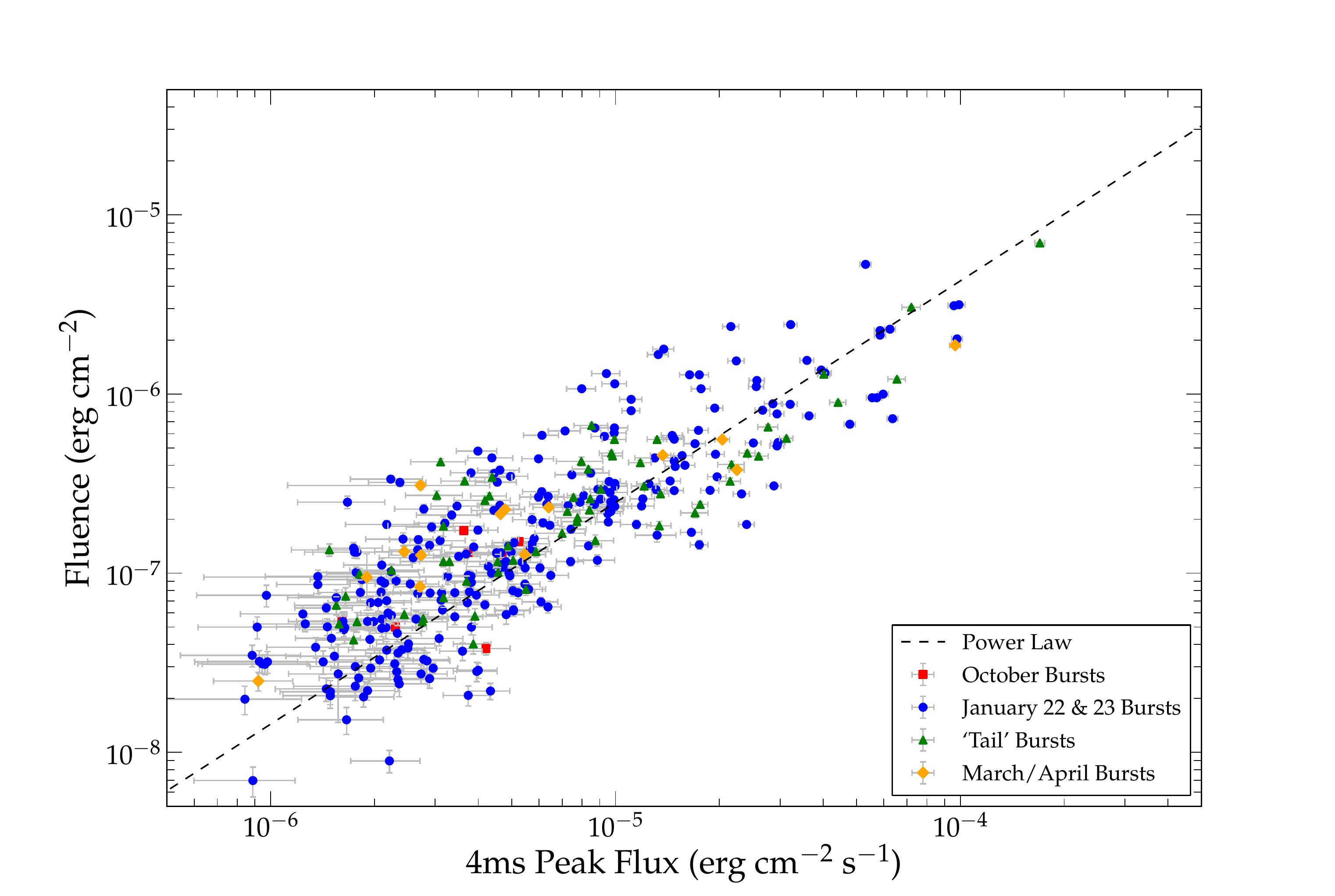}
\end{center}
\caption{Integrated fluence of \Jfifteen bursts versus their peak flux. There is a strong correlation between these two parameters, and a PL fit results in a slope of \valerr{1.2}{0.4}.}
\label{1550_flu_v_pkflux}
\end{figure}

With regards to the spectral properties, the COMP parameters of all known events are very similar, with \Ep centered at $\sim40$~keV (Figure~\ref{allSGRs_comp}). The 19 unknown events have an arithmetic mean of \Ep$\sim37$~keV. The temperatures of the BB$+$BB center around $\sim4.5$~keV and $\sim15$~keV (Figure~\ref{allSGRs_bbbb}). The BB temperatures of the 19 unknown events have arithmetic means of $\sim4.3$~keV and $\sim12.4$~keV, very similar to the ones obtained from the \Jfifteen bursts. For the former we find an average 4~ms peak flux (8$-$200~keV) of \scinum{6.5}{-6}~erg/cm$^2$/s and an average fluence of \scinum{2.3}{-7}~erg/cm$^2$, which places them right in the middle of the flux-fluence diagram of \Jfifteen (see also~Figure \ref{1550_flu_v_pkflux}).

Finally, the location distribution of 8 events of unknown origin strongly indicates an association with the recently discovered magnetar 3XMM\,J$185246.6+0033.7$. A definite association of these bursts is not possible, however, due to the presence of multiple magnetar sources in that region.

\acknowledgements
This publication is part of the GBM/Magnetar Key Project (NASA grant NNH07ZDA001-GLAST, PI: C. Kouveliotou). A.C.C. was supported by an appointment to the NASA Postdoctoral Program at the Marshall Space Flight Center, administered by Oak Ridge Associated Universities through a contract with NASA. C.K. and G.A.Y. acknowledge support from NASA grant NNH07ZDA001-GLAST. D.H. was supported by the Moore-Sloan Data Science Environment at New York University. A.L.W. acknowledges support from a Netherlands Organization for Scientific Research (NWO) Vidi Fellowship. A.v.K. was supported by the Bundesministeriums f\"ur Wirtschaft und
Technologie (BMWi) through DLR grant 50 OG 1101. M.v.d.K. acknowledges support from the Netherlands Organisation for Scientific Research (NWO) and the Royal Netherlands Academy of Arts and Sciences (KNAW).

\clearpage

\clearpage

\begin{deluxetable}{ccc}
\tablewidth{0pc}
\tabletypesize{\scriptsize}
\tablecaption{Summary of GBM magnetar bursts}
\tablehead{\colhead{Source} & \colhead{Burst active periods} & \colhead{Number of bursts} \\
\colhead{} & \colhead{} & \colhead{with TTE data}}
\startdata
\Jfifteen	         &	Oct 2008 $-$ Apr 2009	&	386	\\
\Jzerofive          &	Aug/Sep 2008	&	29      \\
\oneEeight	&	Feb $-$ Jul 2011	&	6	\\
\Jzerofour 	&	Jun 2009	&	2       \\
\SGReighteen  &	Mar 2010	&	1	\\
\Jeighteen	&	Jul 2011	&	1      \\
\AXPfouru 	&	Jul 2011	&	1	\\
\SGRoneE	&	Aug 2011	&	1 	\\
unknown          &	...	&	19     \\
\enddata
\label{summary}
\end{deluxetable}

\begin{landscape}
\begin{deluxetable}{cccccccc}
\tablewidth{0pc}
\tabletypesize{\tiny}
\tablecaption{Durations, phases, peak fluxes, and fluences for \Jfifteen bursts}
\tablehead{\colhead{Burst}  &  \colhead{\Tnin Start}  &  \colhead{\Tnin}  &  \colhead{\Tfif Start}  & \colhead{\Tfif}  &  \colhead{Phase\tablenotemark{c}}  &  \colhead{4ms Peak Flux\tablenotemark{d}\tablenotemark{e}}  &  \colhead{Fluence\tablenotemark{d}\tablenotemark{e}}  \\
(yymmddfff) & (UT) & (ms) & (UT) & (ms) & & (erg/s/cm$^{2}$) & (erg/s/cm$^{2}$) }
\startdata
081003377a\tablenotemark{a}	&	09:02:46.112	&	160 $\pm$ 29	&	09:03:46.128	&	72 $\pm$ 11	&	$-$	&		$-$		&	4.10E-08	$\pm$	5.00E-09	\\
081003377b		&	09:03:05.956	&	276 $\pm$ 53	&	09:03:05.980	&	120 $\pm$ 25	&	$-$	&		$-$		&	1.34E-07	$\pm$	9.00E-09	\\
081003377c\tablenotemark{a}	&	09:03:16.144	&	672 $\pm$ 72	&	09:03:16.176	&	320 $\pm$ 72	&	$-$	&		$-$		&	8.10E-08	$\pm$	9.00E-09	\\
081003377d\tablenotemark{a}	&	09:03:33.172	&	68 $\pm$ 25	&	09:03:33.184	&	28 $\pm$ 11	&	$-$	&		$-$		&	6.50E-08	$\pm$	6.00E-09	\\
081003377e\tablenotemark{a}	&	09:04:35.576	&	128 $\pm$ 23	&	09:04:35.592	&	56 $\pm$ 23	&	$-$	&		$-$		&	7.60E-08	$\pm$	7.00E-09	\\
081003377f\tablenotemark{a}	&	09:04:55.520	&	424 $\pm$ 72	&	09:04:05.544	&	120 $\pm$ 18	&	$-$	&		$-$		&	1.31E-07	$\pm$	1.00E-08	\\
081003377g\tablenotemark{a}	&	09:05:03.504	&	64 $\pm$ 23	&	09:05:03.520	&	24 $\pm$ 18	&	$-$	&		$-$		&	4.50E-08	$\pm$	1.00E-08	\\
081003377h\tablenotemark{a}	&	09:05:03.904	&	104 $\pm$ 34	&	09:05:03.912	&	56 $\pm$ 11	&	$-$	&		$-$		&	3.60E-08	$\pm$	6.00E-09	\\
\enddata
\label{1550_time}
\tablenotetext{a}{Untriggered burst}
\tablenotetext{b}{Saturated burst}
\tablenotetext{c}{Phase of the 4ms peak of the burst}
\tablenotetext{d}{By COMP model}
\tablenotetext{e}{8-200 keV energy range}
\tablenotetext{f}{Used in \citet{Younes14}}
\end{deluxetable}

\begin{deluxetable}{cccccccc}
\tablewidth{0pc}
\tabletypesize{\tiny}
\tablecaption{Time-integrated spectral modeling results for \Jfifteen bursts}
\tablehead{ & OTTB & \multicolumn{2}{c}{COMP} & \multicolumn{2}{c}{BB$+$BB $-$ Low} & \multicolumn{2}{c}{BB$+$BB $-$ High}\\
\colhead{Burst}	 & \colhead{kT} & \colhead{\Ep} & \colhead{$\alpha$} & \colhead{kT} & \colhead{R$^{2}$}	& \colhead{kT} & \colhead{R$^{2}$}\\
 & (keV) & (keV) & & (keV) & (km) & (keV) & (km)}
\startdata
081003377a\tablenotemark{a}	&	44.7$^{+9.1}_{-6.6}$	&	$-$	&	$-$	&	$-$	&		$-$		&	$-$	&		$-$		\\
081003377b	&	52.5$^{+7.2}_{-4.2}$	&	$-$	&	$-$	&	$-$	&		$-$		&	$-$	&		$-$		\\
081003377c\tablenotemark{a}	&	49.5$^{+11.2}_{-8.4}$	&	$-$	&	$-$	&	$-$	&		$-$		&	$-$	&		$-$		\\
081003377d\tablenotemark{a}	&	59.5$^{+10.4}_{-8.8}$	&	$-$	&	$-$	&	$-$	&		$-$		&	$-$	&		$-$		\\
081003377e\tablenotemark{a}	&	56$^{+8.4}_{-6.7}$	&	$-$	&	$-$	&	$-$	&		$-$		&	$-$	&		$-$		\\
081003377f\tablenotemark{a}	&	52.5$^{+7.5}_{-5.4}$	&	$-$	&	$-$	&	$-$	&		$-$		&	$-$	&		$-$		\\
081003377g\tablenotemark{a}	&	54.5$^{+13.4}_{-10}$	&	$-$	&	$-$	&	$-$	&		$-$		&	$-$	&		$-$		\\
081003377h\tablenotemark{a}	&	70.3$^{+30.5}_{-0}$	&	$-$	&	$-$	&	$-$	&		$-$		&	$-$	&		$-$		\\
\enddata
\label{1550_spec}
\tablenotetext{a}{Untriggered burst}
\tablenotetext{b}{Used in \citet{Younes14}}
\end{deluxetable}

\begin{deluxetable}{cccc}
\tablewidth{0pc}
\tabletypesize{\tiny}
\tablecaption{Distribution average and spread for parameters of \Jfifteen bursts}
\tablehead{\colhead{Value} && \colhead{Fit $\mu$} & \colhead{Fit $\sigma$}}
\startdata
\Tnin $\left(\log{\left(\!\unit{ms}\right)}\right)$	&&  	2.19$\pm$0.02 	&  0.40$\pm$0.02 \\
\Tfif   $\left(\log{\left(\!\unit{ms}\right)}\right)$	&&	1.71$\pm$0.04 	& 0.44$\pm$0.03  \\
OTTB $-$ kT  $\left(\!\unit{keV}\right)$		&& 39.8$\pm$0.7  		& 13.7$\pm$0.7  \\
Comp $-$ \Ep $\left(\!\unit{keV}\right)$		&& 	39.6$\pm$ 0.6 		& 12.9$\pm$0.6  \\
Comp $-$ $\alpha$ 						&& -0.93$\pm$0.02 		& 0.45$\pm$0.02  \\
BB$+$BB  low kT 	$\left(\!\unit{keV}\right)$	&& 4.55$\pm$0.05 		& 0.71$\pm$0.05  \\
BB$+$BB high kT 	$\left(\!\unit{keV}\right)$	&& 14.96$\pm$0.16 		& 2.03$\pm$0.16 \\
\enddata
\label{1550_burst_info}
\end{deluxetable}

\begin{deluxetable}{ccc}
\tablewidth{0pc}
\tabletypesize{\tiny}
\tablecaption{Correlation coefficient and probability between parameters of \Jfifteen bursts}
\tablehead{\colhead{Parameters}	&	\colhead{Coefficient}	&	\colhead{Probability}}
\startdata
\Ep $-$ Fluence	&	-0.400	&	\scinum{9.7}{-14}	\\
\Tnin $-$ Fluence	&	0.194	&	\scinum{2.5}{-4}	\\
\Tfif $-$ Fluence	&	0.086	&	\scinum{1.1}{-1}	\\
Comp Index $-$ Fluence	&	-0.082	&	\scinum{1.4}{-1}	\\
\\						
\Ep $-$ Peak Flux	&	-0.463	&	\scinum{1.8}{-18}	\\
\Tnin $-$ Peak Flux	&	-0.084	&	\scinum{1.2}{-1}	\\
\Tfif $-$ Peak Flux	&	-0.240	&	\scinum{7.1}{-6}	\\
Comp Index $-$ Peak Flux	&	-0.010	&	\scinum{8.6}{-1}\\
\\						
Peak Flux $-$ Phase	&	-0.013	&	\scinum{8.1}{-1}	\\	
Fluence $-$ Phase	&	0.039	&	\scinum{4.8}{-1}	\\	
\Ep $-$ Phase	&	0.127	&	\scinum{2.3}{-2}	\\	
\Tnin $-$ Phase	&	0.085	&	\scinum{1.2}{-1}	\\	
\\						
\Tnin $-$ \Ep	&	0.001	&	\scinum{9.9}{-1}	\\
\Tfif $-$ \Ep	&	0.112	&	\scinum{4.6}{-2}	\\
\\						
Fluence $-$ Peak Flux	&	0.831	&	\scinum{3.0}{-89}	\\
\enddata
\label{1550_cor_table}
\end{deluxetable}

\begin{deluxetable}{ccccccc}
\tablewidth{0pc}
\tabletypesize{\tiny}
\tablecaption{Durations, peak fluxes, and fluences of \Jzerofive bursts}
\tablehead{\colhead{Trigger} & \colhead{\Tnin Start} & \colhead{\Tnin} & \colhead{\Tfif Start} & \colhead{\Tfif} & \colhead{8ms Peak Flux\tablenotemark{c}\tablenotemark{d}} & \colhead{Fluence\tablenotemark{c}\tablenotemark{d}}\\
(yymmddfff) & (UT) & (ms) & (UT) & (ms) & (erg/s/cm$^{2}$) & (erg/cm$^{2}$)}
\startdata		
080822529a	&	12:41:56.902	&	80 $\pm$ 16	&	12:41:56.902	&	38 $\pm$ 7	&	1.59E-06	$\pm$	3.20E-07	&	7.05E-08	$\pm$	6.20E-09	\\
080822647a	&	15:41:17.712	&	226 $\pm$ 24	&	15:41:17.712	&	24 $\pm$ 24	&	7.04E-06	$\pm$	6.70E-07	&	1.93E-07	$\pm$	1.42E-08	\\
080822981a	&	23:32:57.728	&	30 $\pm$ 15	&	23:32:57.728	&	12 $\pm$ 4	&	2.37E-06	$\pm$	5.20E-07	&	4.41E-08	$\pm$	6.70E-09	\\
080823020a	&	00:28:09.908	&	48 $\pm$ 7	&	00:28:09.908	&	20 $\pm$ 7	&	6.61E-06	$\pm$	5.90E-07	&	2.50E-07	$\pm$	1.12E-08	\\
080823091a	&	02:11:36.694	&	554 $\pm$ 40	&	02:11:36.694	&	386 $\pm$ 7	&	6.49E-06	$\pm$	7.40E-07	&	8.28E-07	$\pm$	3.04E-08	\\
080823174a	&	04:10:19.222	&	330 $\pm$ 51	&	04:10:19.222	&	142 $\pm$ 7	&	1.63E-06	$\pm$	3.70E-07	&	1.41E-07	$\pm$	1.29E-08	\\
080823248a	&	05:56:31.487	&	276 $\pm$ 34	&	05:56:31.487	&	34 $\pm$ 7	&	3.42E-06	$\pm$	6.40E-07	&	2.22E-07	$\pm$	1.80E-08	\\
080823293a	&	07:01:09.971	&	164 $\pm$ 7	&	07:01:09.971	&	82 $\pm$ 3	&	2.89E-06	$\pm$	3.40E-07	&	2.01E-07	$\pm$	9.00E-09	\\
\enddata
\label{0501_durations}
\tablenotetext{a}{Untriggered Burst}
\tablenotetext{b}{Saturated Burst}
\tablenotetext{b}{by COMP model}
\tablenotetext{c}{8 - 200 keV energy range}
\end{deluxetable}

\begin{deluxetable}{ccccccccc}
\tablewidth{0pc}
\tabletypesize{\tiny}
\tablecaption{Time-integrated spectral modeling results for \Jzerofive bursts}
\tablehead{ & \multicolumn{2}{c}{COMP} & OTTB & BB & \multicolumn{2}{c}{BB$+$BB} \\
\colhead{Trigger ID} & \colhead{\Ep} & \colhead{$\alpha$} & \colhead{kT} & \colhead{kT} & \colhead{kT1} & \colhead{kT2} \\
& (keV) & & (keV) & (keV) & (keV) & (keV) }
\startdata
080822529a	&	40.83 $\pm$ 2.76	&	1.06	$\pm$ 0.70	&	44.65 $\pm$ 7.08	&	10.33 $\pm$ 0.60	&		$-$		         &		$-$		\\	
080822647a	&	39.83 $\pm$ 5.98	&	-1.32 $\pm$ 0.34	&	40.95 $\pm$ 4.32	&	9.85	$\pm$ 0.42	&	4.24	$\pm$ 0.52	&	17.24 $\pm$ 2	\\	
080822981a	&	44.86 $\pm$ 5.02	&	1.48	$\pm$ 1.42	&	50.89 $\pm$ 16.3	&	11.62 $\pm$ 1.17	&		$-$		         &		$-$		\\	
080823020a	&	36.37 $\pm$ 3.28	&	-1.27	 $\pm$ 0.20	&	37.85 $\pm$ 2.35	&	9.53	$\pm$ 0.23	&	4.66	$\pm$ 0.34	&	16.99 $\pm$ 1.21	\\	
080823091a	&	42.09 $\pm$ 2.63	&	-1.17	 $\pm$ 0.17	&	42.53 $\pm$ 2.27	&	10.27 $\pm$ 0.22	&	5.06	$\pm$ 0.39	&	17.40 $\pm$ 1.08	\\	
080823174a	&	57.53 $\pm$ 7.06	&	-0.52	 $\pm$ 0.44	&	61.65 $\pm$ 9.66	&	12.86 $\pm$ 0.76	&		$-$		         &		$-$		\\	
080823248a	&	51.94 $\pm$ 3.41	&	1.03	$\pm$ 0.55	&	68.30 $\pm$ 11.1	&	13.00 $\pm$ 0.68	&		$-$		         &   		$-$		\\	
080823293a	&	48.13 $\pm$ 1.85	&	0.52	$\pm$ 0.29	&	53.43 $\pm$ 4.12	&	11.82 $\pm$ 0.34	&		$-$		         &		$-$		\\
\enddata
\label{0501_Spec}
\tablenotetext{a}{Untriggered Burst}
\end{deluxetable}

\begin{deluxetable}{ccccccc}
\tablewidth{0pc}
\tabletypesize{\tiny}
\tablecaption{Durations, peak fluxes, and fluences of \oneEeight bursts}
\tablehead{\colhead{Trigger} & \colhead{\Tnin Start} & \colhead{\Tnin\tablenotemark{a}} & \colhead{\Tfif Start} & \colhead{\Tfif\tablenotemark{a}} & \colhead{8ms Peak Flux\tablenotemark{b}\tablenotemark{c}} & \colhead{Fluence\tablenotemark{b}\tablenotemark{c}}\\
(yymmddfff) & (UT) & (ms) & (UT) & (ms) & (erg/s/cm$^{2}$) & (erg/cm$^{2}$)}
\startdata
110209218a	&	05:14:25.930	&	72 $\pm$ 23	&	05:14:25.934	&	16 $\pm$ 11	&	3.12E-06	$\pm$  5.34E-07	&	5.1E-08  $\pm$  4E-09	\\
110217330a	&	07:55:55.241	&	72 $\pm$ 30	&	07:55:55.265	&	42 $\pm$ 3	&	2.91E-06	$\pm$  4.62E-07	&	8.4E-08  $\pm$  5E-09	\\
110221029a	&	00:41:16.224	&	28 $\pm$ 11	&	00:41:16.242	&	8 $\pm$ 6		&	7.93E-06	$\pm$  6.46E-07	&	1E-07     $\pm$  1E-08	\\
110616881a	&	21:09:08.420	&	40 $\pm$ 6	&	21:09:08.428	&	20 $\pm$ 6	&	1.29E-05	$\pm$  1.15E-06	&	2.9E-07  $\pm$  1E-08	\\
110623612a	&	14:41:42.752	&	32 $\pm$ 14	&	14:41:42.756	&	12 $\pm$ 6	&	1.11E-05	$\pm$  1.06E-06	&	1.9E-07  $\pm$  1E-08	\\
110625969a	&	23:26:03.165	&	56 $\pm$ 11	&	23:26:03.167	&	8 $\pm$ 11	&	9.50E-06	$\pm$  1.00E-06	&	1.1E-07  $\pm$  1E-08	\\
\enddata
\label{1841_durations}
\tablenotetext{a}{8 - 100 keV energy range.}
\tablenotetext{b}{by COMP model}
\tablenotetext{c}{8 - 200 keV energy range}
\end{deluxetable}

\begin{deluxetable}{cccccccc}
\tablewidth{0pc}
\tabletypesize{\tiny}
\tablecaption{Time-integrated spectral modeling results for \oneEeight bursts}
\tablehead{ & \multicolumn{2}{c}{COMP} & OTTB &  \multicolumn{2}{c}{BB$+$BB} \\
\colhead{Trigger ID} & \colhead{\Ep} & \colhead{$\alpha$} & \colhead{kT} & \colhead{kT1} & \colhead{kT2} \\
& (keV) & & (keV) & (keV) & (keV)}
\startdata
110209218a	&	51$^{+5}_{-4}$	&	-0.19$^{+0.45}_{-0.41}$	&	60.93 $\pm$ 9.80	&	$-$	&	$-$	\\
110217330a	&	45$^{+3}_{-2}$	&	0.44$^{+0.44}_{-0.4}$	&	41.48 $\pm$ 4.35	&	$-$	&	$-$	\\
110221029a	&	41$^{+2}_{-2}$	&	0.11$^{+0.29}_{-0.27}$	&	43.24 $\pm$ 3.36	&	5.6 $\pm$ 1.1	&	13.1 $\pm$ 1.2	\\
110616881a	&	28$^{+2}_{-2}$	&	-0.9$^{+0.2}_{-0.2}$		&	27.94 $\pm$ 1.37	&	$-$	&	$-$	\\
110623612a	&	40$^{+2}_{-2}$	&	-0.11$^{+0.27}_{-0.26}$	&	38.42 $\pm$ 2.92	&	$-$	&	$-$	\\
110625969a	&	37$^{+2}_{-2}$	&	-0.04$^{+0.37}_{-0.35}$	&	34.88 $\pm$ 3.66	&	$-$	&	$-$	\\
\enddata
\label{1841_Spec}
\end{deluxetable}

\begin{deluxetable}{ccccccccccc}
\tablewidth{0pc}
\tabletypesize{\tiny}
\tablecaption{Locations, durations, peak fluxes, and fluences for bursts of known sources}
\tablehead{\colhead{Trigger} & \colhead{Source} & \multicolumn{2}{c}{Location\tablenotemark{a}} & \colhead{\Tnin Start} & \colhead{\Tnin} & \colhead{\Tfif Start} & \colhead{\Tfif} & \colhead{4ms Peak Flux\tablenotemark{b}\tablenotemark{c}} & \colhead{Fluence\tablenotemark{b}\tablenotemark{c}}\\
(yymmddfff) & & RA & Dec. & (UT) & (ms) & (UT) & (ms) & (erg/cm$^{2}$) & (erg/cm$^{2}$)}
\startdata	
090605862a	&	\Jzerofour 	&	04h 18m 33.867s	&	+57$^{\circ}$ 32' 22.91''	&	20:30:48.875	&	40 $\pm$ 7	&	20:30:48.881	&	10 $\pm$ 4	&	1.00E-05	$\pm$	5.80E-07	&	9.80E-08	$\pm$	3.92E-09	\\
090605876a	&	\Jzerofour 	&	04h 18m 33.867s	&	+57$^{\circ}$ 32' 22.91''	&	21:01:35.011	&	80 $\pm$ 6	&	21:01:35.019	&	34 $\pm$ 4	&	1.71E-06	$\pm$	1.50E-07	&	2.94E-08	$\pm$	1.72E-09	\\
100326867a	&	\SGReighteen	&	18h 08m 39.337s	&	-20$^{\circ}$ 24' 39.85'' 	&	20:48:59.836	&	64 $\pm$ 9	&	20:48:59.848	&	32 $\pm$ 6	&	1.99E-06	$\pm$	3.59E-07	&	8.44E-08	$\pm$	5.46E-09	\\
110714533a	&	\Jeighteen	&	18h 22m 18.00s	&	-16$^{\circ}$ 04' 26.8'' 	&	12:47:54.368	&	40 $\pm$ 18	&	12:47:54.368	&	8 $\pm$ 11	&	8.66E-06	$\pm$	6.42E-07	&	5.81E-08	$\pm$	4.52E-09	\\
110729737a	&	\AXPfouru 	&	01h 46m 22.407s	&	+61$^{\circ}$ 45' 03.19''	&	17:40:37.110	&	20 $\pm$ 6	&	17:40:37.114	&	12 $\pm$ 6	&	6.50E-06	$\pm$	1.08E-06	&	1.33E-07	$\pm$	9.72E-09	\\
120421346a	&	\SGRoneE	&	23h 01m 08.295s	&	+58${\circ}$ 52' 44.45'' 	&	08:17:43.694	&	40 $\pm$ 33	&	08:17:43.694	&	8 $\pm$ 11	&	6.71E-06	$\pm$	7.91E-07	&	5.73E-08	$\pm$	5.40E-09	\\
130617411a	&	\Jfifteen	         &	15h 50m 54.12386s	&	-54$^{\circ}$ 18' 24.1141'' 	&	09:52:30.276	&	156 $\pm$ 16	&	09:52:30.300	&	88 $\pm$ 11	&	1.72E-06	$\pm$	3.76E-07	&	1.27E-07	$\pm$	6.66E-09	\\
\enddata
\label{known_table1}
\tablenotetext{a}{Source: \citet{O&K13}}
\tablenotetext{b}{by COMP model}
\tablenotetext{c}{8-200 keV energy range}
\end{deluxetable}

\begin{deluxetable}{cccccc}
\tablewidth{0pc}
\tabletypesize{\tiny}
\tablecaption{Time-integrated spectral modeling results for bursts of known sources}
\tablehead{ & & OTTB & \multicolumn{2}{c}{COMP} & BB \\
\colhead{Trigger ID} & \colhead{Source} & \colhead{kT} & \colhead{\Ep} & \colhead{$\alpha$} & \colhead{kT}\\
(yymmddfff) & (keV) & (keV) & & (keV)}
\startdata		
090605862a	&	\Jzerofour 	&	33.46 $\pm$ 2.23	&	34.72 $\pm$ 1.85	&	-0.510 $\pm$ 0.260	&	8.70 $\pm$ 0.25	\\
090605876a	&	\Jzerofour 	&	19.71 $\pm$ 1.96	&	21.39 $\pm$ 2.55	&	-0.660 $\pm$ 0.520	&	6.06 $\pm$ 0.27	\\
100326867a	&	\SGReighteen	&	32.79 $\pm$ 3.19	&	34.23 $\pm$ 2.84	&	-0.600 $\pm$ 0.368	&	8.64 $\pm$ 0.39	\\
110714533a	&	\Jeighteen	&	50.26 $\pm$ 6.14	&	50.06 $\pm$ 5.80	&	-0.929 $\pm$ 0.357	&	50.26 $\pm$ 6.14	\\	
110729737a	&	\AXPfouru 	&	36.96 $\pm$ 4.17	&	38.46 $\pm$ 2.98	&	-0.324 $\pm$ 0.410	&	9.92 $\pm$ 0.44	\\	
120421346a	&	\SGRoneE	&	85.55 $\pm$ 16.70	&	74.30 $\pm$ 10.70	&	-0.469 $\pm$ 0.412	&	16.48 $\pm$ 0.96	\\	
130617411a	&	\Jfifteen	         &	63.05 $\pm$ 6.78	&	50.78 $\pm$ 1.98	&	1.547 $\pm$ 0.416	&	13.12 $\pm$ 0.47	\\	
\enddata
\label{known_table2}
\end{deluxetable}

\begin{deluxetable}{cccccccccccc}
\tablewidth{0pc}
\tabletypesize{\tiny}
\tablecaption{Locations, durations, peak fluxes, and fluences for bursts of unknown sources}
\tablehead{& \multicolumn{3}{c}{Location $-$ Galactic} & \multicolumn{2}{c}{Location $-$ Galactic} & & & & & & \\
\colhead{Trigger ID} & \colhead{RA} & \colhead{Dec} & \colhead{Err} & \colhead{$ l$} & \colhead{b} & \colhead{\Tnin Start} & \colhead{\Tnin} & \colhead{\Tfif - Start} & \colhead{\Tfif} & \colhead{4ms Peak Flux\tablenotemark{a}\tablenotemark{b}}  &	\colhead{Fluence\tablenotemark{a.b}} \\
(yymmddfff) & (Deg)  & (Deg) & (Deg) & (Deg) & (Deg) & (UT) & (ms) & (UT) & (ms) & (erg/s/cm$^{2}$) & (erg/cm$^{2}$) }
\startdata
081106241a	&	268.0	&	-15.3	&	5.2	&	12.5	&	5.8	&	05:47:04.036	&	56 $\pm$ 11	&	05:47:04.052	&	24 $\pm$ 18	&	1.83E-06	$\pm$	3.47E-07	&	7.08E-08	$\pm$	4.90E-09	\\
081129027a	&	272.3	&	-23.0	&	4.9	&	7.8	&	-1.6	&	00:38:12.655	&	116 $\pm$ 16	&	00:38:12.675	&	40 $\pm$ 6	&	3.06E-06	$\pm$	6.51E-07	&	1.66E-07	$\pm$	8.00E-08	\\
090804852a	&	273.4	&	-19.2	&	3.7	&	11.6	&	-0.7	&	20:26:41.019	&	96 $\pm$ 33	&	20:26:41.027	&	28 $\pm$ 9	&	4.40E-06	$\pm$	4.81E-07	&	1.42E-07	$\pm$	6.43E-09	\\
101114746a	&	264.7	&	-21.8	&	4.2	&	5.3	&	5.1	&	17:54:26.717	&	64 $\pm$ 25	&	17:54:26.725	&	8 $\pm$ 11	&	3.68E-06	$\pm$	4.82E-07	&	6.38E-08	$\pm$	4.99E-09	\\
110107970a	&	271.6	&	-25.6	&	7.6	&	5.2	&	-2.3	&	23:16:28.380	&	80 $\pm$ 36	&	23:16:28.380	&	32 $\pm$ 23	&	1.86E-06	$\pm$	3.42E-07	&	5.02E-08	$\pm$	5.46E-09	\\
110531601a	&	279.6	&	1.0	&	6.4	&	32.3	&	3.3	&	14:25:54.672	&	44 $\pm$ 23	&	14:25:54.680	&	16 $\pm$ 11	&	4.18E-06	$\pm$	4.96E-07	&	8.35E-08	$\pm$	5.32E-09	\\
110807970a	&	276.9	&	-6.3	&	7.0	&	24.6	&	2.4	&	23:16:24.848	&	96 $\pm$ 23	&	23:16:24.864	&	48 $\pm$ 23	&	2.72E-06	$\pm$	5.77E-07	&	1.03E-07	$\pm$	7.38E-09	\\
110810685a	&	285.8	&	-7.4	&	7.4	&	27.6	&	-6.0	&	16:26:28.188	&	40 $\pm$ 18	&	16:26:28.196	&	16 $\pm$ 11	&	2.04E-06	$\pm$	3.86E-07	&	4.27E-08	$\pm$	3.90E-09	\\
\enddata
\label{unknown_table1}
\tablenotetext{a}{By COMP model.}
\tablenotetext{b}{8 - 200 keV energy range}
\tablenotetext{c}{This burst may be associated with the magnetar J1834.9$-$0846.  However, due to large error on the GBM location, and the crowdedness of the region, we cannot definitively associate it with this source.}
\end{deluxetable}

\begin{deluxetable}{cccccccc}
\tablewidth{0pc}
\tabletypesize{\tiny}
\tablecaption{Time-integrated spectral modeling results for bursts of unknown sources}
\tablehead{ & PL & OTTB & \multicolumn{2}{c}{COMP} & BB & \multicolumn{2}{c}{BB$+$BB} \\
\colhead{TRIGGER}	&	\colhead{$\alpha$}	 &	\colhead{kT} &	\colhead{Ep} & \colhead{$\alpha$} & \colhead{kT} &	\colhead{low kT} & \colhead{high kT}			\\
& & (keV) & (keV) & & (keV) & (keV) & (keV) }
\startdata
081106241a	&	-2.380 $\pm$ 0.094	&	25.73 $\pm$ 2.52	&	23.64 $\pm$ 4.63	&	-1.281 $\pm$ 0.354	&	7.34 $\pm$ 0.33	&	$-$	&	$-$	\\
081129027a	&	-2.335 $\pm$ 0.097	&	26.67 $\pm$ 2.65	&	25.20 $\pm$ 4.17	&	-1.188 $\pm$ 0.367	&	7.34 $\pm$ 0.33	&	$-$	&	$-$	\\
090804852a	&	-2.461 $\pm$ 0.071	&	22.58 $\pm$ 1.49	&	24.66 $\pm$ 1.68	&	-0.533 $\pm$ 0.321	&	6.81 $\pm$ 0.21	&	$-$	&	$-$	\\
101114746a	&	-2.280 $\pm$ 0.105	&	31.49 $\pm$ 3.60	&	32.55 $\pm$ 3.56	&	-0.773 $\pm$ 0.443	&	8.70 $\pm$ 0.45	&	$-$	&	$-$	\\
110107970a	&	-2.491 $\pm$ 0.153	&	21.67 $\pm$ 3.15	&	17.33 $\pm$ 8.80	&	-1.446 $\pm$ 0.599	&	6.80 $\pm$ 0.45	&	$-$	&	$-$	\\
110531601a	&	-2.056 $\pm$ 0.087	&	42.00 $\pm$ 4.41	&	43.15 $\pm$ 2.55	&	0.179 $\pm$ 0.388	&	10.98 $\pm$ 0.45	&	$-$	&	$-$	\\
110807970a\tablenotemark{a}	&	-1.770 $\pm$ 0.085	&	63.81 $\pm$ 8.36	&	53.54 $\pm$ 3.41	&	0.532 $\pm$ 0.436	&	13.15 $\pm$ 0.62	&	$-$	&	$-$	\\
110810685a	&	-1.936 $\pm$ 0.120	&	47.71 $\pm$ 7.69	&	44.42 $\pm$ 2.84	&	1.778 $\pm$ 0.284	&	47.71 $\pm$ 7.69	&	$-$	&	$-$	\\
\enddata 
\label{unknown_table2}
\tablenotetext{a}{This burst may be associated with the magnetar J1834.9$-$0846.  However, due to large error on the GBM location, and the crowdedness of the region, we cannot definitively associate it with this source.}
\end{deluxetable}
\end{landscape}

{}

\end{document}